\documentclass[12pt,a4paper,reqno]{article} 
\pdfoutput=1 
\usepackage{amssymb,amsthm}
\usepackage{amsmath}
\usepackage{xcolor}
\usepackage{graphicx}
\usepackage[colorlinks,hyperfootnotes=false]{hyperref} 
\usepackage[all]{hypcap}	
\hypersetup{pdfauthor={Nathan Haouzi and Christian Schmid},pdftitle={Little String Origin of Surface Defects}, pdfkeywords={Surface Defects, Little Strings}, pdfdisplaydoctitle={true},bookmarksopen={true}} 
\usepackage{url}
\usepackage{epstopdf}
\usepackage{microtype}
\usepackage{subcaption}
\usepackage{booktabs}
\usepackage{dsfont} 
\usepackage{tikz}
\usepackage{ytableau} 
\usepackage{mciteplus} 
\usepackage{setspace}
\usepackage[skins]{tcolorbox}

\usetikzlibrary{shapes.misc}
\usetikzlibrary{calc}
\usetikzlibrary{decorations.pathreplacing}

\newcommand{\beq}{\begin{equation}}
\newcommand{\eeq}{\end{equation}}

\newcommand{\RR}{\mathbb{R}}

\newcommand{\1}{\mathds{1}}
\newcommand{\fb}{\mathfrak{b}}
\newcommand{\fg}{\mathfrak{g}}
\newcommand{\fh}{\mathfrak{h}}
\newcommand{\fl}{\mathfrak{l}}
\newcommand{\fm}{\mathfrak{m}}
\newcommand{\fn}{\mathfrak{n}}
\newcommand{\fp}{\mathfrak{p}}
\newcommand{\fs}{\mathfrak{s}}
\newcommand{\fR}{\mathfrak{R}}
\newcommand{\cC}{\mathcal{C}}
\newcommand{\cN}{\mathcal{N}}
\newcommand{\cW}{\mathcal{W}}
\newcommand{\cO}{\mathcal{O}}

\newcommand{\cS}{\mathcal{S}}

\def\normOrd#1{\mathop{:}\nolimits\!#1\!\mathop{:}\nolimits}	
\newcommand{\tikzmark}[1]{\tikz[overlay,remember picture] \node (#1) {};} 

\newcommand*\xbar[1]{%
  \hbox{%
    \vbox{%
      \hrule height 0.5pt 
      \kern0.3ex
      \hbox{%
        \kern-0.1em
        \ensuremath{#1}%
        \kern-0.05em
      }%
    }
  }
}

\newtheoremstyle{fullit}
  {\topsep}      
  {\topsep}      
  {\normalfont}  
  {0pt}          
  {\itshape}     
  {.\ }          
  {0pt}          
  {\thmname{#1} \thmnumber{#2}}             
  
\newtheorem{thm}{Theorem}[section]

\theoremstyle{definition}

\theoremstyle{fullit}
\newtheorem{example}[thm]{Example}

\DeclareCaptionSubType*[alph]{figure}
\tikzset{cross/.style={cross out, draw=black, minimum size=2*(#1-\pgflinewidth), inner sep=0pt, outer sep=0pt},
cross/.default={1pt}}
\definecolor{urlssc}{HTML}{161680}
\definecolor{linksc}{HTML}{80164B}
\definecolor{citesc}{HTML}{16804B}
\hypersetup{
  linkcolor  = linksc!80!black,
  citecolor  = citesc!80!black,
  urlcolor   = urlssc,
  colorlinks = true,
}

\begin{document}
\setlength{\parindent}{0pt}
\vspace*{-1.5cm}

\vspace{1.5cm}
\begin{center}
{\LARGE
Little String Origin of Surface Defects
}
\vspace{0.4cm}

\end{center}

\vspace{0.35cm}
\begin{center}
Nathan Haouzi\footnote{Email: \href{mailto:nathanhaouzi@berkeley.edu}{nathanhaouzi@berkeley.edu}} and Christian Schmid\footnote{Email: \href{mailto:cschmid@berkeley.edu}{cschmid@berkeley.edu}}
\end{center}

\vspace{0.1cm}
\begin{center} 
\emph{Center for Theoretical Physics\\ 
University of California, Berkeley, USA}
\vspace{0.25cm}

\vspace{0.2cm}

 \vspace{0.5cm} 
\end{center} 

\vspace{1cm}


\begin{abstract}
\noindent	
We derive the codimension-two defects  of 4d $\mathcal{N}=4$ Super Yang-Mills (SYM) theory from the $(2,0)$ little string. The origin of the little string is type IIB theory compactified on an $ADE$ singularity. The defects are D-branes wrapping the 2-cycles of the singularity. We use this construction to make contact with the description of SYM defects due to Gukov and Witten \cite{Gukov:2006jk}. Furthermore, we derive from a geometric perspective the complete nilpotent orbit classification of codimension-two defects, and the connection to $ADE$-type Toda CFT. The only data needed to specify the defects is a set of weights of the algebra obeying certain constraints, which we give explicitly. We highlight the differences between the defect classification in the little string theory and its $(2,0)$ CFT limit.
\end{abstract}

\clearpage

\tableofcontents


\newpage

\section{Introduction}
\label{sec:intro}

String theory has been an invaluable tool to study the properties of quantum field theories in various dimensions. In particular, it predicts the existence of the so-called  $(2,0)$ conformal field theory in six dimensions; this CFT has become of great interest in recent years (\cite{Gaiotto:2009we,Gaiotto:2009hg,Witten:2009at}). Most importantly, it gave new insights into lower-dimensional supersymmetric gauge theories, for example in four dimensions (see \cite{Chacaltana:2012zy,Balasubramanian:2014dia}). The $(2,0)$ CFT is labeled by an $ADE$ Lie algebra $\fg$, and arises in string theory after taking a double limit. One sends the string coupling to zero in type IIB string theory on an $ADE$ surface $X$ to decouple the bulk modes; this gives a six-dimensional string theory, called the $(2,0)$ little string theory. Second, one takes the string mass $m_s$ to infinity, while keeping the moduli of the $(2,0)$ string fixed.

After taking these limits, there  is no scale left in the theory, and we are left with the $(2,0)$ CFT. The lack of a good mathematical definition of the CFT, however, is a limitation in the predictions we can make about gauge theories. Instead, it proves fruitful to keep $m_s$ finite and study the $(2,0)$ little string. In \cite{Aganagic:2015cta}, the $(2,0)$ little string was studied on a Riemann surface $\cC$ with defects. Specifically, the $(2,0)$ little string is placed on $\cC$, and codimension-two defects are introduced as D5 branes at points on $\cC$ and wrapping non-compact 2-cycles of the $ADE$ surface $X$.

In a different context, Gukov and Witten analyze the surface defects of 4d $\cN=4$ SYM from a gauge theory perspective, by studying the singular behavior of the gauge and Higgs fields in SYM near the defect  \cite{Gukov:2006jk}. In this paper, we explain the origin of these defects using the $(2,0)$ little string on $\cC$, compactified on an additional torus $T^2$. At energies below the Kaluza--Klein scale of compactification and the string scale, this becomes the 4d $\cN=4$ SYM theory. The defects come from D5 branes wrapping the $T^2$, or equivalently, from D3 branes at points on $T^2$. In particular, the S-duality of 4d SYM theory with defects is realized in little string theory as T-duality on the torus; the case without defects was studied already some time ago in \cite{Vafa:1997mh}. In the CFT limit, the resolution \cite{Springer:1969,*Steinberg:1976,*Fu:2003} of the (singular) Coulomb branch of the theory on the D3 branes is in fact described by the cotangent bundle $T^*(G/\mathcal{P})$, where $\mathcal{P}$ is a parabolic subgroup of the gauge group $G$. This space already appeared in \cite{Gukov:2006jk} as an alternate way to describe surface defects. It comes about as a moduli space of solutions to Hitchin's equations, which are precisely the equations obeyed by the brane defects in the low energy limit. We will see here that the space $T^*(G/\mathcal{P})$ arises from the geometry; indeed, a given set of D3 branes wrapping 2-cycles of $X$ will determine a unique parabolic subalgebra of $\fg$ at low energies.

Everywhere except at the origin of the moduli space, which is singular, the $(2,0)$ little string theory with the D5 brane defects is in effect described by the theory on the branes themselves. Specifically, it has a description at low energy as a 4d $\cN=2$ superconformal quiver gauge theory of Dynkin shape.
In particular, the gauge theory description of the so-called $T_N$ theory (when viewed as a 5d $\cN=1$ theory), or full puncture, was derived for any simply-laced Lie algebra $\fg=A, D, E$ in \cite{Aganagic:2015cta} (see also \cite{Bergman:2014kza,Hayashi:2014hfa} for the $A_n$ case). In this paper, we give the full classification of punctures of the $(2,0)$ little string theory on  $\cC$. Each ``class'' of defects will be given by a collection of certain weights of $\fg$, from which one can  read off a superconformal quiver gauge theory. Taking the string mass $m_s$ to infinity, we generically lose the low energy quiver gauge theory description of the defects. Indeed, if $\tau$ is the D-brane gauge coupling, it must go to zero, as the combination $\tau m_s^2$ is a modulus of the $(2,0)$ theory, kept fixed in the limit. Nonetheless, we obtain the full list of punctures given in the literature in terms of nilpotent orbits \cite{Chacaltana:2012zy} (see also \cite{Tachikawa:2009rb} for an M-theory approach in the specific case of $D_n$).

Finally, the AGT correspondence \cite{Alday:2009aq} relates 4d $\cN=2$ theories compactified on a Riemann surface to 2d Toda conformal field theory on the surface. In the little string setup, the precise statement is that the partition function of the $(2,0)$ little string on $\cal{C}$ with brane defects  is in fact equal to a $q$-deformation of the Toda CFT conformal block on $\cal{C}$. The vertex operators are determined by positions and types of defects. So in particular, the codimension-two defects of the 6d $(2,0)$ CFT are expected to be classified from the perspective of the 2d Toda theory. This can be done by studying the Seiberg-Witten curve of the 4d $\cN=2$ quiver gauge theory on the D5 branes, or equivalently, after $T^2$ compactification, the curve of the 2d $\cN=(4,4)$ theory on the D3 branes. At the root of the Higgs branch, and in the $m_s$ to infinity limit, the curve develops poles at the puncture locations. The residues at each pole obey relations which describe the level 1 null states of the Toda CFT; this was previously studied in the $A_n$ case in \cite{Kanno:2009ga}. We argue that this characterization of defects as null states of the CFT naturally gives the same parabolic subalgebra classification obtained in this note, for $\fg=A, D, E$.\\

The paper is organized as follows. In section 2, we review the description of surface defects of $\cN=4$ SYM given in \cite{Gukov:2006jk}, and give its $(2,0)$ little string theory origin. We further derive the action of S-duality from T-duality on $T^2$. In section 3, we explain how to extract a parabolic subalgebra and characterize the sigma model $T^*(G/\mathcal{P})$ from the D3 brane defect data of the little string. In section 4, we make contact with the nilpotent orbit classification of defects given in the literature \cite{Chacaltana:2012zy}. In section 5, we explain how the parabolic subalgebras determined in section 3 can also be recovered from  null states of the $\fg$-type Toda CFT, and how this is related to the nilpotent orbit classification. In section 6, we explain the differences between the defects of the little string proper, and its $(2,0)$ CFT limit. In order to give the exhaustive list of defects of the little string, we will need to extend our definition of defects to characterize punctures on $\cC$ that do not specify a definite parabolic subalgebra. In section 7, we provide a plethora of detailed examples for $\fg=A,D,E,$ and illustrate all the statements made in the rest of the paper.

\newpage

\section{SYM Surface Defects From Little String Theory}
\label{sec:review}

In this section, we begin by recalling the  description of two-dimensional surface defects in 4d $\cN=4$ SYM given by Gukov and Witten in \cite{Gukov:2006jk}. We then review the analysis of little string theory on a Riemann surface \cite{Aganagic:2015cta}, use it to describe these surface defects and derive their S-duality transformation.

\subsection{Gukov--Witten Surface Defects of $\cN=4$ SYM}

Surface defects of $\cN=4$ SYM are $\tfrac{1}{2}$-BPS operators; to describe them, one starts with a four-dimensional manifold $M$, which is locally $M=D\times D'$, where $D$ is two-dimensional, and $D'$ is a fiber to the normal bundle to $D$. Surface defects are then  codimension-two objects living on $D$, and located at a point on $D'$; they are introduced by specifying the singular behavior of the gauge field near this defect. A surface operator naturally breaks the gauge group $G$ to a subgroup $\mathbb{L}\subset G$, called a Levi subgroup.\\

The story so far is in fact valid for $\cN=2$ SUSY, but $\cN=4$ SUSY has additional parameters $\vec{\beta}$ and $\vec{\gamma}$, which describe the singular behavior of the Higgs field $\phi$ near the surface operator; choosing $D'=\mathbb{C}$ with coordinate $z=r e^{i\theta}=x_2+i x_3$, we have:
\begin{align}\label{GWfields}
A &=\vec\alpha d\theta+\ldots,\\
\phi &= \frac{1}{2}\left(\vec{\beta}+i\vec{\gamma}\right)\frac{dz}{z}+\ldots\label{eq:betagamma},
\end{align}
which solve the Hitchin equations \cite{Hitchin:1986vp}:
\begin{align}\label{Hitch}
F &=[\phi,\overline{\phi}],\\
\overline{D}_z\phi &=0=D_z\overline{\phi}.
\end{align}
As written above, we have chosen a complex structure which depends holomorphically on $\beta+i\gamma$, while the K\"ahler structure depends on $\alpha$. Quantum mechanics also requires the consideration of the Theta angle, denoted by $\eta$; by supersymmetry, it will complexify the K\"ahler parameter $\alpha$.\\

S-duality is the statement that this theory is equivalent to $\cN=4$ gauge theory with a dual gauge group and coupling constant 
\[
g'_{4d}=1/g_{4d}.
\]
The action of S-duality on the surface defect parameters is a rescaling of the Higgs field residue 
\begin{equation}
\label{eq:bgsduality}
(\beta,\gamma)\rightarrow\left(\frac{4\pi}{g^2_{4d}}\right)(\beta,\gamma),
\end{equation}
and an exchange of the gauge field and Theta angle parameters \cite{Gukov:2006jk}
\begin{equation}
\label{eq:aesduality}
(\alpha,\eta)\rightarrow(\eta,-\alpha).
\end{equation}

The analysis of \cite{Gukov:2006jk} gives a second description of the surface operators of $\cN=4$ SYM, which will be of great relevance to us; one couples the 4d theory to a 2d non-linear sigma model on $D$. In the $\cN=4$ case, the 2d theory is a sigma model to $T^*(G/\mathcal{P})$, where $\mathcal{P}\subset G$ is a parabolic subgroup of the gauge group. The quotient describes a partial flag manifold when the Lie algebra $\fg$ is $A_n$. In the case of a general Lie algebra, the quotient is  a generalized flag variety. This target space is in fact the moduli space of solutions to the Hitchin equations \eqref{Hitch}.\\

Then, to describe a surface operator, one can either specify the parameters $(\beta,\gamma,\alpha)$ for the singular Higgs and gauge fields, or spell out the sigma model $T^*(G/\mathcal{P})$. It turns out that both of these descriptions have an origin in  string theory, and we will now show this explicitly; our starting point will be the $(2,0)$ little string theory.

\subsection{Little String on a Riemann surface and D5 Branes}

We first review some basic facts about the little string (\cite{Seiberg:1997zk,Witten:1995zh,Losev:1997hx}; see \cite{Aharony:1999ks} for a review), and discuss the role of D5 branes in the theory.

\vskip 0.5cm
{\it --(2,0) $ADE$ Little String Theory--}
\vskip 0.5cm

The $ADE$ little string theory with $(2,0)$ supersymmetry is a six dimensional string theory, and therefore has 16 supercharges. It is obtained by sending the string coupling $g_s$ to zero in type IIB string theory on an $ADE$ surface $X$; this has the effect of decoupling the bulk modes of the full type IIB string theory. $X$ is a hyperk\"ahler manifold, obtained by resolving a ${\mathbb C}^2/\Gamma$ singularity where $\Gamma$ is a discrete subgroup of $SU(2)$, related to ${\bf g}$ by the McKay correspondence \cite{Reid:1997zy}. The little string is not a local QFT, as the strings have a tension $m_s^2$. The $(2,0)$ little string reduces to a $(2,0)$ 6d conformal field theory at energies well below the string scale $m_s$. The moduli space of the little string is $\left(\mathbb{R}^4\times S^1\right)^{\mbox{rk}(\fg)}/W$, with $W$ the Weyl group of $\fg$. The scalars parametrizing this moduli space come from the periods of the NS B-field $m_s^2/g_s\int_{S_a}B_{NS}$, the RR B-field $m_s^2\int_{S_a}B_{RR}$, and a triplet of self-dual two-forms obtained from deformations of the metric on $X$, $m_s^4/g_s\int_{S_a}\omega_{I,J,K}$. Here, $S_a$ are two-cycles generating the homology group $H_2(X,\mathbb{Z})$. The $(S^1)^{\mbox{rk}(\fg)}$ have radius $m_s^2$ and are parametrized by the periods of $B_{RR}$. When $g_s$ is sent to zero, we keep  the above periods fixed in that limit. We set for all $a$'s
\beq\label{FI}
 \int_{S_a} \omega_{J,K} =0,\;   \int_{S_a} B_{NS}=0, 
\eeq
and let
\beq\label{taua}
\tau_a =   \int_{S_a} \, ( m_s^2 \,\omega_I/g_s + i  \, B_{RR} )
\eeq
be arbitrary complex numbers with ${\rm Re}(\tau_a)>0$.\\

We start by compactifying the $(2,0)$ little string theory on a fixed Riemann surface $\cal{C}$, which is chosen to have a flat metric. This guarantees $X\times \cal{C}$ to be a solution of type IIB string theory. We want to introduce codimension-two defects in the little string, at points on $\cal{C}$ and filling the four remaining directions $\mathbb{C}^2$. These correspond to D5 branes in IIB string theory, wrapping non-compact 2-cycles in $X$ and  $\mathbb{C}^2$ \cite{Aganagic:2015cta}. Their tension remains finite in the little string limit, so they are the correct objects to study (D3 branes also keep finite tension, but they do not describe the codimension-two defects we are after; other objects of type IIB  either decouple or get infinite tension when $g_s\rightarrow 0$).\\

In \cite{Aganagic:2015cta}, it is argued that the dynamics of the $(2,0)$ little string theory on ${\cal C}\times  {\mathbb C}^2$, with an arbitrary collection of D5 brane defects at points on $\cal C$, is captured by the theory on the branes themselves. 
Because the Riemann surface $\cal{C}$, which is transverse to the D5 branes, has a flat metric, the theory on the D5 branes is four dimensional at low energies. In fact, it has 4d $\cN=2$ super Poincare invariance, since the D5 branes break half the supersymmetry. We will focus specifically on the class of D5 branes that retain some conformal invariance in the resulting low energy 4d theory. This corresponds to a very specific choice of non-compact 2 cycles of $X$ wrapped by the D5 branes, which we review here.

\vskip 0.5cm
{\it --D5 Branes  and $ADE$ quiver gauge theories--}
\vskip 0.5cm

For definiteness, we will choose the Riemann surface $\cC$ to be the complex plane in what follows (one could  equally choose to work on the cylinder as in \cite{Aganagic:2015cta}, or on the torus.)
The four-dimensional theory on the D5 branes is a quiver gauge theory, of shape the Dynkin diagram of $\fg$ \cite{Douglas:1996sw}. 
The 4d gauge couplings are the $\tau_a$ defined in equation \eqref{taua}, which are the moduli of the $(2,0)$ theory in 6d. The masses of  fundamental hypermultiplets are the positions of the D5 branes on $\cal C$ wrapping  non-compact two-cycles  of $X$. Finally, the Coulomb moduli  are the positions of the D5 branes on $\cal C$ wrapping  compact two-cycles  of $X$.\\

In order to specify a defect D5 brane charge, we pick a class $[S^*]$ corresponding to non-compact two-cycles in the relative homology $H_2(X, \partial X; {\mathbb Z}) = \Lambda_*$, which we identify with the (co-)weight lattice of $\bf g$:
\beq\label{ncomp}
[S^*] = -\sum_{a=1}^n \, m_a \, w_a \;\;  \in \Lambda_*
\eeq
with non-negative integers  $m_a$ and fundamental weights $w_a$.  A necessary condition for conformal invariance in 4d is that the net D5 brane flux vanishes at infinity. This constrains the form of the coefficients $m_a$. To satisfy the condition, we add D5 branes that wrap a compact homology class $[S]$ in $H_2(X, {\mathbb Z})=\Lambda$, which we identify with the root lattice of $\bf g$:
\beq\label{comp}
[S] = \sum_{a=1}^n  \,d_a\,\alpha_a\;\;  \in  \Lambda
\eeq
with non-negative integers $d_a$ and the simple roots  $\alpha_a$, 
such that
\beq\label{conf}
[S+S_*] =0.
\eeq
The vanishing of $S+S^*$ in homology is equivalent to vanishing of $\# (S_a \cap (S+S_*))$ for all $a$.  We can therefore rewrite \eqref{conf} as
\beq\label{conformal}
\sum_{b=1}^n C_{ab} \;d_b = m_a
\eeq
where $C_{ab}$ is the Cartan matrix of $\fg$.
On the Higgs branch of the low energy gauge theory,  the gauge group $\prod_{a=1}^n U(d_a)$ is broken to its $U(1)$ centers, one for each node. There, the D5 branes wrapping the compact cycles $S$ and the non-compact cycles $S^*$ recombine to form D5 branes wrapping a collection of non-compact cycles $S_i^*$, whose homology classes are elements $\omega_i$ of the weight lattice $\Lambda^* = H_2(X, \partial X; {\mathbb Z})$:
\beq\label{weightsfr}
\omega_i  = [S_i^*] \qquad \in\Lambda^*.
\eeq
It is these weights $\omega_i$ that will classify the defects of the little string. 
Each of the $\omega_i$'s comes from one of the non-compact D5 branes on $S^*$.
For the branes to bind, the positions on $\mathcal{C}$ of the compact branes must coincide with the positions of one of the non-compact D5 branes. Recall that the positions of non-compact D5 branes are mass parameters of the quiver gauge theory, while the positions of compact D5 branes on ${\cal C}$ are Coulomb moduli; when a Coulomb modulus coincides with one of the masses, the corresponding fundamental hypermultiplet becomes massless and can get expectation values, describing the root of the Higgs branch.  One can reasonably worry that the binding of the D5 branes will break supersymmetry, but it is in fact preserved when one turns on the FI terms, which are the periods $ \int_{S_a} \omega_{J,K},\;   \int_{S_a} B_{NS}$.\\

Then, the $\omega_i$'s can always be written as a negative fundamental weight $-w_a$ plus the sum of positive simple roots $\alpha_a$, from bound compact branes. Not any such combination will correspond to truly bound branes: a sufficient condition is that  $\omega_i $ is in the Weyl orbit of $-w_a = [S_a^*]$ (we will relax this condition in section \ref{sec:types} and end up with a new class of defects of the little string). Furthermore, the collection of weights 
\beq\label{WS}
{\cal W}_{\cal S} = \{ \omega_i\}
\eeq
we get must be such that it accounts for all the D5 brane charges in $[S^*]$ and in $[S]$. One simple consequence is that the number of $\omega_i$'s is the total rank of the 4d flavor group, $\sum_{a=1}^n m_a$. The fact that the net D5 charge is zero, $[S+S^*]=0$, implies that
\[
\sum_{\omega_i\in{\cal W}_{\cal S}}\omega_i=0,
\]
which is equivalent to \eqref{conformal}.\\

The most canonical type of defect is the one analyzed in \cite{Aganagic:2015cta}, which makes use of the fact that the weight lattice of a Lie algebra of rank  $n$ is $n$-dimensional. Then we can construct a set ${\cal W}_{\cal S}$ by picking any $n+1$ weight vectors which lie in the Weyl group orbits of the fundamental weights $-w_a$ such that they sum up to zero and $n$ of them span $\Lambda_*$. This leads to a full puncture defect of the $(2,0)$ little string on ${\cal C}$. The example below features $\fg=A_3$.

\begin{example}
Let us look  at the set of all the weights in the antifundamental representation  of $A_3$; these weights all add up to 0, and all weights are in the orbit of (minus) the fundamental weight $[-1,0,0]$, written in Dynkin labels, so this set defines a valid set ${\cal W}_{\cal S}$.
Writing $w_i$ for the  $i$-th fundamental weight, we note that:
\begin{alignat*}{2}
\omega_1&=[-1,\phantom{-}0,\phantom{-}0]& &=-w_1,\\
\omega_2&=[\phantom{-}1,-1,\phantom{-}0]& &=-w_1+\alpha_1,\\
\omega_3&=[\phantom{-}0,\phantom{-}1,-1]& &=-w_1+\alpha_1+\alpha_2,\\
\omega_4&=[\phantom{-}0,\phantom{-}0,\phantom{-}1]& &=-w_1+\alpha_1+\alpha_2+\alpha_3.
\end{alignat*}
Written in this fashion, the set  ${\cal W}_{\cal S}$ defines a 4d superconformal quiver gauge theory, shown in Figure \ref{fig:a3full}. This is called the full puncture.
\begin{figure}[htbp]
\begin{center}
\begin{tikzpicture}
 \begin{scope}[auto, every node/.style={minimum size=0.75cm}]
\def \spac {1cm}

\node[circle, draw](k1) at (0,0) {$3$};
\node[circle, draw](k2) at (1*\spac,0) {$2$};
\node[circle, draw](k3) at (2*\spac,0) {$1$};

\node[draw, inner sep=0.1cm,minimum size=0.67cm](N1) at (0*\spac,\spac) {$4$};

\draw[-] (k1) to (k2);
\draw[-] (k2) to (k3);

\draw (k1) -- (N1);

\end{scope}
\end{tikzpicture}
\end{center}
\caption{The quiver theory describing a full puncture for $\fg=A_3$}
\label{fig:a3full}
\end{figure}
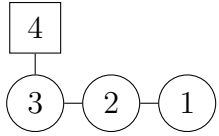
\end{example}

The full classification of defects for simply-laced $\fg$ is obtained by constructing the set  ${\cal W}_{\cal S}$ to have size $n+1$ \textit{or less}. As we will explain in later sections, this is where the rich structure of the parabolic subalgebras of $\fg$ will emerge, and it will be our main object of study.\\

When the string scale $m_s$ is taken to infinity, the $(2,0)$ little string reduces to the $(2,0)$ CFT of type $\fg$ compactified on $\cal C$; we lose the Lagrangian description in general, and the Coulomb branch dimension, previously equal to $\sum_{a=1}^n d_a$, generically decreases. This loss of Coulomb moduli is expected, as the theory loses degrees of freedom in the limit. This distinction in counting Coulomb moduli between the little string and CFT cases will be important to keep in mind throughout the rest of our discussion.

\subsection{Little String Theory origin of SYM surface defects and S-duality}

\vskip 0.5cm
{\it --Hitchin System and Higgs Field Data--}
\vskip 0.5cm

In the little string theory, the brane defects we are studying are solutions to Bogomolny equations on  $\cC$ times an extra circle $S^1(R_1)$ \cite{Aganagic:2015cta}:
\beq\label{bogomolny}
D\phi=*F.
\eeq
Little string theory enjoys T-duality, so in particular, the $(2,0)$ $ADE$ Little String of type IIB compatified on  $S^1(R_1)$ is dual to the (1,1) $ADE$ Little String of type IIA compatified on  $S^1(\hat{R}_1)$ of radius  $\hat{R}_1=1/m_s^2 R_1$.
The defects are then D4 branes after T-dualizing, and are points on ${\cal C} \times S^1(\hat{R}_1)$. These are monopoles, magnetically charged under the gauge field coming from the (1,1) little string. The $n$ scalars are $\phi_a=\int_{S^2_a}m_s^3\omega^I/g'_s$, where $g'_s$ is the IIA string coupling, related to the IIB one by $1/g'_s={R_1}m_s/g_s$. $F$ is the curvature of the gauge field coming from the (1,1) little string.\\

If we want to recover the original  description of the defects as D5 branes, we can take the dual circle size $\hat{R}_1$ to be very small; the upshot is that the Bogomolny equations simplify and we recover the Hitchin equations \eqref{Hitch} we considered previously:
\begin{align}
F &=[\phi,\overline{\phi}],\\
\overline{D}_z\phi &=0=D_z\overline{\phi}.
\end{align}
A subtlety here is that the field $\phi$ got complexified in passing from D4 branes back to D5 branes. The imaginary part of $\phi$ is  the holonomy of the (1,1) gauge field around $S^1(\hat{R}_1)$; this comes from the fact that the D4 branes are magnetically charged under the RR 3-form: $R_1 \int_{S^2_a\times S^1(R_1)}m_s^2 \; C^{(3)}_{RR}$. In type IIB language, after T-duality, the D5 branes are charged under the RR 2-form instead: $1/\hat{R}_1 \int_{S^2_a} B_{RR}$.
All in all, the Higgs field is then written in IIB variables as
\beq\label{higgsIIB}
\phi_a = (\alpha_a, \phi) = 1/{\hat{R}}_1 \int_{S^2_a} (m_s^2 \omega_I/g_s + i B_{RR})=\tau_a/{\hat{R}}_1.
\eeq

The Seiberg-Witten curve of the quiver gauge theory on the D5 branes arises as the spectral curve of the Higgs field $\phi$, taken in some representation $\fR$ of $\fg$:
\begin{equation}
\label{eq:swcurve5d}
\det{}_\fR(e^{\hat{R}_1\phi}-e^{\hat{R}_1 p})=0.
\end{equation}
In the absence of monopoles,  $\phi$ is constant:  the vacuum expectation value of the Higgs field is $\hat{R}_1 \phi = \tau$.\\

By construction, then, the Coulomb branch of the $ADE$ quiver theory on the D5 branes is the moduli space of monopoles  on ${\cal C} \times S^1({\hat R_1})$. As we described in the previous section, we ultimately want to go on the Higgs branch of the theory, where we get a description of quiver theories as a fixed set of weights ${\cal W}_{\cal S}$ in $\fg$; there, all the non-abelian monopoles reduce to  Dirac monopoles. The effect on $\phi$ of adding a Dirac monopole of charge $\omega_i^{\vee}$, at a point $x_i=\hat{R}_1 \hat\beta_i$ on ${\cal C}$,  is to shift:
\beq\label{addone}
e^{\hat{R}_1\phi} \rightarrow e^{\hat{R}_1\phi}  \cdot (1-z\, e^{-\hat{R}_1\hat\beta_i})^{-w_i^{\vee}}.
\eeq
Here, $z$ is the complex coordinate on $\cC=\mathbb{C}$.
Thus, the Higgs field solving the Hitchin equations at the point where the Higgs and the Coulomb branches meet is
\beq\label{Higgs}
e^{\hat{R}_1\phi(x)} = e^{\tau} \prod_{\omega^V_i \in {\cal W}_{{\cal S}} }\;(1-z\, e^{-\hat{R}_1\hat\beta_i})^{-\omega^\vee_{i}}.
\eeq
To take the string mass $m_s$ to infinity, we relabel $e^{\hat R_1 \hat \beta_i}=z_{\mathcal P}\, e^{\hat{R}_1\beta_{i,\mathcal P}}$. We can then safely take the limit $\hat{R}_1\to 0$; the imaginary part of $\phi$ decompactifies, and equation \eqref{eq:swcurve5d} becomes the spectral curve of the Hitchin integrable system \cite{Gaiotto:2009hg}:
\begin{equation}
\label{eq:swcurve4d}
\det{}_\fR(\phi-p)=0.
\end{equation}
In this limit, the Higgs field near a puncture of $\cC$ has a pole of order one, and takes the form
\beq\label{Higgs2}
\phi(z) = {\beta_0\over z} +  \sum_{{\cal P}} \sum_{\omega_i \in {\cal W}_{{\cal P}} }\;{ \beta_{i,{\cal P}} \, \omega_i^{\vee}\over z_{\cal P} - z},
\eeq
with $\beta_0=\tau/\hat{R}_1$ and ${\cal P}$ the  set of punctures.
Therefore, in the $(2,0)$ CFT, we have poles on ${\cal C}$ at $z= z_{\cal P}$, with residues 
\[
\beta_{\cal P} =  \sum_{\omega_i \in {\cal W}_{{\cal P}}} \beta_{i,{\cal P}} \, \omega_i^{\vee}.
\]
These residues are what we called $\beta+i\gamma$ in the $\cN=4$ SYM setup of eq. \eqref{eq:betagamma}.

\vskip 0.5cm
{\it -- 4d S-duality is T-duality of the Little String--}
\vskip 0.5cm

To provide evidence that the surface defects of $\cN=4$ SYM really are branes at points on $\cC$ in the $(2,0)$ little string, we now derive four-dimensional S-duality from  T-duality of the string theory, compactified on an additional torus $T^2$. Here, $T^2$ is the product of two $S^1$'s, one from each of the two complex planes $\mathbb{C}^2$. We label those circles as $S^1(R_1)$ and $S^1(R_2)$, of radius $R_1$ and $R_2$ respectively.\\

\begin{figure}[htbp]
\tcbset{enhanced,size=small,nobeforeafter,tcbox raise=-2.3mm,colback=white,colframe=white}
\centering
\begin{tikzpicture}
\node at (0,0) {\tcbox[borderline={0.2mm}{0mm}{violet!70!black}]{$(1,1)$ string on $S^1(\hat R_1)\times S^1(R_2)\times \RR^2\times\cC$ with $(\text{D}4,\text{D}4)$ branes}};
\draw (0,-0.5)[<->, thick] -- (0,-1.1);
\node at (1.2,-0.8) {$T_1$-duality};
\node at (0,-1.55) {\tcbox[borderline={0.2mm}{0mm}{green!30!black}]{$(2,0)$ string on $S^1(R_1)\times S^1(R_2)\times \RR^2\times\cC$ with $(\text{D}3,\text{D}5)$ branes}};
\draw (0,-2.05)[<->, thick] -- (0,-2.65);
\node at (1.2,-2.35) {$T_2$-duality};
\node at (0,-3.1) {\tcbox[borderline={0.2mm}{0mm}{violet!70!black}]{$(1,1)$ string on $S^1(R_1)\times S^1(\hat R_2)\times \RR^2\times\cC$ with $(\text{D}4,\text{D}4)$ branes}};
\end{tikzpicture}
\caption{One starts with the $(1,1)$ little string theory on $T^2\times\RR^2\times\cC$. After doing two T-dualities in the torus directions, we get the $(1,1)$ little string theory on the T-dual torus; in the low energy limit, the pair of $(1,1)$ theories gives an S-dual pair of $\cN=4$ SYM theories. D3 branes at a point on  $T^2$ map to D4 branes in either $(1,1)$ theory, while D5 branes wrapping  $T^2$ map to another set of D4 branes.}
\label{fig:sdual}
\end{figure}
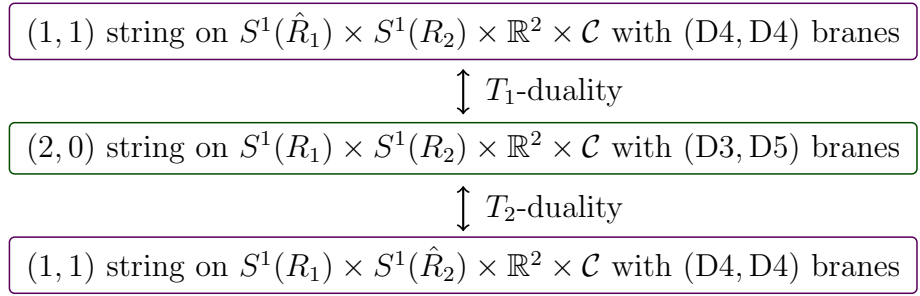

First, without any D5 branes, S-duality was derived in \cite{Vafa:1997mh}, and the line of reasoning went as follows:
suppose we first compactify on, say, $S^1(R_1)$; this is what we just did in the previous section to make contact with D4 branes as magnetic monopoles.
Then we are equivalently studying the (1,1) little string on $S^1(\hat{R}_1)$. Compactifying further on $S^1(R_2)$, this theory is the same as the (1,1) little string on $S^1(R_1)\times S^1(\hat{R}_2)$, by  $T^2$-duality.
4d SYM S-duality then naturally follows from the $T^2$-duality of this pair of (1,1) theories. Indeed, at low energies, both $(1,1)$ little string theories become the maximally supersymmetric 6d SYM, with gauge group dictated by $\fg$ and gauge coupling $1/g_{6d}^2=m_s^2$. We wish to take the string scale $m_s$ to infinity; in the case of the $(1,1)$ string on $S^1(\hat R_1)$, since $m_s^2 \hat{R}_1=1/R_1$, the  radius $\hat{R}_1$ goes to 0 in that limit. The theory then becomes 5d $\cN=2$ SYM, with inverse gauge coupling $1/g_{5d}^2=1/R_1$. After the further compactification on $S^1(R_2)$, we obtain at low energies 4d $\cN=4$ SYM, with inverse gauge coupling $1/g_{4d}^2=R_2/g_{5d}^2=R_2/R_1$.

Now, the same reasoning applied to the $T^2$-dual theory $S^1(R_1)\times S^1(\hat{R}_2)$ gives  4d $\cN=4$ SYM in the $m_s$ to infinity limit, with inverse gauge coupling $1/g_{4d}'^2=R_1/R_2$.

Note that $1/g'_{4d}=g_{4d}$. This is just the action of S-duality on the gauge coupling of $\cN=4$ SYM.
Writing $R_2/R_1\equiv \mbox{Im}(\tau')$, with $\tau'$ the modular parameter of the $T^2$, we see that S-duality is  a consequence of $T^2$-duality for the pair of $(1,1)$ little string theories. An illustration of the dualities is shown in Figure \ref{fig:sdual}.\\

Now, we extend this argument and introduce the D5 brane defects;
since the D5 branes were initially wrapping $T^2\times\mathbb{C}$, note that we can equivalently consider the defects to be D3 branes at a point on $T^2$.  We now argue that the S-duality action on the half BPS surface defects of SYM has its origin in the same $T^2$-duality of $(1,1)$ theories we presented in the previous paragraph.

First, recall that after $S^1(R_1)$ compactification, the D5 branes are charged magnetically, with period:
\[
\phi_a = 1/\hat{R}_1 \int_{S_a} (m_s^2 \omega_I/g_s + i B_{RR}).
\]
In type IIB variables, we call this period $\beta+i\gamma$. By T-dualizing along $S^1(R_1)$ we obtain D4 branes wrapping $S^1(R_2)$ in the $(1,1)$ little string. Now suppose we T-dualize the D5 branes along $S^1(R_2)$ instead; then we have D4 branes wrapping $S^1(R_1)$, in the $T^2$-dual $(1,1)$ little string. The D4 brane tensions in both $(1,1)$ theories are proportional to each other, with factor $R_2/R_1$. But then $(\beta,\gamma) \rightarrow R_2/R_1 \, (\beta,\gamma)$ after $T^2$-duality. The D4 branes are then heavy, magnetic objects in one $(1,1)$ theory, while they are light, electric objects in the other. In the $m_s\to\infty$ limit, $(\beta,\gamma)$ are the parameters of the Higgs field in 4d SYM. This is precisely the action of S-duality for the Higgs field data: $(\beta,\gamma)\rightarrow\mbox{Im}(\tau')(\beta,\gamma)$ \eqref{eq:bgsduality}.\\

Second, after $T^2$ compactification, the D3 branes, which are points on $T^2$, are  charged under the RR 4-form: $\int_{S_a\times \widetilde{S^1}\times S^1(R_1)} C^{(4)}_{RR}$, where $\widetilde{S^1}$ is a circle around the point defect on $\cal{C}$. As before, $S^1(R_1)$ is one of the  1-cycles of $T^2$, and $S_a$ is a compact 2-cycle in the ALE space $X$. We call this period $\alpha$. The D3 branes are also charged under $\int_{S_a\times \widetilde{S^1}\times S^1(R_2)} C^{(4)}_{RR}$, where $S^1(R_2)$ is the other  1-cycle of $T^2$; we call this period $\eta$.

Suppose we T-dualize  in the $S^1(R_1)$ direction. Then $\alpha$ becomes the period of the RR 3-form on $S_a\times \widetilde{S^1}$; this period is in fact an electric coupling for the holonomy of the $(1,1)$ gauge field around $\widetilde{S^1}$. Also, $\eta$ becomes the period of the RR 5-form on $S_a\times \widetilde{S^1}\times S^1(R_2)\times S^1(\hat{R}_1)$;  this period is in fact a magnetic coupling for the holonomy of the $(1,1)$ gauge field around $\widetilde{S^1}$. T-dualizing on $S^1(R_2)$ instead, we reach the $T^2$-dual $(1,1)$ theory. We see that $\alpha$ gets mapped to $\eta$, while $\eta$ gets mapped to $-\alpha$ (the minus sign arises because the 5-form is antisymmetric). So in the end, under $T^2$-duality, the periods change as $(\alpha,\eta)\rightarrow(\eta,-\alpha)$. 

Note that because the 1-cycles generating the $T^2$ appear explicitly in the definition of these periods, $T^2$-duality does not amount to a simple rescaling of $(\alpha,\eta)$, as \ was the case for $(\beta,\gamma)$. In the low energy limit, we recover the S-duality of the gauge field and Theta angle parameters of 4d SYM $\alpha$ and $\eta$ in the presence of a defect \eqref{eq:aesduality}.

\vskip 0.5cm
{\it -- $T^*(G/\mathcal{P})$ sigma model and Coulomb branch of the Defect Theory--}
\vskip 0.5cm

We made contact with the surface defects of Gukov and Witten after compactifying the $(2,0)$ little string on $T^2$ and T-dualizing the D5 branes to D3 branes. In this process, as long as $m_s$ is kept finite, the 4d $ADE$ quiver gauge theory that describes the D5 branes at low energies simply becomes a two-dimensional quiver theory for the D3 branes, with the same gauge gauge groups and fundamental matter. In the rest of this paper, we will label this low energy $ADE$ quiver theory on the D3 branes, together with the set of weights $\cal{W}_{\cal{S}}$ that specified it, as $T^{2d}$.  In the CFT limit, we will label the theory as $T^{2d}_{m_s\rightarrow \infty}$. As we mentioned already, unlike $T^{2d}$, the theory $T^{2d}_{m_s\rightarrow \infty}$ generically has no Lagrangian description.\\

Now, Gukov and Witten showed that surface operators of $\cN=4$ SYM can also be described by a 2d sigma model $T^*(G/\mathcal{P})$, which is a moduli space of solutions to the Hitchin equations \eqref{Hitch}. After taking the CFT limit of the little string theory, we saw that this moduli space is also the Coulomb branch of the $(2,0)$ CFT theory on the Riemann surface  $\cC$ times a circle $S^1(R_1)$ (the radius $R_1$ here being very big).
As an algebraic variety, this Coulomb branch is singular, while $T^*(G/\mathcal{P})$ is smooth. The statement is then that the (resolution of the) Coulomb branch of the 2d $ADE$ quiver gauge theories on the D3 branes we presented, in the appropriate $m_s$ to infinity limit, is expected to be the sigma model to $T^*(G/\mathcal{P})$. In other terms, the Coulomb branch of $T^{2d}_{m_s\rightarrow \infty}$ can be identified with $T^*(G/\mathcal{P})$.\\

A natural question arises: how do parabolic subgroups $\mathcal{P}$ in $T^*(G/\mathcal{P})$ arise from the point of view of the defects of the $(2,0)$ little string? 

We will now see that to every $ADE$  theory $T^{2d}$ on the D3 branes, we will be able to associate a unique parabolic subalgebra from the geometry (specifically, the non-compact 2-cycles of $X$), or equivalently, from the representation theory of $\fg$ (the Higgs field we introduced is valued in the Lie algebra $\fg$, so we will speak of parabolic subalgebras rather than parabolic subgroups); in particular, after taking the CFT limit, we will be able to read it from the data of the weight system ${\cal W}_{\cal S}$ that defines the theory $T^{2d}_{m_s\rightarrow \infty}$.\\

As a side note, it is known (\cite{Chacaltana:2012zy, Gaiotto:2008ak,Hanany:2011db,Cremonesi:2014uva}) that $T^*(G/\mathcal{P})$ is the resolution of the Higgs branch of different theories from the ones we have been considering. In the little string setup, as we reviewed, the moduli space of monopoles naturally arises as a Coulomb branch instead of a Higgs branch. A natural guess is that those two descriptions could be related by mirror symmetry, and this is indeed the case in all the cases we could explicitly check (all defects in the $A_n$ case, and some low rank defects in the $D_n$ case; see also \cite{Hanany:2016gbz}). We will not investigate this point further here, but it would be important to get a clear understanding of the mirror map.

\section{From Brane Defects to Parabolic Subalgebra Classification}
\label{sec:parastrings}
We now explain how the $ADE$ quiver theories $T^{2d}$ determine the parabolic subalgebras of $\fg$.
\subsection{Mathematics Preliminaries}
\label{ssec:levi}
Because they will be so crucial to our story, we review here the mathematics of parabolic and Levi subalgebras of a Lie algebra  $\fg$.\\

A \emph{Borel subalgebra} of $\fg$ is a maximal solvable subalgebra of $\fg$, and always has the form $\fb=\fh\oplus\fm$, where $\fh$ is a Cartan subalgebra of $\fg$ and $\fm=\sum_{\alpha\in\Phi^+}\fg_\alpha$ for some choice of positive roots $\Phi^+$. A \emph{parabolic subalgebra} $\fp$ is defined to be a subalgebra of $\fg$ that contains a Borel subalgebra $\fb$, so $\fb\subseteq\fp\subseteq\fg$.\\

There are many different choices of Borel subalgebras of $\fg$, but we will choose one for each $\fg$ and keep it fixed. Since the Borel subalgebra is the sum of all the positive root spaces, we can get any $\fp$ by adding the root spaces associated to any closed system of negative roots.\\

Let us extend our notations to differentiate between distinct parabolic subalgebras: 
We denote the set of positive simple roots by  $\Delta$.
Take an arbitrary subset $\Theta\subset\Delta$. We define $\fp_\Theta$ to be the subalgebra of $\fg$ generated by $\fb$ and all of the root spaces $\fg_\alpha$, with $\alpha\in\Delta$ or $-\alpha\in\Theta.$ Then $\fp_\Theta$ is a parabolic subalgebra of $\fg$ containing $\fb$, and every parabolic subalgebra of $\fg$ containing $\fb$ is of the form $\fp_\Theta$ for some $\Theta\subset\Delta$. In fact, every parabolic subalgebra of  $\fg$ is conjugate to one of the form $\fp_\Theta$ for some $\Theta\subset\Delta$. We state the important result:\\

Let $\langle\Theta\rangle$ denote the subroot system generated by $\Theta$ and write  $\langle\Theta\rangle^+= \langle\Theta\rangle\cap\Phi^+.$
There is a \emph{direct sum decomposition} $\fp_\Theta=\fl_\Theta\oplus\fn_\Theta$, where $\fl_\Theta=\fh\ \oplus\sum_{\alpha\in\langle\Theta\rangle}\fg_\alpha$ is a reductive subalgebra  (a reductive Lie algebra is a direct sum of a semi-simple and an abelian Lie algebra), called a Levi subalgebra, and  $\fn_\Theta=\sum_{\alpha\in\Phi^+ \backslash \langle\Theta\rangle^+}\fg_\alpha$, is called the nilradical of $\fp_\Theta$. Here, $\alpha\in\Phi^+ \backslash \langle\Theta\rangle^+$ means that $\alpha$ is a positive root not in $\langle\Theta\rangle^+$.
Note that $\fn_{\Theta}\cong \sum_{\alpha\in\Phi^- \backslash \langle\Theta\rangle^-}\fg_\alpha\cong\fg/\fp_{\Theta}$.

Furthermore, all Levi subalgebras of a given parabolic subalgebra are conjugate to each other \cite{Malcev:1942}. We illustrate the above statements in the examples below:

\begin{example}
Consider $\fg=A_2$ in the fundamental, three-dimensional representation. Then the elements in the Cartan subalgebra have the form
\begin{equation}
\fh=\begin{pmatrix}
*&0&0\\
0&*&0\\
0&0&*
\end{pmatrix}.
\end{equation}
We associate to a root $\alpha_{ij}=h_i-h_j$ the space $\mathbb{C}E_{ij}$, where $E_{ij}$ is the matrix that has a $+1$ in the $i$-th row and $j$-th column, and zeroes everywhere else. Thus, we see that
\begin{equation}
\fb=\begin{pmatrix}
*&*&*\\
0&*&*\\
0&0&*
\end{pmatrix},
\end{equation}
and the parabolic subalgebras are
\begin{align}
\fp_\varnothing=\fb&=\begin{pmatrix}
*&*&*\\
0&*&*\\
0&0&*
\end{pmatrix},\\
\fp_{\{\alpha_1\}}&=\begin{pmatrix}
*&*&*\\
*&*&*\\
0&0&*
\end{pmatrix},\\
\fp_{\{\alpha_2\}}&=\begin{pmatrix}
*&*&*\\
0&*&*\\
0&*&*
\end{pmatrix},\\
\fp_{\{\alpha_1,\alpha_2\}}=\fg&=\begin{pmatrix}
*&*&*\\
*&*&*\\
*&*&*
\end{pmatrix}.
\end{align}
\end{example}

Let us look at the Levi decompositions of the above:
\begin{example}
\label{ex:levisl3}
For $\fg=A_2$, we get the following decompositions:
\begin{align}
\fp_\varnothing&=\begin{pmatrix}
*&*&*\\
0&*&*\\
0&0&*
\end{pmatrix}=\begin{pmatrix}
*&0&0\\
0&*&0\\
0&0&*
\end{pmatrix}\oplus
\begin{pmatrix}
0&*&*\\
0&0&*\\
0&0&0
\end{pmatrix}=\fl_\varnothing\oplus\fn_\varnothing,\\
\fp_{\{\alpha_1\}}&=\begin{pmatrix}
*&*&*\\
*&*&*\\
0&0&*
\end{pmatrix}=\begin{pmatrix}
*&*&0\\
*&*&0\\
0&0&*
\end{pmatrix}\oplus
\begin{pmatrix}
0&0&*\\
0&0&*\\
0&0&0
\end{pmatrix}=\fl_{\{\alpha_1\}}\oplus\fn_{\{\alpha_1\}},\\
\fp_{\{\alpha_2\}}&=\begin{pmatrix}
*&*&*\\
0&*&*\\
0&*&*
\end{pmatrix}=\begin{pmatrix}
*&0&0\\
0&*&*\\
0&*&*
\end{pmatrix}\oplus
\begin{pmatrix}
0&*&*\\
0&0&0\\
0&0&0
\end{pmatrix}=\fl_{\{\alpha_2\}}\oplus\fn_{\{\alpha_2\}},\\
\fp_{\{\alpha_1,\alpha_2\}}&=\begin{pmatrix}
*&*&*\\
*&*&*\\
*&*&*
\end{pmatrix}=\begin{pmatrix}
*&*&*\\
*&*&*\\
*&*&*
\end{pmatrix}\oplus
\begin{pmatrix}
0&0&0\\
0&0&0\\
0&0&0
\end{pmatrix}=\fl_{\{\alpha_1,\alpha_2\}}\oplus\fn_{\{\alpha_1,\alpha_2\}}.
\end{align}
\end{example}
\begin{example} In the table below, we show the root spaces that the Borel subalgebra of $A_3$ is made of:

\tcbset{enhanced,size=fbox,nobeforeafter,tcbox raise=-2.3mm,colback=white,colframe=white}
\begin{table}[htp]
\renewcommand{\arraystretch}{1.2}
\begin{center}
\begin{tabular}{c|c|c|c}
$\Theta$&$\fp_\Theta$&$\fl_\Theta$&$\fn_\Theta$\\
\hline
$\varnothing$&$\begin{pmatrix}
*&\tcbox[colframe=lime]{*}&\tcbox[borderline={0.2mm}{0mm}{green!95!black,dashed}]{*}&\tcbox[borderline={0.2mm}{0mm}{blue,dotted}]{*}\\
0&*&\tcbox[colframe=brown]{*}&\tcbox[borderline={0.2mm}{0mm}{gray!80!black,dashed}]{*}\\
0&0&*&\tcbox[colframe=purple!80!black]{*}\\
0&0&0&*
\end{pmatrix}$
&
$\begin{pmatrix}
*&0&0&0\\
0&*&0&0\\
0&0&*&0\\
0&0&0&*
\end{pmatrix}$&$\begin{pmatrix}
0&*&*&*\\
0&0&*&*\\
0&0&0&*\\
0&0&0&0
\end{pmatrix}$
\end{tabular}\\
\end{center}
\hspace*{5em}\tcbox[colframe=lime]{\phantom{*}}: $\alpha_1$\hspace{3em} \tcbox[borderline={0.2mm}{0mm}{green!95!black,dashed}]{\phantom{*}}: $(\alpha_1+\alpha_2)$\hspace{3em}\tcbox[borderline={0.2mm}{0mm}{blue,dotted}]{\phantom{*}}:$(\alpha_1+\alpha_2+\alpha_3)$\\
\hspace*{5em}\tcbox[colframe=brown]{\phantom{*}}: $\alpha_2$\hspace{3em} \tcbox[borderline={0.2mm}{0mm}{gray!80!black,dashed}]{\phantom{*}}: $(\alpha_2+\alpha_3)$\\
\hspace*{5em}\tcbox[colframe=purple!80!black]{\phantom{*}}: $\alpha_3$

	\caption{This table illustrates the Levi decomposition of $\fp_\Theta$, when $\Theta$ is the empty set and $\fg=A_3$. $\fp_\Theta$ consists of all the matrices in $A_3$ with zeroes in the indicated places and the other entries are arbitrary.  The color code shows which positive root is denoted by which nonzero entry.}
\label{tab:a3ex}
\end{table}

\end{example}

\subsection{Parabolic Subalgebras from Weight Data}
\label{ssec:3d}

We reviewed in section 2 how we could specify a defect of the little string from a set of weights 
\beq
{\cal W}_{\cal S} = \{ \omega_i\},
\eeq
all in the orbit of some (possibly different) fundamental weights, and adding up to 0. We make the claim that to each set $\cW_\cS$ we can associate a parabolic subalgebra $\fp$. This map is not injective, as many different sets of weights will typically determine the same $\fp$.\\

As reviewed in the last section, all parabolic subalgebras of $\fg$ are determined by a subset $\Theta$ of the simple positive roots $\Delta$ of $\fg$. Thus, our strategy will be to extract such a set $\Theta$ from the weights in $\cW_\cS$.

We do so by first computing the inner product $\langle \alpha_i, \omega_j\rangle$, for all weights $\omega_j$ in ${\cal W}_{\cal S}$, and for all positive simple roots $\alpha_i$ of $\fg$. Then all the $\alpha_i$ which satisfy

\beq
\langle \alpha_i, \omega_j\rangle =0
\eeq
for all weights $\omega_j$ in $\cW_{\cS}$ will make up the set $\Theta$. There is one caveat to the above procedure: The set $\Theta$ defined as such is not invariant under the global action of the Weyl group on $\cW_{\cS}$. Thus, we modify the above prescription and define $\Theta$ as the maximal such set in the Weyl group orbit of $\cW_{\cS}$.\footnote{Note that the Weyl group acts on all the weights in $\cW_{\cS}$ simultaneously.}

Moreover, the positive roots $e_\gamma$ for which
\beq
\langle e_{\gamma}, \omega_i\rangle <0,\text{\footnotemark}
\eeq
\footnotetext{Or equivalently, $\langle e_{\gamma}, \omega_i\rangle >0$.}
for at least one $\omega_i\in\cW_{\mathcal S}$, form a nilradical $\fn$; this nilradical specifies the Coulomb branch of  $T^{2d}_{m_s\rightarrow \infty}$.

This $\fn$ can always be obtained from the Levi decomposition $\fp_{\Theta}=\fl_{\Theta}\oplus\fn_{\Theta}$ of the parabolic subalgebra $\fp_{\Theta}$.\\

As  mentioned already, we emphasize here that the Coulomb branch of $T^{2d}$ is generically bigger than the Coulomb branch of $T^{2d}_{m_s\rightarrow \infty}$. In the little string case, the Coulomb branch of $T^{2d}$  has dimension $\sum_{a=1}^{n}d_a$,  where $d_a$ are the ranks of the gauge groups (here, we include the $U(1)$ centers of the $U(d_a)$ gauge groups). In the CFT limit, the space $X\times\cC$ can be reinterpreted as a Calabi--Yau manifold. Thus, one can use the techniques of complex geometry to count the Coulomb moduli of $T^{2d}_{m_s\rightarrow \infty}$ as the complex structure deformations of this Calabi--Yau \cite{Cachazo:2001gh}. For instance, for $\cC$ a sphere with 3 full punctures, meaning the residues of the Higgs fields $\phi(z)$ are generic, the dimension of the Coulomb branch of  $T^{2d}_{m_s\rightarrow \infty}$ is the number $|\Phi^{+}|$ of positive roots of $\fg$. Note that for $A_n$, the full puncture of  $T^{2d}$ has Coulomb branch dimension $\sum_{a=1}^{n}d_a=|\Phi^+|$, so in that specific case the CFT counting is the same as the little string counting. This is generally not so for $\fg=D_n$ and $E_n$.\\

The dimension of $T^{2d}_{m_s\rightarrow \infty}$ can be conveniently recovered from the representation theory of $\fg$. Indeed,  by just keeping track of which positive roots  satisfy $\langle e_{\gamma}, \omega_i\rangle <0$ for an $\omega_i\in\cW_\mathcal{S}$, and not recording the actual value of the inner product, the positive roots are counting Coulomb moduli of the defect theory in the CFT limit. This point is crucial  in the $D_n$ and $E_n$ cases, where higher positive root multiplicity has to be ignored to identify a nilradical of $\fg$.\\

\begin{example}[$A_3$ example]
From Table \ref{tab:a3ex} above, we will read off a nilradical from a set of weights  ${\cal W}_{\cal S}$ for an $A_3$  theory. We choose  ${\cal W}_{\cal S}$ to be the set of all four weights in the antifundamental representation (note that they add up to 0, as they should); they make up the full puncture of $A_3$. Next, we note the following:\\

$[-1,\phantom{-}0,\phantom{-}0]$ has a negative inner product with $h_1-h_2,  h_1-h_3, h_1-h_4$.

$[\phantom{-}1,-1,\phantom{-}0]$ has a negative inner product with $h_2-h_3,  h_2-h_4.$

$[\phantom{-}0,\phantom{-}1,-1]$  has a negative inner product with $h_3-h_4$.

$[\phantom{-}0,\phantom{-}0,\phantom{-}1]$ has no negative inner product with any of the positive roots.\\

We see that all positive roots of $\fg$ are accounted for, so the nilradical $\fn_{\Theta}$  is constructed using all the positive roots, and thus, $\Theta=\varnothing.$ From the Levi decomposition, we therefore identify the parabolic subalgebra as $\fp_\varnothing$. This is consistent with the fact that no simple root $\alpha_i$ has a vanishing inner product $\langle \alpha_i, \omega_j\rangle$  with all the weights $\omega_j$ in $\cW_\cS$. The discussion is summarized in Figure \ref{fig:weightreading3} below.
\end{example}

\begin{figure}[htpb]
	\begin{center}
        \tcbset{enhanced,size=fbox,nobeforeafter,tcbox raise=-2.3mm,colback=white,colframe=white}
		\begin{tikzpicture}[baseline]
		\node at (-0.5,0) {\tcbox[colframe=violet!70!black]{$\Theta=\varnothing$}};
		\draw[->, -stealth,  line width=0.4em, postaction={draw,-stealth,white,line width=0.2em,
                shorten <=0.10em,shorten >=0.26em}](2,0) -- (1,0);
		\node[align=justify] at (4,0) {$\omega_1:[-1,\phantom{-}0,\phantom{-}0]$\\$\omega_2:[\phantom{-}1,-1,\phantom{-}0]$\\$\omega_3:[\phantom{-}0,\phantom{-}1,-1]$\\$\omega_4:[\phantom{-}0,\phantom{-}0,\phantom{-}1]$};
		\draw[->, -stealth,  line width=0.4em](6,0) -- (7,0);
		\node at (9,0) {\begin{tikzpicture}
 \begin{scope}[auto, every node/.style={minimum size=0.75cm}]
\def \spac {1cm}

\node[circle, draw](k1) at (0,0) {$3$};
\node[circle, draw](k2) at (1*\spac,0) {$2$};
\node[circle, draw](k3) at (2*\spac,0) {$1$};

\node[draw, inner sep=0.1cm,minimum size=0.67cm](N1) at (0*\spac,\spac) {$4$};

\draw[-] (k1) to (k2);
\draw[-] (k2) to (k3);

\draw (k1) -- (N1);

\end{scope}
\end{tikzpicture}};
\node[text width=9em,align=center] at (-0.5,-2) {Simple root subset of $T^{2d}_{m_s\to\infty}$};
\node[text width=9em] at (5,-2) {Weights};
\node[text width=9em] at (10,-2) {2d Gauge Theory};
		\end{tikzpicture}
	\end{center}
	\caption{From the set of weights  ${\cal W}_{\cal S}$, we read off the parabolic subalgebra $\fp_\varnothing$ of $A_3$ (in this case, the choice of weights is unique up to global $\mathbb{Z}_2$ action on the set). Reinterpreting each weight as a sum of ``minus a fundamental weight and simple roots,'' we obtain the 2d quiver gauge theory shown on the right. The white arrow implies we take the CFT limit.}
	\label{fig:weightreading3}
\end{figure}
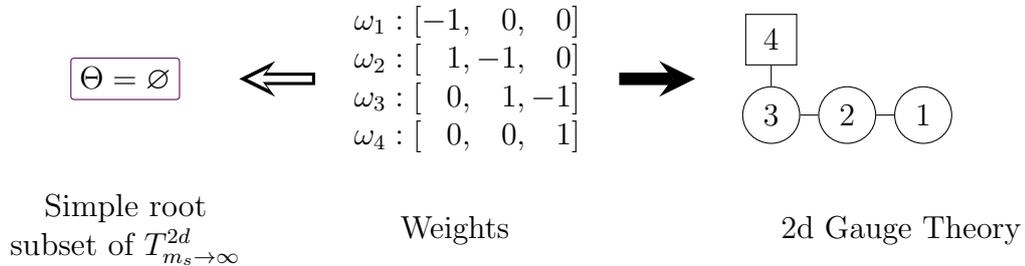


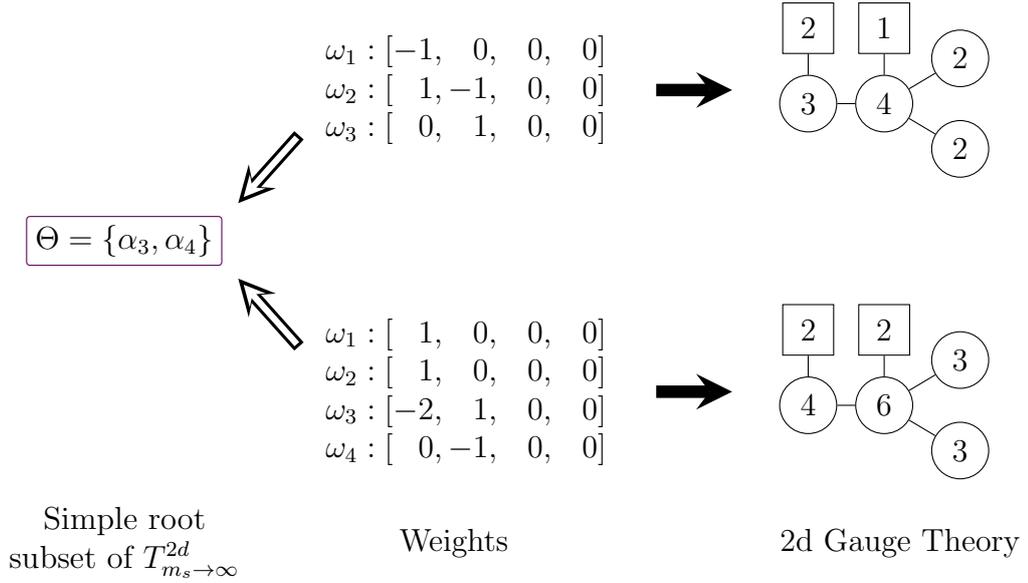
\begin{figure}[htpb]
	\begin{center}
	    \tcbset{enhanced,size=fbox,nobeforeafter,tcbox raise=-2.3mm,colback=white,colframe=white}
		\begin{tikzpicture}[baseline]
		\node at (-0.5,0) {\tcbox[colframe=violet!70!black]{$\Theta=\{\alpha_3,\alpha_4\}$}};
		\draw[->, -stealth,  line width=0.4em, postaction={draw,-stealth,white,line width=0.2em,
                shorten <=0.10em,shorten >=0.26em}](1.8,1.4) -- (1,0.5);
		\node[align=justify] at (4,2) {$\omega_1:[-1,\phantom{-}0,\phantom{-}0,\phantom{-}0]$\\$\omega_2:[\phantom{-}1,-1,\phantom{-}0,\phantom{-}0]$\\$\omega_3:[\phantom{-}0,\phantom{-}1,\phantom{-}0,\phantom{-}0]$};
		\draw[->, -stealth,  line width=0.4em](6.5,2) -- (7.5,2);
		\node at (9.5,2) {\begin{tikzpicture}[baseline]
 \begin{scope}[auto, every node/.style={minimum size=0.75cm}]
\def \spac {1cm}

\node[circle, draw](k1) at (0,0) {$3$};
\node[circle, draw](k2) at (1*\spac,0) {$4$};
\node[circle, draw](k3) at (2*\spac,0.6*\spac) {$2$};
\node[circle, draw](k4) at (2*\spac,-0.6*\spac) {$2$};

\node[draw, inner sep=0.1cm,minimum size=0.67cm](N1) at (0*\spac,\spac) {$2$};
\node[draw, inner sep=0.1cm,minimum size=0.67cm](N2) at (1*\spac,\spac) {$1$};

\draw[-] (k1) to (k2);
\draw[-] (k2) to (k3);
\draw[-] (k2) to (k4);

\draw (k1) -- (N1);
\draw (k2) -- (N2);

\end{scope}
\end{tikzpicture}};
\draw[->, -stealth,  line width=0.4em, postaction={draw,-stealth,white,line width=0.2em,
                shorten <=0.10em,shorten >=0.26em}](1.8,-1.4) -- (1,-0.5);
		\node[align=justify] at (4,-2) {$\omega_1:[\phantom{-}1,\phantom{-}0,\phantom{-}0,\phantom{-}0]$\\$\omega_2:[\phantom{-}1,\phantom{-}0,\phantom{-}0,\phantom{-}0]$\\$\omega_3:[-2,\phantom{-}1,\phantom{-}0,\phantom{-}0]$\\$\omega_4:[\phantom{-}0,-1,\phantom{-}0,\phantom{-}0]$};
		\draw[->, -stealth,  line width=0.4em](6.5,-2) -- (7.5,-2);
		\node at (9.5,-2) {\begin{tikzpicture}[baseline]
 \begin{scope}[auto, every node/.style={minimum size=0.75cm}]
\def \spac {1cm}

\node[circle, draw](k1) at (0,0) {$4$};
\node[circle, draw](k2) at (1*\spac,0) {$6$};
\node[circle, draw](k3) at (2*\spac,0.6*\spac) {$3$};
\node[circle, draw](k4) at (2*\spac,-0.6*\spac) {$3$};

\node[draw, inner sep=0.1cm,minimum size=0.67cm](N1) at (0*\spac,\spac) {$2$};
\node[draw, inner sep=0.1cm,minimum size=0.67cm](N2) at (1*\spac,\spac) {$2$};

\draw[-] (k1) to (k2);
\draw[-] (k2) to (k3);
\draw[-] (k2) to (k4);

\draw (k1) -- (N1);
\draw (k2) -- (N2);

\end{scope}
\end{tikzpicture}};
\node[text width=9em, align=center] at (-0.5,-4) {Simple root subset of $T^{2d}_{m_s\to\infty}$};
\node[text width=9em] at (5,-4) {Weights};
\node[text width=9em] at (10,-4) {2d Gauge Theory};
		\end{tikzpicture}
	\end{center}
	\caption{From the two sets of weights  ${\cal W}_{\cal S}$, we read off the parabolic subalgebra $\fp_{\{\alpha_3,\alpha_4\}}$ of $D_4$. Reinterpreting each weight as a sum of ``minus a fundamental weight and simple roots,'' we obtain two different 2d quiver gauge theories shown on the right. The white arrows imply we take the CFT limit.}
	\label{fig:weightreading4}
\end{figure}


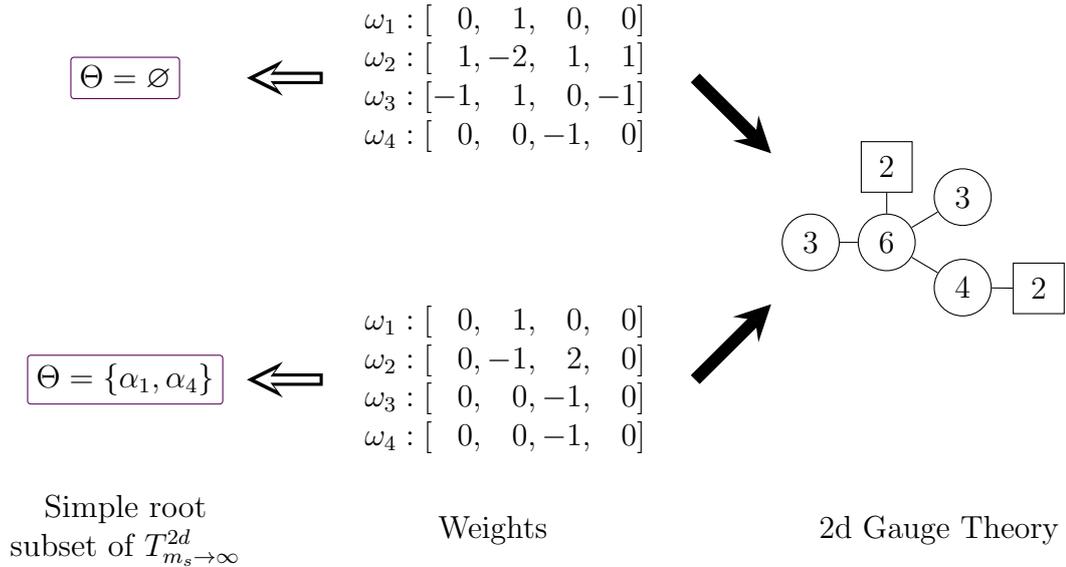
\begin{figure}[htb]
\begin{center}
\begin{tikzpicture}[baseline]
		\tcbset{enhanced,size=fbox,nobeforeafter,tcbox raise=-2.3mm,colback=white,colframe=white}
\node at (-1,2) {\tcbox[colframe=violet!70!black]{$\Theta=\varnothing$}};
		\draw[->, -stealth,  line width=0.4em, postaction={draw,-stealth,white,line width=0.2em,
                shorten <=0.10em,shorten >=0.26em}](1.6,2) -- (0.6,2);

		\node[align=justify] at (4,2) {$\omega_1:[\phantom{-}0,\phantom{-}1,\phantom{-}0,\phantom{-}0]$\\$\omega_2:[\phantom{-}1,-2,\phantom{-}1,\phantom{-}1]$\\$\omega_3:[-1,\phantom{-}1,\phantom{-}0,-1]$\\$\omega_4:[\phantom{-}0,\phantom{-}0,-1,\phantom{-}0]$};
		\draw[->, -stealth,  line width=0.4em](6.5,2) -- (7.5,1);
\node at (-1,-2) {\tcbox[colframe=violet!70!black]{$\Theta=\{\alpha_1,\alpha_4\}$}};
		\draw[->, -stealth,  line width=0.4em, postaction={draw,-stealth,white,line width=0.2em,
                shorten <=0.10em,shorten >=0.26em}](1.6,-2) -- (0.6,-2);
	\node[align=justify] at (4,-2) {$\omega_1:[\phantom{-}0,\phantom{-}1,\phantom{-}0,\phantom{-}0]$\\$\omega_2:[\phantom{-}0,-1,\phantom{-}2,\phantom{-}0]$\\$\omega_3:[\phantom{-}0,\phantom{-}0,-1,\phantom{-}0]$\\$\omega_4:[\phantom{-}0,\phantom{-}0,-1,\phantom{-}0]$};
		\draw[->, -stealth,  line width=0.4em](6.5,-2) -- (7.5,-1);
		\node at (9.5,0) {\begin{tikzpicture}[baseline]
 \begin{scope}[auto, every node/.style={minimum size=0.75cm}]
\def \spac {1cm}

\node[circle, draw](k1) at (0,0) {$3$};
\node[circle, draw](k2) at (1*\spac,0) {$6$};
\node[circle, draw](k3) at (2*\spac,0.6*\spac) {$3$};
\node[circle, draw](k4) at (2*\spac,-0.6*\spac) {$4$};

\node[draw, inner sep=0.1cm,minimum size=0.67cm](N2) at (1*\spac,\spac) {$2$};
\node[draw, inner sep=0.1cm,minimum size=0.67cm](N4) at (3*\spac,-0.6\spac) {$2$};

\draw[-] (k1) to (k2);
\draw[-] (k2) to (k3);
\draw[-] (k2) to (k4);

\draw (k2) -- (N2);
\draw (k4) -- (N4);

\end{scope}
\end{tikzpicture}};
\node[text width=9em,align=center] at (-1,-4) {Simple root subset of $T^{2d}_{m_s\to\infty}$};
\node[text width=9em] at (5,-4) {Weights};
\node[text width=9em] at (10,-4) {2d Gauge Theory};
		\end{tikzpicture}
\end{center}
\caption{Two sets of weights ${\cal W}_{\cal S}$ which spell out the same quiver, but denote two different defects; we see it is really the weights, and not quivers, that define a defect. This is clear in the CFT limit, where two distinct parabolic subalgebras are distinguished.}
\label{fig:samequiverd4}
\end{figure}

\begin{example}[$D_4$ example]
As a nontrivial example, let us first study the set  at the top of Figure \ref{fig:weightreading4} for  $\fg=D_4$:  ${\cal W}_{\cal S}=\{[-1,0,0,0],[1,-1,0,0],[0,1,0,0]\}$.  Except for the two simple roots $\alpha_3$ and $\alpha_4$, all the other positive roots  $e_{\gamma}$ satisfy $\langle e_{\gamma}, \omega_i\rangle <0$ for at least one $\omega_i\in{\cal W}_{\cal S}$. Indeed,  it is easy to check that $\langle\alpha_3, \omega_i\rangle=0=\langle\alpha_4, \omega_i\rangle$ for all the $\omega_i\in{\cal W}_{\cal S}$;
the set of positive roots we obtain defines the nilradical $\fn_{\{\alpha_3,\alpha_4\}}$. We then conclude from the Levi decomposition that ${\cal W}_{\cal S}$ characterizes the parabolic subalgebra $\fp_{\{\alpha_3,\alpha_4\}}$.

Now, in this example, we could have very well studied a different set:  
\[
{\cal W}_{\cal S}=\{[1,0,0,0],[1,0,0,0],[-2,1,0,0],[0,-1,0,0]\},
\]
shown at the bottom of Figure \ref{fig:weightreading4}. It is an easy exercise to show that one identifies the same nilradical  $\fn_{\{\alpha_3,\alpha_4\}}$ as previously, so the same parabolic subalgebra $\fp_{\{\alpha_3,\alpha_4\}}$.
This illustrates that theories $T^{2d}$ that have different quiver descriptions can end up determining the same parabolic subalgebra after taking $m_s$ to infinity.

In particular, the two 2d theories of Figure \ref{fig:weightreading4} have different Coulomb branch dimensions. In the CFT limit, we lose the quiver description of the theories, and the complex Coulomb branch dimension of both theories reduces to 10, which is the dimension of $\fn_{\{\alpha_3,\alpha_4\}}$.
\end{example}

We want to emphasize that throughout this discussion,  it really is the set of weights  ${\cal W}_{\cal S}$, not the resulting quiver, that characterizes a defect, since two different defects in the CFT limit can have the same quiver origin in the little string; see Figure \ref{fig:samequiverd4} for an illustration.\\

Note that the case of  $\fg=A_n$ is special, in  that one can start from a parabolic subalgebra of $A_n$ and obtain a 2d quiver theory from it, without any explicit reference to a set of weights; see Figure \ref{fig:nilradical} for an illustration. 
All the resulting quivers obey equation \ref{conformal}, as a consequence of the Levi decomposition. Indeed, the nilradical gives the exact Coulomb moduli of the quiver, read in a diagonal fashion from a matrix representative. The masses are read off from the Levi subalgebra, since the latter specifies a partition.\\

\begin{figure}[htpb]
	\begin{tabular}{lc}
$\Theta$&$\fl_\Theta$\\[0.3cm]
\noalign{\vskip-0.2cm}\toprule\noalign{\vskip0.2cm}
$\varnothing$&$\begin{pmatrix}
			*&0&0&0\\
			0&*&0&0\\
			0&0&*&0\\
			0&0&0&*
			\end{pmatrix}$\tikzmark{empt}\\[1.0cm]
$\{\alpha_1\}$&$\begin{pmatrix}
			*&*&0&0\\
			*&*&0&0\\
			0&0&*&0\\
			0&0&0&*
			\end{pmatrix}$\tikzmark{a1}\\[1.0cm]
$\{\alpha_2\}$&$\begin{pmatrix}
			*&0&0&0\\
			0&*&*&0\\
			0&*&*&0\\
			0&0&0&*
			\end{pmatrix}$\tikzmark{a2}\\[1.0cm]
$\{\alpha_3\}$&$\begin{pmatrix}
			*&0&0&0\\
			0&*&0&0\\
			0&0&*&*\\
			0&0&*&*
			\end{pmatrix}$\tikzmark{a3}\\[1.0cm]
$\{\alpha_1,\alpha_2\}$&$\begin{pmatrix}
			*&*&*&0\\
			*&*&*&0\\
			*&*&*&0\\
			0&0&0&*
			\end{pmatrix}$\tikzmark{a12}\\[1.0cm]
$\{\alpha_2,\alpha_3\}$&$\begin{pmatrix}
			*&0&0&0\\
			0&*&*&*\\
			0&*&*&*\\
			0&*&*&*
			\end{pmatrix}$\tikzmark{a23}\\[1.0cm]
$\{\alpha_1,\alpha_3\}$&$\begin{pmatrix}
			*&*&0&0\\
			*&*&0&0\\
			0&0&*&*\\
			0&0&*&*
			\end{pmatrix}$\tikzmark{a13}\\[1.0cm]
$\{\alpha_1,\alpha_2,\alpha_3\}$&$\begin{pmatrix}
			*&*&*&*\\
			*&*&*&*\\
			*&*&*&*\\
			*&*&*&*
			\end{pmatrix}$\tikzmark{a123}
	\end{tabular}
	\begin{tikzpicture}[overlay, remember picture]
\draw [decoration={brace,amplitude=0.5em},decorate,ultra thick,gray]
 ($(empt)+(0,1.1)$) --  ($(empt)+(0,-1)$);
 \draw[dotted] ($(empt)+(1.1,0)$) node {[1,1,1,1]};
\draw [decoration={brace,amplitude=0.5em},decorate,ultra thick,gray]
 ($(a1)+(0,1.1)$) --  ($(a3)+(0,-1)$);
 \draw[dotted] ($(a2)+(1,0)$) node {[2,1,1]};
\draw [decoration={brace,amplitude=0.5em},decorate,ultra thick,gray]
 ($(a12)+(0,1.1)$) --  ($(a23)+(0,-1)$);
 \draw[dotted] ($0.5*(a12)+0.5*(a23)+(0.9,0)$) node {[3,1]};
\draw [decoration={brace,amplitude=0.5em},decorate,ultra thick,gray]
 ($(a13)+(0,1.1)$) --  ($(a13)+(0,-1)$);
 \draw[dotted] ($(a13)+(0.9,0)$) node {[2,2]};
\end{tikzpicture}
\hspace*{2cm}
	\begin{tabular}{c}
$\fn_\Theta$\\[0.3cm]
\noalign{\vskip-0.2cm}\toprule\noalign{\vskip0.2cm}
$\begin{pmatrix}
			0&*&*&*\\
			0&0&*&*\\
			0&0&0&*\\
			0&0&0&0
			\end{pmatrix}$\tikzmark{emptnil}\\[1.0cm]
$\begin{pmatrix}
			0&0&*&*\\
			0&0&*&*\\
			0&0&0&*\\
			0&0&0&0
			\end{pmatrix}$\tikzmark{a1nil}\\[1.0cm]
$\begin{pmatrix}
			0&*&*&*\\
			0&0&0&*\\
			0&0&0&*\\
			0&0&0&0
			\end{pmatrix}$\tikzmark{a2nil}\\[1.0cm]
$\begin{pmatrix}
			0&*&*&*\\
			0&0&*&*\\
			0&0&0&0\\
			0&0&0&0
			\end{pmatrix}$\tikzmark{a3nil}\\[1.0cm]
$\begin{pmatrix}
			0&0&0&*\\
			0&0&0&*\\
			0&0&0&*\\
			0&0&0&0
			\end{pmatrix}$\tikzmark{a12nil}\\[1.0cm]
$\begin{pmatrix}
			0&*&*&*\\
			0&0&0&0\\
			0&0&0&0\\
			0&0&0&0
			\end{pmatrix}$\tikzmark{a23nil}\\[1.0cm]
$\begin{pmatrix}
			0&0&*&*\\
			0&0&*&*\\
			0&0&0&0\\
			0&0&0&0
			\end{pmatrix}$\tikzmark{a13nil}\\[1.0cm]
$\begin{pmatrix}
			0&0&0&0\\
			0&0&0&0\\
			0&0&0&0\\
			0&0&0&0
			\end{pmatrix}$\tikzmark{a123nil}
	\end{tabular}
\begin{tikzpicture}[overlay, remember picture,font=\small]
	\draw[blue!50!green] ($(emptnil)+(-1.7,1)$) -- ($(emptnil)+(-0.3,-0.3)$) node {\hspace*{0.6cm} 3};
	\draw[blue!50!green] ($(emptnil)+(-1.15,1)$) -- ($(emptnil)+(-0.3,0.2)$) node {\hspace*{0.6cm} 2};
	\draw[blue!50!green] ($(emptnil)+(-0.6,1)$) -- ($(emptnil)+(-0.3,0.7)$) node {\hspace*{0.6cm} 1};
	\draw[blue!50!green] ($(a1nil)+(-1.7,1)$) -- ($(a1nil)+(-0.3,-0.3)$) node {\hspace*{0.6cm} 2};
	\draw[blue!50!green] ($(a1nil)+(-1.15,1)$) -- ($(a1nil)+(-0.3,0.2)$) node {\hspace*{0.6cm} 2};
	\draw[blue!50!green] ($(a1nil)+(-0.6,1)$) -- ($(a1nil)+(-0.3,0.7)$) node {\hspace*{0.6cm} 1};
	\draw[blue!50!green] ($(a2nil)+(-1.7,1)$) -- ($(a2nil)+(-0.3,-0.3)$) node {\hspace*{0.6cm} 2};
	\draw[blue!50!green] ($(a2nil)+(-1.15,1)$) -- ($(a2nil)+(-0.3,0.2)$) node {\hspace*{0.6cm} 2};
	\draw[blue!50!green] ($(a2nil)+(-0.6,1)$) -- ($(a2nil)+(-0.3,0.7)$) node {\hspace*{0.6cm} 1};
	\draw[blue!50!green] ($(a3nil)+(-1.7,1)$) -- ($(a3nil)+(-0.3,-0.3)$) node {\hspace*{0.6cm} 2};
	\draw[blue!50!green] ($(a3nil)+(-1.15,1)$) -- ($(a3nil)+(-0.3,0.2)$) node {\hspace*{0.6cm} 2};
	\draw[blue!50!green] ($(a3nil)+(-0.6,1)$) -- ($(a3nil)+(-0.3,0.7)$) node {\hspace*{0.6cm} 1};
	\draw[blue!50!green] ($(a12nil)+(-1.7,1)$) -- ($(a12nil)+(-0.3,-0.3)$) node {\hspace*{0.6cm} 1};
	\draw[blue!50!green] ($(a12nil)+(-1.15,1)$) -- ($(a12nil)+(-0.3,0.2)$) node {\hspace*{0.6cm} 1};
	\draw[blue!50!green] ($(a12nil)+(-0.6,1)$) -- ($(a12nil)+(-0.3,0.7)$) node {\hspace*{0.6cm} 1};
	\draw[blue!50!green] ($(a23nil)+(-1.7,1)$) -- ($(a23nil)+(-0.3,-0.3)$) node {\hspace*{0.6cm} 1};
	\draw[blue!50!green] ($(a23nil)+(-1.15,1)$) -- ($(a23nil)+(-0.3,0.2)$) node {\hspace*{0.6cm} 1};
	\draw[blue!50!green] ($(a23nil)+(-0.6,1)$) -- ($(a23nil)+(-0.3,0.7)$) node {\hspace*{0.6cm} 1};
	\draw[blue!50!green] ($(a13nil)+(-1.7,1)$) -- ($(a13nil)+(-0.3,-0.3)$) node {\hspace*{0.6cm} 1};
	\draw[blue!50!green] ($(a13nil)+(-1.15,1)$) -- ($(a13nil)+(-0.3,0.2)$) node {\hspace*{0.6cm} 2};
	\draw[blue!50!green] ($(a13nil)+(-0.6,1)$) -- ($(a13nil)+(-0.3,0.7)$) node {\hspace*{0.6cm} 1};
	\draw [decoration={brace,amplitude=0.5em},decorate,ultra thick,gray]
 ($(emptnil)+(0.5,1.1)$) --  ($(emptnil)+(0.5,-1)$);
 \draw [decoration={brace,amplitude=0.5em},decorate,ultra thick,gray]
 ($(a1nil)+(0.5,1.1)$) --  ($(a3nil)+(0.5,-1)$);
 \draw [decoration={brace,amplitude=0.5em},decorate,ultra thick,gray]
 ($(a12nil)+(0.5,1.1)$) --  ($(a23nil)+(0.5,-1)$);
 \draw [decoration={brace,amplitude=0.5em},decorate,ultra thick,gray]
 ($(a13nil)+(0.5,1.1)$) --  ($(a13nil)+(0.5,-1)$);
 
 \node at ($(emptnil)+(2.5,0.3)$) {\begin{tikzpicture}
 \begin{scope}[auto, every node/.style={minimum size=0.5cm}]
\def \spac {1cm}

\node[circle, draw](k1) at (0,0) {$3$};
\node[circle, draw](k2) at (1*\spac,0) {$2$};
\node[circle, draw](k3) at (2*\spac,0) {$1$};

\node[draw, inner sep=0.1cm,minimum size=0.6cm](N1) at (0*\spac,\spac) {$4$};

\draw[-] (k1) to (k2);
\draw[-] (k2) to (k3);

\draw (k1) -- (N1);

\end{scope}
\end{tikzpicture}};
 \node at ($(a2nil)+(2.5,0.3)$) {\begin{tikzpicture}
 \begin{scope}[auto, every node/.style={minimum size=0.5cm}]
\def \spac {1cm}

\node[circle, draw](k1) at (0,0) {$2$};
\node[circle, draw](k2) at (1*\spac,0) {$2$};
\node[circle, draw](k3) at (2*\spac,0) {$1$};

\node[draw, inner sep=0.1cm,minimum size=0.6cm](N1) at (0*\spac,\spac) {$2$};
\node[draw, inner sep=0.1cm,minimum size=0.6cm](N2) at (1*\spac,\spac) {$1$};

\draw[-] (k1) to (k2);
\draw[-] (k2) to (k3);

\draw (k1) -- (N1);
\draw (k2) -- (N2);

\end{scope}
\end{tikzpicture}};
 \node at ($0.5*(a23nil)+0.5*(a12nil)+(2.5,0.3)$) {\begin{tikzpicture}
 \begin{scope}[auto, every node/.style={minimum size=0.5cm}]
\def \spac {1cm}

\node[circle, draw](k1) at (0,0) {$1$};
\node[circle, draw](k2) at (1*\spac,0) {$1$};
\node[circle, draw](k3) at (2*\spac,0) {$1$};

\node[draw, inner sep=0.1cm,minimum size=0.6cm](N1) at (0*\spac,\spac) {$1$};
\node[draw, inner sep=0.1cm,minimum size=0.6cm](N3) at (2*\spac,\spac) {$1$};

\draw[-] (k1) to (k2);
\draw[-] (k2) to (k3);

\draw (k1) -- (N1);
\draw (k3) -- (N3);

\end{scope}
\end{tikzpicture}};
  \node at ($(a13nil)+(2.5,0.3)$) {\begin{tikzpicture}
 \begin{scope}[auto, every node/.style={minimum size=0.5cm}]
\def \spac {1cm}

\node[circle, draw](k1) at (0,0) {$1$};
\node[circle, draw](k2) at (1*\spac,0) {$2$};
\node[circle, draw](k3) at (2*\spac,0) {$1$};

\node[draw, inner sep=0.1cm,minimum size=0.6cm](N2) at (1*\spac,\spac) {$2$};

\draw[-] (k1) to (k2);
\draw[-] (k2) to (k3);

\draw (k2) -- (N2);

\end{scope}
\end{tikzpicture}};
 \end{tikzpicture}
	\caption{How to read off $A_n$ quiver theories directly from the Levi decomposition of a parabolic subalgebra; here we show $n=3$. The matter content is written as a partition, specified by the Levi subalgebra. The nilradical, read off in diagonal fashion in the upper triangular matrix, gives the Coulomb content. Note the resulting quivers automatically obey the condition \ref{conformal}. This way of reading off a quiver gauge theory directly from a parabolic subalgebra is a peculiarity of the $\fg=A_n$ case.}
	\label{fig:nilradical}
	
\end{figure}

Lastly, there are extra ``special'' punctures which cannot be obtained from the sets of weights  ${\cal W}_{\cal S}$ as defined so far. They are special in the sense that they do not determine a  parabolic subalgebra of $\fg$. We defer the analysis of these extra theories to section \ref{sec:types}. 

\subsection{Parabolic subalgebras from Higgs field data}
\label{sec:higgstopara}
The characterization of defects so far has relied on identifying a subset of simple roots $\Theta$ of the algebra $\fg$. There is yet another way  the above classification can be recovered, which relies on identifying a Levi subalgebra of $\fg$ instead. This Levi subalgebra appears in the Levi decomposition of $\fp_\Theta$ as $\fp_{\Theta}=\fl_{\Theta}\oplus\fn_{\Theta}$. Either way, we obtain the same parabolic subalgebra $\fp_{\Theta}$. Let us  derive this  explicitly.

Recall that the Seiberg--Witten curve of the quiver gauge theory  on the D5 branes  is the spectral curve of the Higgs field $\phi$, taken in some representation $\fR$ of $\fg$ (\cite{Nekrasov:2012xe,Nekrasov:2013xda,Nanopoulos:2009uw}). We described the $m_s$ to infinity limit after which the Seiberg--Witten curve of the theory becomes the spectral curve of the Hitchin integrable system
\begin{equation*}
\det{}_\fR(\phi-p)=0.
\end{equation*} 
After $T^2$ compactification, the same equation is solved by D3 branes instead, so we can say that the above spectral curve is the Seiberg--Witten curve of the two-dimensional theory $T^{2d}_{m_s\rightarrow\infty}$.
At the root of the Higgs branch, where the Coulomb and Higgs branches meet, this expression simplifies: the Higgs field near a puncture of $\cC$ has a pole of order one. After shifting this pole to $z=0$, we get
\beq\label{Scurve}
0=\det\left(p\cdot \mathds{1}-\frac{\sum_{\omega_i\in \cal{W}_S}\hat\beta_i \omega_i}{z} +\text{reg.}\right),
\eeq
where $\cal{W}_S$ is the set of weights introduced in section \ref{sec:review}.
The $\hat\beta_i$ are mass parameters of the gauge theory, which correspond to insertion points of the D3 branes on $\cC$.\\

Thus, the residue at the pole diagonalizes, and the diagonal entries can be interpreted as hypermultiplet masses. So at the root of the Higgs branch, the Higgs field is described by an honest semi-simple element of $\fg$. From this semi-simple element, we can once again recover a parabolic subalgebra  $\fp$.
Indeed, given a semi-simple (diagonalizable) element $S$ (in our cases, we'll always have $S\in \fh$), its centralizer
\begin{equation}
\fg^S:=\{X\in \fg\,\big|\,[X,S]=0\}
\end{equation}
is reductive and is in fact a Levi subalgebra $\fl_S$ of some parabolic subalgebra $\fp_S$.\\

Since the Higgs field at a puncture of $\cC$ has a pole with semi-simple residue, we can use this construction to associate a Levi subalgebra $\fl$ to a defect. The smallest parabolic subalgebra containing $\fl$ is then the parabolic subalgebra defining the theory. Thus, we achieved our goal of building a parabolic subalgebra, starting from a given Higgs field of a quiver theory $T^{2d}$.

\begin{example}
For $\fg=A_2$, assume that the Higgs field has a pole with semi-simple residue $\phi=\frac{S}{z}$ near $z=0$. In the fundamental representation of $\mathfrak{sl}_3$, a possible choice for $S$ is
\begin{equation}
S=\begin{pmatrix}
\beta&0&0\\
0&\beta&0\\
0&0&-2\beta
\end{pmatrix}.
\end{equation}
The Levi subalgebra of $\mathfrak{sl}_3$ associated to this semi-simple element is the centralizer of $S$, which has the form
\begin{equation}
\fg^S=\begin{pmatrix}
*&*&0\\
*&*&0\\
0&0&*
\end{pmatrix}=\fl_{\{\alpha_1\}}
\end{equation}
The parabolic subalgebra associated to this $S$ is then $\fp_{\{\alpha_1\}}$ from example \ref{ex:levisl3}.
\end{example}

\section{Surface defects and Nilpotent Orbits}
\label{sec:nil}
We now explain how the classification of surface defects presented here is connected to the classification of codimension-two defects via nilpotent orbits.
\subsection{A short review}
The characterization of a puncture as studied in the 6d $(2,0)$ CFT literature \cite{Chacaltana:2012zy} is given in terms of a \emph{nilpotent orbit} of the algebra: An element $X\in \fg$ is nilpotent if the matrix representative (in some faithful representation) is a nilpotent matrix. If $X$ is nilpotent, then the whole orbit $\cO_X$ of $X$ under the adjoint action of $G$ is nilpotent -- we call this a nilpotent orbit. For readers interested in details and applications, the textbook \cite{Collingwood:1993} serves as an excellent introduction.

For a simple Lie algebra, the number of such nilpotent orbits is finite, and studying their properties leads to many connections to different branches of representation theory.
For $\fg=A_n$, these orbits are labeled by Young diagrams with $n+1$ boxes; for $\fg=D_n$, they are classified by Young diagrams with $2n$ boxes which satisfy some conditions (see \cite{Collingwood:1993} for details.)

An important fact is that for any nilpotent orbit $\mathcal{O}$, the closure $\xbar{\mathcal{O}}$ is always a union of nilpotent orbits. Furthermore, there is a maximal orbit $\mathcal{O}_{\text{max}}$ whose union contains all other nilpotent orbits of $\fg$. This allows us to define an ordering on these orbits:

Given two nilpotent orbits $\mathcal{O}_1,\mathcal{O}_2\subset\fg$, we define the relation
\begin{equation}
\mathcal{O}_1\preceq\mathcal{O}_2:\Leftrightarrow \mathcal{O}_1\subseteq\xbar{\mathcal{O}_2},
\end{equation}
where $\xbar{\mathcal{O}}$ is the closure in the Zariski topology. This turns the set of all nilpotent orbits into a partially ordered set.

For $A_n$ and $D_n$, this order corresponds to the dominance order of the Young diagrams used to label the orbits.

\begin{example}[$A_3$]
For an $A_n$ nilpotent orbit labeled by a partition $[d_1,\ldots,d_k]$, a matrix representative is given by $k$ Jordan blocks of size $d_i\times d_i$. Taking the example of $n=3$, there are five different nilpotent orbits. Their Hasse diagram can be found below in Figure \ref{fig:hassea3}. For instance, the sub-dominant diagram $[3,1]$ labels the orbit of
\beq
X_{[3,1]}=\begin{pmatrix}
0&1&0&0\\
0&0&1&0\\
0&0&0&1\\
0&0&0&0
\end{pmatrix}.
\eeq

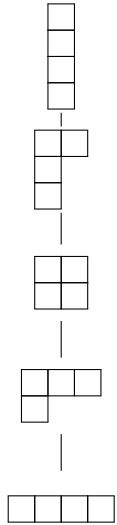
\begin{figure}[htpb]
\centering
\ytableausetup{smalltableaux}
\begin{tikzpicture}[align=center,font=\small]
\begin{scope}[auto, every node/.style={minimum size=1cm,inner sep=1}]
\def \spac {1.5cm}
\def \spacmat {1cm}
\def \outu {0}
\def \inu {180}
\def \outl {200}
\def \inl {340}

\node(k1) at (0,0) {\ydiagram{1,1,1,1}};
\node(k2) at (0,-1*\spac) {\ydiagram{2,1,1}};
\node(k3) at (0,-2*\spac) {\ydiagram{2,2}};
\node(k4) at (0,-3*\spac) {\ydiagram{3,1}};
\node(k5) at (0,-4*\spac) {\ydiagram{4}};

\draw (k1) -- (k2);
\draw (k2) -- (k3);
\draw (k3) -- (k4);
\draw (k4) -- (k5);

\end{scope}

\end{tikzpicture}
\caption{This diagram represents the inclusion relations between the nilpotent orbits of $A_3$.}
\label{fig:hassea3}
\end{figure}
\end{example}

In \cite{Chacaltana:2012zy}, boundary conditions of the 6d $(2,0)$ CFT are determined by solutions to Nahm's equations. These equations admit singular solutions near a puncture which are labeled by embeddings $\rho:\mathfrak{sl}_2\to\fg$. Since $\sigma_+\in \mathfrak{sl}_2$ is nilpotent, its image $\rho(\sigma_+)$ is as well, and defines a nilpotent orbit. By the Jacobson--Morozov theorem, this gives a one-to-one correspondence between such embeddings and nilpotent orbits.
Thus, by dimensional reduction, $\frac12$-BPS surface defects of 4d $\cN=4$ super Yang--Mills are typically labeled by nilpotent orbits.
\subsection{Nilpotent orbits from Levi subalgebras}
\label{sec:nillevi}
Since we now have two different constructions of surface defects, we should explain how we can relate them (a related discussion can be found in \cite{Chacaltana:2012zy}):

Given a parabolic subalgebra $\fp=\fl\oplus\fn$, the nilpotent orbit $\mathcal{O}_\fp$ associated to it is the maximal orbit that has a representative $X\in \mathcal{O}_\fp$ for which $X\in \fn$.
This induced orbit agrees with what is referred to as the Richardson orbit of $\fp$.

If $\fg$ is $A_n$ or $D_n$, this map can be most easily described using the semi-simple pole of the Higgs field. We represent the pole in the first fundamental representation, and assign a Young diagram (with $n+1$ or $2n$ boxes, respectively) to it by counting the multiplicities of the eigenvalues. For $A_n$, these Young diagrams are given by the sizes of the blocks making up the Levi subalgebra $\fl$ (see Figure \ref{fig:nilradical}).

To this Young diagram, we can apply the so-called Spaltenstein map \cite{Spaltenstein:1982}, which gives another Young diagram of the same size \cite{Collingwood:1993}. For $A_n$, this map is just the transposition.

This Young diagram labels the nilpotent orbit describing a defect according to \cite{Chacaltana:2012zy}; adding this nilpotent element to the Higgs field describes a Coulomb deformation of the theory $T^{2d}_{m_s\rightarrow\infty}$, meaning we are moving away from the root of the Higgs branch.

Young diagrams are not available for exceptional Lie algebras, but this correspondence can be described at any rate by using the so-called Bala-Carter labels \cite{Bala:1976msaa,*Bala:1976msab, Haouzi:2016}.

Thus, we get a map which associates one of the theories in \cite{Chacaltana:2012zy} to the 2d theory $T^{2d}_{m_s\rightarrow\infty}$. This was checked explicitly by comparing to the data in \cite{Chacaltana:2011ze,Chacaltana:2010ks,Chacaltana:2014jba}. Furthermore, we will revisit this correspondence when considering the Seiberg--Witten curves of our theories in section \ref{sec:nullstaterels}.

\begin{example}
Let us show how to get the nilpotent orbits of $A_3$ in Figure \ref{fig:hassea3} from parabolic subalgebras. To assign the right nilpotent orbit to them, we take the transpose of the partition describing the Levi subalgebra. The resulting Young diagram labels a nilpotent orbit, which describes a Coulomb deformation of the theory. Since this partition is the same one that is assigned to the pole of the Higgs field (in the first fundamental representation), we can also directly get the nilpotent orbit from the Higgs field data.

The correspondence we get can be read off from Table \ref{tab:spalta3} below.

\begin{table}[htp]
\renewcommand{\arraystretch}{1.2}
\centering
\begin{tabular}{cc}
$\Theta$&$\cO$\\
\hline
$\varnothing$&[4]\\
$\{\alpha_i\}\,\scriptstyle i=1,2,3$&[3,1]\\
$\{\alpha_1,\alpha_2\}$&[2,2]\\
$\{\alpha_1,\alpha_3\}$&[2,1,1]\\
$\{\alpha_1,\alpha_2,\alpha_3\}$&[1,1,1,1]
\end{tabular}
\caption{In this table, we read off which parabolic subalgebras of $A_3$ (labelled by a subset $\Theta$ of positive simple roots) induce which nilpotent orbits $\cO$ (labelled by Young diagrams).}
\label{tab:spalta3}
\end{table}

\end{example}

\clearpage

\section{Surface Defect classification and  $\cW(\fg)$-algebras}
\label{sec:toda}

In \cite{Aganagic:2015cta}, the partition function of the $(2,0)$ $\fg=ADE$ little string on $\cal{C}$ with certain D5 brane defects is shown to be equal to a $q$-deformation of the $\fg$-Toda CFT conformal block on $\cal{C}$, with vertex operators determined by positions and types of defects. 
In this section, we analyze the previous classification of defects of the little string and its relation to parabolic subalgebras from the point of view of the dual $\fg$-type Toda CFT.
Strictly speaking, the theory dual to the little string  is a  $q$-deformation of $\fg$-type Toda, which has a deformed $\cW(\fg)$-algebra symmetry, and is therefore not a CFT \cite{Frenkel:1998}; for an analysis in this deformed setting, see \cite{kimura:2015rgi}. For our purposes, it will be enough to turn off that deformation and work with the usual Toda CFT and its $\cW(\fg)$-algebra symmetry; this is the counterpart to the $m_s$ to infinity limit in the $(2,0)$ little string description, which gives the $(2,0)$ 6d CFT.

\subsection{Levi subalgebras from level-1 null states of Toda CFT}
\label{sec:nullstates}

In free field formalism, the $ADE$ Toda field theory can be written in terms of $n={\rm rk}({\bf g})$ free bosons in two dimensions with a background charge contribution and the Toda potential that couples them:
\beq\label{Todaaction}
S_{Toda} =  \int dz d{\bar z} \;\sqrt g \; g^{z{\bar z}}[\left(\partial_z \vec\varphi\cdot  \partial_{\bar z} \vec\varphi\right)+  \left(\vec{\rho}\cdot \vec\varphi\right)\,Q R + \sum_{a=1}^n e^{\vec{\alpha}_a\cdot\vec\varphi/b} ].
\eeq
The field $\varphi$ is a vector in the $n$-dimensional (co-)weight space, the inner product is the Killing form on the Cartan subalgebra of $\bf g$,  $\vec\rho$ is the Weyl vector, and $Q=b+ 1/b$.  The $\vec \alpha_a$ label the simple positive roots.\\

The Toda CFT has an extended conformal symmetry, a ${\cW}({{\fg}})$-algebra symmetry.
The elements of the Cartan subalgeba $\fh\subset \fg$  define the highest weight states $|\vec\beta\rangle$ of the  $\cW(\fg)$-algebra. It turns out that null states of this algebra  play a crucial role in classifying the defects we have identified from the gauge theory perspective. Indeed, as shown in \cite{Kanno:2009ga} for $\fg=A_n$, punctures can be classified via level 1 null states of the Toda CFT. This is also true for $D_n$ and $E_n$; in this section, we will review how to construct these null states, and we will see that they distinguish the same parabolic subalgebras $\fp_\Theta$ of $\fg$ we encountered before. As we will explain, the set of simple roots $\Theta$ plays a very clear role in the $\cW(\fg)$-algebra null state condition.\\

We can use the vertex operators to construct highest weight states $|\vec\beta\rangle$ of the  $\cW(\fg)$-algebra by acting on the vacuum, $|\vec\beta\rangle=\lim_{z\to0} e^{\vec{\beta}\cdot \vec{\phi}(z)}|0\rangle$. These give rise to a Verma module over $|\vec\beta\rangle$ by acting with $\cW(\fg)$-algebra generators. For some of the $|\vec\beta\rangle$, these representations are degenerate, because they contain a null state; we say that $|\chi\rangle$, in the Verma module over $|\vec\beta\rangle$, is a \emph{level $k$ null state} of the  $\cW(\fg)$-algebra if for all spins $s$:
\begin{align}
W^{(s)}_{n}|\chi\rangle&=0,\quad \forall n>0,\\
W^{(2)}_{0}|\chi\rangle&=(E_{\beta}+k)|\chi\rangle,
\end{align}
where $W^{(2)}_{0}|\vec\beta\rangle=E_{\beta}|\vec\beta\rangle$.\\

The Verma module over $|\vec\beta\rangle$ contains such a null state at level $k$ if the Ka\v c determinant at level $k$ vanishes. For any semi-simple $\fg$, this determinant at level $k$ is a non-zero factor times
\begin{equation}
\label{eq:kac}
\prod_{\substack{\vec\alpha\in\Phi\\m,n\leq k}} \left((\vec{\beta}+\alpha_+\vec{\rho}+\alpha_- \vec{\rho}^{\,\vee})\cdot \vec{\alpha}-(\tfrac12\vec\alpha^2\,m\alpha_++n\alpha_-)\right)^{p_N(k-mn)},
\end{equation}
where $p_N(l)$ counts the partitions of $l$ with $N$ colours and $\Phi$ is the set of all roots of $\fg$ \cite{Bouwknegt:1992wg}. For us, $(\alpha_+,\alpha_-)=(b,1/b)$.

Note that this determinant is invariant only under the shifted action of the Weyl group, 
\begin{equation}
\vec{\beta}\mapsto w(\vec{\beta}+\alpha_+\vec{\rho}+\alpha_- \vec{\rho}^{\,\vee})-(\alpha_+\vec{\rho}+\alpha_- \vec{\rho}^{\,\vee}),
\end{equation}
where $w$ is the ordinary Weyl action.

If $\fg$ is simply laced, and $\vec\alpha=\vec\alpha_i$ is a simple root, the condition that this determinant vanishes can be phrased as
\begin{equation}
\vec\beta\cdot\vec\alpha_i=(1-m)\alpha_++(1-n)\alpha_-.
\end{equation}

We see that any $\vec{\beta}$ with $\vec\beta\cdot \vec\alpha_i=0$ for a \emph{simple root} $\vec\alpha_i$ gives rise to a level 1 null state, and if $Q:=(\alpha_++\alpha_-)\to 0$, a null state at level 1 occurs if $\vec\beta\cdot \vec\alpha=0$ for any $\vec\alpha\in\Phi$. Furthermore, in this limit, the shift in the Weyl group action disappears. It is enough to work in this ``semi-classical'' limit for our purposes, so we will set $Q$ to 0 in what follows.\\ 

We can explicitly construct these null states: Consider the \emph{screening charge operators}
\begin{equation}
Q_i^{\pm}=\oint \frac{dz}{2\pi i} \exp(i\alpha_\pm \vec\alpha_i\cdot \vec\phi)
\end{equation}
and observe that 
\begin{equation}
[W^{(k)}_n,Q^\pm_i]=0.
\end{equation}

The level 1 null state is then
\begin{equation}
S_i^+|\vec{\beta}-\alpha_+\vec{\alpha}_i\rangle.
\end{equation}
Explicit forms of these null states for $\fg=A_n$ or $D_n$ are shown in the examples of section \ref{sec:examples}. The relation to the parabolic subalgebras introduced in section \ref{sec:parastrings} is immediate: we simply associate a generic null state $|\vec\beta\rangle$ satisfying
\[
\vec\beta\cdot\vec\alpha_i=0 \quad \forall \vec\alpha_i\in\Theta
\]
with the parabolic subalgebra $\fp_\Theta$.

We also note that this $\vec\beta$ defines a semi-simple element in $\fg$; this is just the residue of the Higgs field at the puncture, as explained in section \ref{sec:higgstopara}.\\

We  show  next that these these null states induce relations in the Seiberg--Witten curve of the theory $T^{2d}_{m_s\rightarrow\infty}$. Indeed, the Seiberg--Witten curve of $T^{2d}_{m_s\rightarrow\infty}$ \eqref{Scurve} can be obtained from a free field realization of the  $\cW(\fg)$-algebra.  We will simply read off the null states  as relations between the curve coefficients. Generically, these relations only involve semi-simple elements of the algebra $\fg$. In \ref{sec:nullstaterels}, we will see these relations are still preserved when one additionally introduces certain nilpotent deformations.\\

When working in the $q$-deformed setting, the formula for the Ka\v c determinant is an exponentiated version of \eqref{eq:kac} \cite{Bouwknegt:1998da}. This implies that the null states can be defined analogously for the $q$-deformed  $\cW(\fg)$-algebra. \\

\subsection{Seiberg--Witten curves from $\cW(\fg)$-algebras}
\label{sec:curves}

As we reviewed previously, the Seiberg--Witten curve of $T^{2d}_{m_s\rightarrow\infty}$ is the spectral curve equation 
\begin{equation}
\label{eq:swcurve}
\det{}_\fR(\phi-p)=0.
\end{equation}

In our case, $\phi$ has a simple pole such that the residue is a semi-simple element of $\fg$, which we can write as 
\begin{equation}
\vec\beta=\sum\limits_{\omega_i\in W_S}\hat\beta^i \omega_i.
\end{equation}

To find the curve near the pole, which we assume to be at $z=0$, we can just choose some convenient representation $\fR$, where the residue of $\phi$ is diagonal, and given by $\text{diag}(\beta_1,\beta_2,\ldots)=:M$. Then $\phi=\frac{M}{z}+A$, with $A$ a generic element in $\fg$.

We now expand eq.\ \eqref{eq:swcurve} and write the curve as
\begin{equation}
0=\det\left(-p\cdot\1+\frac{M}{z}+A\right)=(-p)^{\dim(\fR)}+\sum_{s}p^{\dim(\fR)-s}\varphi^{(s)},
\end{equation}
where $\varphi^{(s)}$ is a meromorphic differential, i.e.\ $\varphi^{(s)}=\sum\limits_{k=0}^s \frac{\varphi^{(s)}_{k}}{z^k}$, where the $\varphi^{(s)}_k$ are regular functions of $\beta^i$ and $a_{ij}$ (the entries of $A$).

Since $M$ is diagonal, this determinant just picks up the diagonal terms $a_{ii}$ of $A$, which we identify with the gauge couplings of the quiver theory.\\

Now, we can also construct the Seiberg--Witten curve of $T^{2d}_{m_s\rightarrow\infty}$ from the  $\cW(\fg)$-algebra \cite{Kanno:2009ga,Keller:2011ek}: For this, we need to perform a Drinfeld--Sokolov reduction to obtain explicit  $\cW(\fg)$-algebra generators in the free field realization\footnote{We thank  Kris Thielemans for sending us his \texttt{OPEDefs.m} package \cite{Thielemans:1991uw}, which allowed us to do these calculations}.
Setting $Q=0$ gives us a direct connection to the two dimensional quiver  defined by the semi-simple element $\vec\beta\in\fg$ (cf.\ section \ref{sec:higgstopara}): We can identify the poles of the Seiberg-Witten differentials with expectation values of these $\cW(\fg)$ algebra generators in the state $|\vec\beta\rangle$:
\begin{equation}
\varphi^{(s)}=\langle\vec\beta|W^{(s)}|\vec\beta\rangle .
\end{equation}
We checked this relation explicitly for $A_n$ and $D_n$ theories.

\begin{example}
\label{ex:swcurvedet}
Let us look at the curve describing the full puncture for $\fg=A_2$:

Take the fundamental three-dimensional representation of $\mathfrak{sl}_3$ and write
\begin{equation}
M=\begin{pmatrix}
\beta_1&0&0\\
0&\beta_2&0\\
0&0&-\beta_1-\beta_2
\end{pmatrix},\quad
A=\begin{pmatrix}
a_{11}&a_{12}&a_{13}\\
a_{21}&a_{22}&a_{23}\\
a_{31}&a_{32}&-a_{11}-a_{22}\\
\end{pmatrix}.
\end{equation}
Then the curve can be expanded, and we read off the differentials. For example, $\varphi^{(2)}$, the coefficient multiplying $p$, has the form
\begin{equation}
\varphi^{(2)}=\frac{\varphi^{(2)}_2}{z^2}+\frac{\varphi^{(2)}_1}{z}+\varphi^{(2)}_0,
\end{equation}
where \begin{align}
\varphi^{(2)}_2&=\frac12\left(\beta_1^2+\beta_2^2+(-\beta_1-\beta_2)^2\right):=\frac12(\vec\beta)^2,\\ 
\varphi^{(2)}_1&=a_{11}(2\beta_1+\beta_2)+a_{22}(\beta_1+2\beta_2).
\end{align}
Furthermore,
\begin{equation}
\begin{split}
\label{eq:a2l1det}
\varphi^{(3)}_3&=-\beta_1^2\beta_2-\beta_2^2\beta_1,\\
\varphi^{(3)}_2&=a_{11}(-2\beta_1\beta_2-\beta_2^2)+a_{22}(-2\beta_1\beta_2-\beta_1^2).
\end{split}
\end{equation}

Now from the CFT side, for $\fg=A_2$, define $X^j=i\partial\phi^j$. In the fundamental representation, $X^1+X^2+X^3=0$. Then the generators are just the energy momentum tensor
\begin{equation*}
T(z)=W^{(2)}(z)=\frac13 (\normOrd{X^1X^1} +\normOrd{X^2X^2}+\normOrd{X^3X^3} -\normOrd{X^1X^2} - \normOrd{X^1X^3} -\normOrd{X^2X^3})
\end{equation*}
and the spin 3 operator
\begin{equation*}
\begin{split}
W^{(3)}(z)=\normOrd{&\left(\frac23 X^1 - \frac13 X^2 - \frac13 X^3\right)\cdot\left(-\frac13 X^1 + \frac23 X^2 - \frac13 X^3\right)\cdot\\
&\cdot\left(-\frac13 X^1 - \frac13 X^2 + \frac23 X^3\right)}\, .
\end{split}
\end{equation*}

For the full puncture, we find at once that $\langle \vec\beta|L_0|\vec\beta\rangle$ is equal to  $\varphi^{(2)}_2$ from above, while  $\langle \vec\beta|W^{(3)}_{0}|\vec\beta\rangle$ is equal to $\varphi^{(3)}_3$, as expected.
For the level 1 modes, one finds 
\begin{align}
\langle\vec\beta|W^{(2)}_{-1}|\vec\beta\rangle&=(2\beta_1+\beta_2)\langle\vec\beta|j^1_{-1}|\vec\beta\rangle+(\beta_1+2\beta_2)\langle\vec\beta|j^2_{-1}|\vec\beta\rangle,\\
\langle\vec\beta|W^{(3)}_{-1}|\vec\beta\rangle&=(-2\beta_1\beta_2-\beta_2^2)\langle\vec\beta|j^1_{-1}|\vec\beta\rangle+(-\beta_1^2-2\beta_1\beta_2)\langle\vec\beta|j^2_{-1}|\vec\beta\rangle,
\end{align}
where $j_k^i$ denotes the $k$-th mode of $X^i$. 

Observe that this has the form \eqref{eq:a2l1det} if we identify $\langle\vec\beta|j^i_{-1}|\vec\beta\rangle$ with the $i$-th gauge coupling constant.
\end{example}

For more complicated defects, the $\cW(\fg)$-algebra generators will have terms that are derivatives of $X$ --- these are set to zero in the semiclassical $Q\rightarrow 0$ limit we are considering; after doing so, the reasoning is as above.

\subsection{Null state relations}
\label{sec:nullstaterels}
Punctures that are not fully generic are determined by semi-simple elements $\vec\beta\in\fg$ whose Verma modules contain null states at level one. Since the eigenvalues of the level one $\cW(\fg)$-algebra generators appear as coefficients in the curve, the existence of these null states induces some relations between these coefficients.

For $\fg=A_n$ and $D_n$ in the fundamental representation, the pattern is easy to see. The condition $\vec\beta\cdot\vec\alpha=0$ for some positive root $\vec\alpha$ will cause some of the entries of $M=\text{diag}(\beta_1,\beta_2,\ldots)$ to be equal to each other; if the entry $\beta_i$ occurs $k$ times, we get null states by letting the operator
\begin{equation}
\sum_{s} \beta_i^s W_{-1}^{(\dim(\fR)-s)},
\end{equation}
and its $k-1$ derivatives with respect to $\beta_i$, act on $|\vec\beta\rangle$. Thus, each theory induces some characteristic null state relations which are realized in the Seiberg--Witten curve.

We now use this observation to connect these curves to nilpotent orbits: note that all the curves considered so far were written as
\begin{equation}
\label{eq:swsemi}
\det\left(-p\cdot\1+\frac{M}{z}+A\right)=0
\end{equation}
for some diagonal $M$ and a generic $A$ in $\fg$. In the nilpotent orbit literature, the curves considered in \cite{Chacaltana:2010ks,Chacaltana:2011ze,Chacaltana:2014jba} have the form
\begin{equation}
\det\left(-p\cdot\1+\frac{X}{z}+A\right)=0,
\end{equation}
where, again, $A$ is a generic element in $\fg$, and $X$ is a representative of a nilpotent orbit $\cO_X$.

We can now simply  combine these two poles and form a curve of the form
\begin{equation}
\label{eq:swall}
\det\left(-p\cdot\1+e \; \frac{X}{z}+\frac{M}{z}+A\right)=0,
\end{equation}
where $M$ is semi-simple, $X\in\cO_X$ is nilpotent and $e$ is a parameter.
We will test the correspondence between theories defined by nilpotent orbits and theories defined by semi-simple elements from this vantage point.
Recall from section \ref{sec:nil} that the semi-simple element $M\in\fg$ induces a nilpotent orbit $\cO$. We observe the following facts:

\begin{itemize}
\item Whenever an orbit $\cO^\prime\preceq\cO$, it is \emph{always} possible to find an $X\in\cO^\prime$ such that all the null state relations of the curve \eqref{eq:swsemi} are still satisfied by the curve \eqref{eq:swall}.
\item Whenever an orbit $\cO^\prime\npreceq\cO$, it is \emph{never} possible find an $X\in\cO^\prime$ such that all the null state relations of the curve \eqref{eq:swsemi} are still satisfied by the curve \eqref{eq:swall}.
\end{itemize}
This gives a prescription for allowed deformations; from the perspective of the theory $T^{2d}_{m_s\rightarrow\infty}$, this corresponds to leaving the root of the Higgs branch by turning on certain Coulomb moduli.
\begin{example} For $\fg=A_2$, the only interesting state is $\vec\beta=(\beta_1,\beta_1,-2\beta_1)$; we can get the level one coefficients of the curve by setting $\beta_1=\beta_2$ in example \ref{ex:swcurvedet}:
\begin{equation}
\begin{split}
\phi^{(2)}_1=\langle W^{(2)}_{-1}\rangle&=3\beta_1(a_{11}+a_{22}),\\
\phi^{(3)}_2=\langle W^{(3)}_{-1}\rangle&=-3\beta_1^2(a_{11}+a_{22}),
\end{split}
\end{equation}
so we  see that
\begin{equation}
\label{eq:nullstatea2}
\langle W^{(3)}_{-1}\rangle+\beta_1\langle W^{(2)}_{-1}\rangle=0.
\end{equation}
If we now add the nilpotent element $X=\begin{pmatrix}
0&0&1\\
0&0&0\\
0&0&0
\end{pmatrix},$ then
\begin{equation}
\begin{split}
\phi^{(2)}_1&=3\beta_1(a_{11}+a_{22})+e \; a_{31},\\
\phi^{(3)}_2&=-3\beta_1^2(a_{11}+a_{22})-e \; \beta_1 a_{31},
\end{split}
\end{equation}
and the null state relation \eqref{eq:nullstatea2} is still satisfied.
\end{example}

\section{Defects of (2,0) Little String versus Defects of (2,0) CFT}

Up until now, we have been been using the little string theory as a tool to derive codimension-two defects of the  $(2,0)$ CFT, and in particular exhibit the parabolic subalgebras that arise in that limit. In this section, we keep $m_s$ finite, and comment on the classification of defects of the $(2,0)$ little string proper. In particular, as we have emphasized in section 3, when we work with the little string and not its conformal field theory limit, parabolic subalgebras are in general not visible (exceptions are when $\fg=A_n$, as we had illustrated in Figure \ref{fig:nilradical}, and in a few low rank cases when $\fg=D_n$ and $\fg=E_n$.)\\

We also address a question that was not answered so far:  certain nilpotent orbits of $\fg$ are not induced from any parabolic subalgebra. The simplest example would be the minimal nilpotent orbit of $D_4$. These denote nontrivial defects of the $(2,0)$ CFT, so one should ask if they arise at all from the little string, since so far all the quiver theories we constructed distinguished a parabolic subalgebra. We will see that these exotic defects do indeed originate from the little string. To properly analyze them, we must first understand how flowing on the Higgs branch of a defect is realized from representation theory.

\subsection{$T^{2d}$ and Higgs flow as Weight Addition}
\label{ssec:weightadd}

In this section, we describe an effective and purely group-theoretical way to flow on the Higgs branch of different 2d quiver theories  $T^{2d}$, for any simple $\fg$. We show that in the $A_n$ case, this agrees with standard brane engineering and Hanany-Witten transitions \cite{Hanany:1996ie}. As an application, this procedure will  be used to analyze the punctures that fall outside of the parabolic subalgebra classification we have spelled out so far.

Our setup will be the usual one in this paper: we consider the quiver gauge theory $T^{2d}$ that describes the low energy limit of D3 branes  wrapping  2-cycles of the ALE space $X$ times $\mathbb{C}$. The D3 branes are points on the Riemann surface $\cC$ and on the torus $T^2$.\\

We claim that moving on the Higgs branch of  $T^{2d}$ translates to a weight addition procedure in the algebra: this makes use of the fact that a weight belonging to a fundamental representation can always be written as the sum of new weights. Each of them should be in the orbit of some fundamental weight (the two orbits do not have to be the same, here), while obeying the rule that no subset adds up to the zero weight.

After moving on the Higgs branch of $T^{2d}$, we obtain a new 2d theory  $T^{2d'}$ with a new set of weights, but the same curve. When the gauge theory can be engineered using branes, going from $T^{2d}$ to $T^{2d'}$ is called a Hanany--Witten transition \cite{Hanany:1996ie}.   There, as we will see, a D5 brane passing an NS5 brane creates or removes D3 branes stretching between the two. When a brane construction is not available, the weight description we give is still valid, for an arbitrary simply laced Lie algebra..\\

Note that this weight addition formalism also gives a generalization of the S-configuration \cite{Hanany:1996ie}:
No weight in a fundamental representation can ever be written as the sum of two identical weights.

In the $A_n$ case, where we have a brane picture, this statement translates immediately to the S-rule, which is then automatically satisfied. This argument is however applicable to $D_n$ and $E_n$ theories as well, so this gives an $ADE$-type S-rule.

\subsection{Brane Engineering and Weights}

\begin{figure}[h!]
\begin{center}
\resizebox{0.8\textwidth}{!}{%
\begin{tikzpicture}[font=\small]

\node at (0,1) {\begin{tikzpicture}[baseline]
\draw [-] (0,1) -- (0,-1);
\draw [-] (0.4,1) -- (0.4,-1);
\draw [-] (0.8,1) -- (0.8,-1);
\draw [-] (1.2,1) -- (1.2,-1);

\draw[-,red,thick] (0.4,0.6)--(1.6,0.6);

\draw (1.6,0.6) node[cross=3pt,red,thick]{};

\node[align=center] at (0.6,-1.7) {$[-1,1,0]$\\\footnotesize{\color{red}$-w_3+\alpha_2+\alpha_3$}};
\end{tikzpicture}
};


\node at (3,1) {\begin{tikzpicture}[baseline]
\draw [-] (0,1) -- (0,-1);
\draw [-] (0.4,1) -- (0.4,-1);
\draw [-] (0.8,1) -- (0.8,-1);
\draw [-] (1.2,1) -- (1.2,-1);
\draw [-] (1.6,1) -- (1.6,-1);


\draw (0.6,0.6) node[cross=3pt,red,thick]{};

\node[align=center] at (0.8,-1.7) {$[0,-1,0,0]$\\\footnotesize{\color{red}$-w_2$}};
\end{tikzpicture}
};


\node at (6,1) {\begin{tikzpicture}[baseline]
\draw [-] (0,1) -- (0,-1);
\draw [-] (0.4,1) -- (0.4,-1);
\draw [-] (0.8,1) -- (0.8,-1);

\draw[-,red,thick] (0.4,0.6)--(0.6,0.6);

\draw (0.6,0.6) node[cross=3pt,red,thick]{};

\node[align=center] at (0.4,-1.7) {$[-1,0]$\\\footnotesize{\color{red}$-w_1$}};
\end{tikzpicture}
};


\node at (0,-3) {\begin{tikzpicture}[baseline]
\draw [-] (0,1) -- (0,-1);
\draw [-] (0.4,1) -- (0.4,-1);
\draw [-] (0.8,1) -- (0.8,-1);

\draw[-,red,thick] (0.4,0.63)--(1.2,0.63);
\draw[-,red,thick] (0.8,0.57)--(1.2,0.57);

\draw (1.2,0.6) node[cross=3pt,red,thick]{};

\node[align=center] at (0.4,-1.7) {$[-1,0]$\\\footnotesize{\color{red}$-w_1$}};
\end{tikzpicture}
};


\node at (3,-3) {\begin{tikzpicture}[baseline]
\draw [-] (0,1) -- (0,-1);
\draw [-] (0.4,1) -- (0.4,-1);
\draw [-] (0.8,1) -- (0.8,-1);
\draw [-] (1.2,1) -- (1.2,-1);
\draw [-] (1.6,1) -- (1.6,-1);

\draw[-,red,thick] (0.8,0.6)--(1.4,0.6);

\draw (1.4,0.6) node[cross=3pt,red,thick]{};

\node[align=center] at (0.8,-1.7) {$[0,-1,1,-1]$\\\footnotesize{\color{red}$-w_3+\alpha_3$}};
\end{tikzpicture}
};


\node at (6,-3) {\begin{tikzpicture}[baseline]
\draw [-] (0,1) -- (0,-1);
\draw [-] (0.4,1) -- (0.4,-1);
\draw [-] (0.8,1) -- (0.8,-1);
\draw [-] (1.2,1) -- (1.2,-1);
\draw [-] (1.6,1) -- (1.6,-1);

\draw[-,red,thick] (0.6,0.6)--(1.6,0.6);

\draw (0.6,0.6) node[cross=3pt,red,thick]{};

\node[align=center] at (0.8,-1.7) {$[0,-1,0,1]$\\\footnotesize{\color{red}$-w_3+\alpha_3+\alpha_4$}};
\end{tikzpicture}
};
\end{tikzpicture}%
}
\end{center}
\label{fig:branereading}
\caption{How to read off weights from a system of D3, D5, and NS5 branes.}
\end{figure}
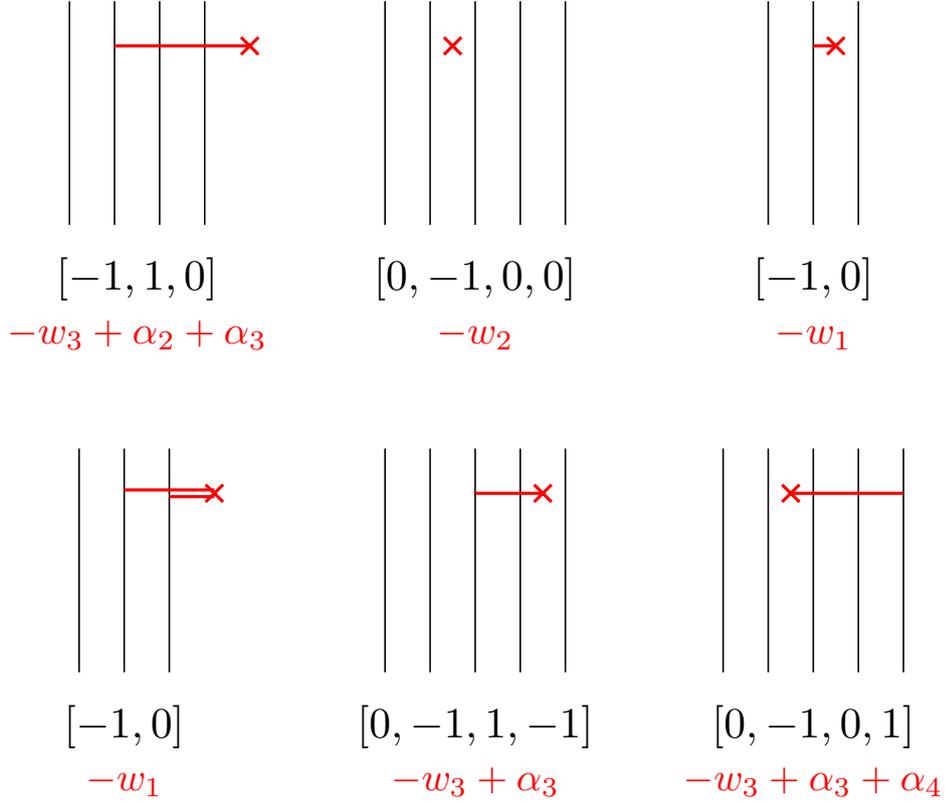

For $A_n$ theories and $D_n$ theories obtainable by an orbifolding procedure, the above discussion can be realized by brane engineering of the theory. 
We can conveniently represent the weights of the algebra, and in particular, their Dynkin labels, using a configuration of D3 branes stretching between NS5's and D5 branes. To see how this works, let us focus on the i-th Dynkin label of a weight:

\begin{itemize}
	\item
	A D3 brane coming from the left ending on the $i-th$ NS5 contributes $-1$ to the weight's i-th label. 
	\item
	A D3 brane coming from the right ending on the $i$-th NS5 contributes $+1$ to the weight's $i$-th label. 
	\item
	A D3 brane coming from the left ending on the $i+1$-th NS5 contributes $+1$ to the weight's $i$-th label.  
	\item
	A D3 brane coming from the right ending on the $i+1$-th NS5 contributes $-1$ to the weight's $i$-th label. 
	\item
	Finally, a D5 brane present between the $i$-th and $i+1$-th NS5's contributes $-1$ to the weight's $i$-th label.  
\end{itemize}

All in all, a D3 brane stretching between a D5 brane and an NS5 brane (while possibly going through some other NS5 branes) produces a weight, whose Dynkin labels are a combination of 1's, $-1$'s, and 0's. The map is not injective: for a given weight, there can be many brane configurations. 

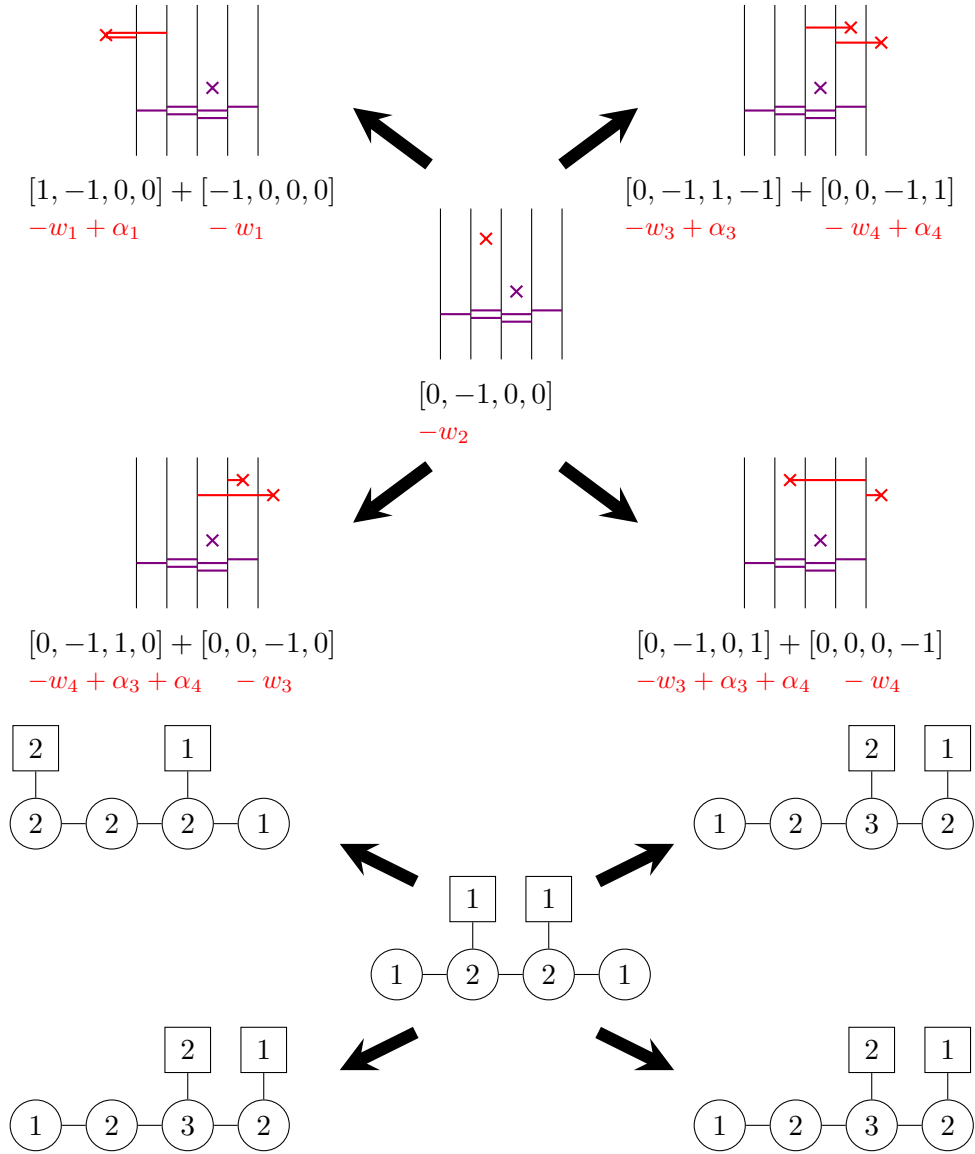
\begin{figure}[h!]
\begin{subfigure}{1.0\textwidth}
\centering
\begin{tikzpicture}[font=\small]

\node at (0,3) {\begin{tikzpicture}[baseline]
\draw [-] (0,1) -- (0,-1);
\draw [-] (0.4,1) -- (0.4,-1);
\draw [-] (0.8,1) -- (0.8,-1);
\draw [-] (1.2,1) -- (1.2,-1);
\draw [-] (1.6,1) -- (1.6,-1);

\draw[-,red,thick] (-0.4,0.57)--(0,0.57);
\draw[-,red,thick] (-0.4,0.63)--(0.4,0.63);

\draw[-,violet,thick] (0,-0.4)--(0.4,-0.4);
\draw[-,violet,thick] (0.4,-0.35)--(0.8,-0.35);
\draw[-,violet,thick] (0.4,-0.45)--(0.8,-0.45);
\draw[-,violet,thick] (0.8,-0.4)--(1.2,-0.4);
\draw[-,violet,thick] (0.8,-0.5)--(1.2,-0.5);
\draw[-,violet,thick] (1.2,-0.35)--(1.6,-0.35);

\draw (-0.4,0.6) node[cross=3pt,red,thick]{};
\draw (1.0,-0.1) node[cross=3pt,violet,thick]{};

\node[align=left] at (0.6,-1.7) {$[1,-1,0,0]+[-1,0,0,0]$\\\footnotesize{\color{red}$-w_1+\alpha_1\qquad\; -w_1$}};
\end{tikzpicture}
};

\node at (8,3) {\begin{tikzpicture}[baseline]
\draw [-] (0,1) -- (0,-1);
\draw [-] (0.4,1) -- (0.4,-1);
\draw [-] (0.8,1) -- (0.8,-1);
\draw [-] (1.2,1) -- (1.2,-1);
\draw [-] (1.6,1) -- (1.6,-1);

\draw[-,red,thick] (0.8,0.7)--(1.4,0.7);
\draw[-,red,thick] (1.2,0.5)--(1.8,0.5);

\draw[-,violet,thick] (0,-0.4)--(0.4,-0.4);
\draw[-,violet,thick] (0.4,-0.35)--(0.8,-0.35);
\draw[-,violet,thick] (0.4,-0.45)--(0.8,-0.45);
\draw[-,violet,thick] (0.8,-0.4)--(1.2,-0.4);
\draw[-,violet,thick] (0.8,-0.5)--(1.2,-0.5);
\draw[-,violet,thick] (1.2,-0.35)--(1.6,-0.35);

\draw (1.4,0.7) node[cross=3pt,red,thick]{};
\draw (1.8,0.5) node[cross=3pt,red,thick]{};
\draw (1.0,-0.1) node[cross=3pt,violet,thick]{};

\node[align=left] at (0.6,-1.7) {$[0,-1,1,-1]+[0,0,-1,1]$\\\footnotesize{\color{red}$-w_3+\alpha_3\qquad\quad -w_4+\alpha_4$}};
\end{tikzpicture}
};

\node at (4,0.3) {\begin{tikzpicture}[baseline]
\draw [-] (0,1) -- (0,-1);
\draw [-] (0.4,1) -- (0.4,-1);
\draw [-] (0.8,1) -- (0.8,-1);
\draw [-] (1.2,1) -- (1.2,-1);
\draw [-] (1.6,1) -- (1.6,-1);


\draw[-,violet,thick] (0,-0.4)--(0.4,-0.4);
\draw[-,violet,thick] (0.4,-0.35)--(0.8,-0.35);
\draw[-,violet,thick] (0.4,-0.45)--(0.8,-0.45);
\draw[-,violet,thick] (0.8,-0.4)--(1.2,-0.4);
\draw[-,violet,thick] (0.8,-0.5)--(1.2,-0.5);
\draw[-,violet,thick] (1.2,-0.35)--(1.6,-0.35);

\draw (0.6,0.6) node[cross=3pt,red,thick]{};
\draw (1.0,-0.1) node[cross=3pt,violet,thick]{};

\node[align=left] at (0.6,-1.7) {$[0,-1,0,0]$\\\footnotesize{\color{red}$-w_2$}};
\end{tikzpicture}
};

\node at (0,-3) {\begin{tikzpicture}[baseline]
\draw [-] (0,1) -- (0,-1);
\draw [-] (0.4,1) -- (0.4,-1);
\draw [-] (0.8,1) -- (0.8,-1);
\draw [-] (1.2,1) -- (1.2,-1);
\draw [-] (1.6,1) -- (1.6,-1);

\draw[-,red,thick] (1.2,0.7)--(1.4,0.7);
\draw[-,red,thick] (0.8,0.5)--(1.8,0.5);

\draw[-,violet,thick] (0,-0.4)--(0.4,-0.4);
\draw[-,violet,thick] (0.4,-0.35)--(0.8,-0.35);
\draw[-,violet,thick] (0.4,-0.45)--(0.8,-0.45);
\draw[-,violet,thick] (0.8,-0.4)--(1.2,-0.4);
\draw[-,violet,thick] (0.8,-0.5)--(1.2,-0.5);
\draw[-,violet,thick] (1.2,-0.35)--(1.6,-0.35);

\draw (1.4,0.7) node[cross=3pt,red,thick]{};
\draw (1.8,0.5) node[cross=3pt,red,thick]{};
\draw (1.0,-0.1) node[cross=3pt,violet,thick]{};

\node[align=left] at (0.6,-1.7) {$[0,-1,1,0]+[0,0,-1,0]$\\\footnotesize{\color{red}$-w_4+\alpha_3+\alpha_4\quad -w_3$}};
\end{tikzpicture}
};

\node at (8,-3) {\begin{tikzpicture}[baseline]
\draw [-] (0,1) -- (0,-1);
\draw [-] (0.4,1) -- (0.4,-1);
\draw [-] (0.8,1) -- (0.8,-1);
\draw [-] (1.2,1) -- (1.2,-1);
\draw [-] (1.6,1) -- (1.6,-1);

\draw[-,red,thick] (0.6,0.7)--(1.6,0.7);
\draw[-,red,thick] (1.6,0.5)--(1.8,0.5);

\draw[-,violet,thick] (0,-0.4)--(0.4,-0.4);
\draw[-,violet,thick] (0.4,-0.35)--(0.8,-0.35);
\draw[-,violet,thick] (0.4,-0.45)--(0.8,-0.45);
\draw[-,violet,thick] (0.8,-0.4)--(1.2,-0.4);
\draw[-,violet,thick] (0.8,-0.5)--(1.2,-0.5);
\draw[-,violet,thick] (1.2,-0.35)--(1.6,-0.35);

\draw (0.6,0.7) node[cross=3pt,red,thick]{};
\draw (1.8,0.5) node[cross=3pt,red,thick]{};
\draw (1.0,-0.1) node[cross=3pt,violet,thick]{};

\node[align=left] at (0.6,-1.7) {$[0,-1,0,1]+[0,0,0,-1]$\\\footnotesize{\color{red}$-w_3+\alpha_3+\alpha_4\quad -w_4$}};
\end{tikzpicture}
};

\draw[->, -stealth,  line width=0.4em](3.25,2.5) -- (2.25,3.25);
\draw[->, -stealth,  line width=0.4em](5,2.5) -- (6,3.25);
\draw[->, -stealth,  line width=0.4em](3.25,-1.5) -- (2.25,-2.25);
\draw[->, -stealth,  line width=0.4em](5,-1.5) -- (6,-2.25);
\end{tikzpicture}
\label{fig:FourTransitions}
\end{subfigure}
\\
\begin{subfigure}{1.0\textwidth}
\centering
\begin{tikzpicture}[font=\small]
\node at (-1,2) {\begin{tikzpicture}[baseline]
\begin{scope}[auto, every node/.style={minimum size=0.5cm}]
\def \spac {1cm}

\node[circle, draw](k1) at (0,0) {$2$};
\node[circle, draw](k2) at (1*\spac,0) {$2$};
\node[circle, draw](k3) at (2*\spac,0) {$2$};
\node[circle, draw](k4) at (3*\spac,0) {$1$};

\node[draw, inner sep=0.1cm,minimum size=0.6cm](N1) at (0*\spac,\spac) {$2$};
\node[draw, inner sep=0.1cm,minimum size=0.6cm](N3) at (2*\spac,\spac) {$1$};

\draw[-] (k1) to (k2);
\draw[-] (k2) to (k3);
\draw[-] (k3) to (k4);

\draw (k1) -- (N1);
\draw (k3) -- (N3);

\end{scope}
\end{tikzpicture}
};
\node at (8,2) {\begin{tikzpicture}[baseline]
\begin{scope}[auto, every node/.style={minimum size=0.5cm}]
\def \spac {1cm}

\node[circle, draw](k1) at (0,0) {$1$};
\node[circle, draw](k2) at (1*\spac,0) {$2$};
\node[circle, draw](k3) at (2*\spac,0) {$3$};
\node[circle, draw](k4) at (3*\spac,0) {$2$};

\node[draw, inner sep=0.1cm,minimum size=0.6cm](N3) at (2*\spac,\spac) {$2$};
\node[draw, inner sep=0.1cm,minimum size=0.6cm](N4) at (3*\spac,\spac) {$1$};

\draw[-] (k1) to (k2);
\draw[-] (k2) to (k3);
\draw[-] (k3) to (k4);

\draw (k3) -- (N3);
\draw (k4) -- (N4);

\end{scope}
\end{tikzpicture}
};
\node at (3.75,0) {\begin{tikzpicture}[baseline]
\begin{scope}[auto, every node/.style={minimum size=0.5cm}]
\def \spac {1cm}

\node[circle, draw](k1) at (0,0) {$1$};
\node[circle, draw](k2) at (1*\spac,0) {$2$};
\node[circle, draw](k3) at (2*\spac,0) {$2$};
\node[circle, draw](k4) at (3*\spac,0) {$1$};

\node[draw, inner sep=0.1cm,minimum size=0.6cm](N2) at (1*\spac,\spac) {$1$};
\node[draw, inner sep=0.1cm,minimum size=0.6cm](N3) at (2*\spac,\spac) {$1$};

\draw[-] (k1) to (k2);
\draw[-] (k2) to (k3);
\draw[-] (k3) to (k4);

\draw (k2) -- (N2);
\draw (k3) -- (N3);

\end{scope}
\end{tikzpicture}
};
\node at (-1,-2) {\begin{tikzpicture}[baseline]
\begin{scope}[auto, every node/.style={minimum size=0.5cm}]
\def \spac {1cm}

\node[circle, draw](k1) at (0,0) {$1$};
\node[circle, draw](k2) at (1*\spac,0) {$2$};
\node[circle, draw](k3) at (2*\spac,0) {$3$};
\node[circle, draw](k4) at (3*\spac,0) {$2$};

\node[draw, inner sep=0.1cm,minimum size=0.6cm](N3) at (2*\spac,\spac) {$2$};
\node[draw, inner sep=0.1cm,minimum size=0.6cm](N4) at (3*\spac,\spac) {$1$};

\draw[-] (k1) to (k2);
\draw[-] (k2) to (k3);
\draw[-] (k3) to (k4);

\draw (k3) -- (N3);
\draw (k4) -- (N4);

\end{scope}
\end{tikzpicture}
};
\node at (8,-2) {\begin{tikzpicture}[baseline]
\begin{scope}[auto, every node/.style={minimum size=0.5cm}]
\def \spac {1cm}

\node[circle, draw](k1) at (0,0) {$1$};
\node[circle, draw](k2) at (1*\spac,0) {$2$};
\node[circle, draw](k3) at (2*\spac,0) {$3$};
\node[circle, draw](k4) at (3*\spac,0) {$2$};

\node[draw, inner sep=0.1cm,minimum size=0.6cm](N3) at (2*\spac,\spac) {$2$};
\node[draw, inner sep=0.1cm,minimum size=0.6cm](N4) at (3*\spac,\spac) {$1$};

\draw[-] (k1) to (k2);
\draw[-] (k2) to (k3);
\draw[-] (k3) to (k4);

\draw (k3) -- (N3);
\draw (k4) -- (N4);

\end{scope}
\end{tikzpicture}
};

\draw[->, -stealth,  line width=0.4em](2.5,0.75) -- (1.5,1.25);
\draw[->, -stealth,  line width=0.4em](4.9,0.75) -- (5.9,1.25);
\draw[->, -stealth,  line width=0.4em](2.5,-1.25) -- (1.5,-1.75);
\draw[->, -stealth,  line width=0.4em](4.9,-1.25) -- (5.9,-1.75);
\end{tikzpicture}
\label{fig:FourQuivers}
\end{subfigure}
\caption{Flowing on the Higgs branch of $T^{2d}$: starting from the theory in the middle, these are all the theories one can obtain by replacing the weight on node 2 by a sum of two weights. The top picture shows the detailed brane picture for each of the quivers.  These all have a low-energy 2d quiver gauge theory description (the ones shown below). At the root of the Higgs branch, the partition functions of all 5 theories are equal.}
\label{fig:quivtrans}
\end{figure}

So the Dynkin labels record the total charge of the D3 brane configuration. The statement that the sum of weights is 0 is then a statement about vanishing of D3 brane flux. Note that the configuration of branes spells out a quiver gauge theory at low energies, which is the expected theory $T^{2d}$ we would write based on the weight data $\cW_S$. See Figure  \ref{fig:quivtrans}  for some examples.\\

\subsection{Polarized and Unpolarized Punctures of the Little String}
\label{sec:types}

\begin{figure}[htb]
	\begin{subfigure}{1.0\textwidth}
	\centering
\begin{tikzpicture}[font=\footnotesize]
\draw[dashed, line width=1.5pt, green!70!black] (0,3.5) -- (0,-3.5);
\draw (-3.6,3) -- (-3.6,-3);
\draw (-2.8,3) -- (-2.8,-3);
\draw (-2,3) -- (-2,-3);
\draw (-1.2,3) -- (-1.2,-3);
\draw (-0.4,3) -- (-0.4,-3);
\draw (0.4,3) -- (0.4,-3);
\draw (1.2,3) -- (1.2,-3);
\draw (2,3) -- (2,-3);
\draw (2.8,3) -- (2.8,-3);
\draw (3.6,3) -- (3.6,-3);

\draw[white, line width=10pt] (-3.7,0) -- (3.7,0);

\node(c1) at (1.6,-1.5) {};
\node at ($(c1)+(0,0.5)$) {\color{red!70!yellow}$\omega_1$};

\draw[-,line width=5pt, red!70!yellow] ($(c1)+(-0.2,0.2)$) -- ($(c1)+(0.2,-0.2)$);
\draw[-,line width=5pt,red!70!yellow] ($(c1)+(-0.2,-0.2)$) -- ($(c1)+(0.2,0.2)$);

\node(c3) at (-1.6,-1.5) {};
\node at ($(c3)+(0,0.5)$) {\color{red!70!yellow}$\omega_1$};

\draw[-,line width=5pt, red!70!yellow] ($(c3)+(-0.2,0.2)$) -- ($(c3)+(0.2,-0.2)$);
\draw[-,line width=5pt, red!70!yellow] ($(c3)+(-0.2,-0.2)$) -- ($(c3)+(0.2,0.2)$);

\node(c2) at (3.2,2) {};
\node at ($(c2)+(0,0.5)$) {\color{red}$\omega_1$};

\draw[-,line width=5pt, red] ($(c2)+(-0.2,0.2)$) -- ($(c2)+(0.2,-0.2)$);
\draw[-,line width=5pt, red] ($(c2)+(-0.2,-0.2)$) -- ($(c2)+(0.2,0.2)$);

\node(c4) at (-3.2,2) {};
\node at ($(c4)+(0,0.5)$) {\color{red}$\omega_1$};

\draw[-,line width=5pt, red] ($(c4)+(-0.2,0.2)$) -- ($(c4)+(0.2,-0.2)$);
\draw[-,line width=5pt, red] ($(c4)+(-0.2,-0.2)$) -- ($(c4)+(0.2,0.2)$);

\draw[red!70!yellow] (-2.8,-1.4) -- (2,-1.4);
\draw[red!70!yellow] (-2,-1.6) -- (2.8,-1.6);

\node[align=left,text width=6.8cm] at (7.5,2) {$\omega_1: [-1,0,0,0,0]=\color{red}-w_1$};
\node[align=left,text width=6.8cm] at (7.5,-1.5) {$\omega_1: [-1,0,0,0,0]=\color{red!70!yellow}-w_3+\alpha_2+2\alpha_3+\alpha_4+\alpha_5$};

\end{tikzpicture}
	\caption{The brane realization of the weight $[-1,0,0,0,0]$ of $D_5$. We started with $A_9$ theory and performed a $\mathbb{Z}_2$ orbifold to obtain the picture. This weight can be written in two ways: by placing the D5 brane between the first two NS5 branes (top), the weight is written in an ``appropriate way'' . By placing the D5 brane between the ``wrong'' set of NS5 branes (bottom), the resulting quiver will be unpolarized and will not distinguish a parabolic subalgebra.}
	\label{fig:cases1and2}
\end{subfigure}
\\
\begin{subfigure}{1.0\textwidth}
	\centering
		\begin{tikzpicture}[font=\footnotesize]
\draw[dashed, line width=1.5pt, green!70!black] (0,3.5) -- (0,-0.5);
\draw (-2.8,3) -- (-2.8,0);
\draw (-2,3) -- (-2,0);
\draw (-1.2,3) -- (-1.2,0);
\draw (-0.4,3) -- (-0.4,0);
\draw (0.4,3) -- (0.4,0);
\draw (1.2,3) -- (1.2,0);
\draw (2,3) -- (2,0);
\draw (2.8,3) -- (2.8,0);

\node(c2) at (1.6,1.5) {};
\node at ($(c2)+(0,0.5)$) {\color{red}$\omega_1$};

\draw[-,line width=5pt, red!70!yellow] ($(c2)+(-0.2,0.2)$) -- ($(c2)+(0.2,-0.2)$);
\draw[-,line width=5pt, red!70!yellow] ($(c2)+(-0.2,-0.2)$) -- ($(c2)+(0.2,0.2)$);

\node(c4) at (-1.6,1.5) {};
\node at ($(c4)+(0,0.5)$) {\color{red}$\omega_1$};

\draw[-,line width=5pt, red!70!yellow] ($(c4)+(-0.2,0.2)$) -- ($(c4)+(0.2,-0.2)$);
\draw[-,line width=5pt, red!70!yellow] ($(c4)+(-0.2,-0.2)$) -- ($(c4)+(0.2,0.2)$);

\draw[red!70!yellow] (-2.8,1.6) -- (2,1.6);
\draw[red!70!yellow] (-2,1.4) -- (2.8,1.4);

\node[align=left,text width=6.8cm] at (7,1.5) {$\omega_1: [0,0,0,0]=\color{red!70!yellow}-w_2+\alpha_1+2\alpha_2+\alpha_3+\alpha_4$};

\end{tikzpicture}
	\caption{Simplest unpolarized defect: the null weight $[0,0,0,0]$ of $D_4$, realized here with branes. We started with $A_7$ theory and performed a $\mathbb{Z}_2$ orbifold to obtain the picture. This theory does not distinguish a parabolic subalgebra.}
	\label{fig:case3}
\end{subfigure}
\end{figure}
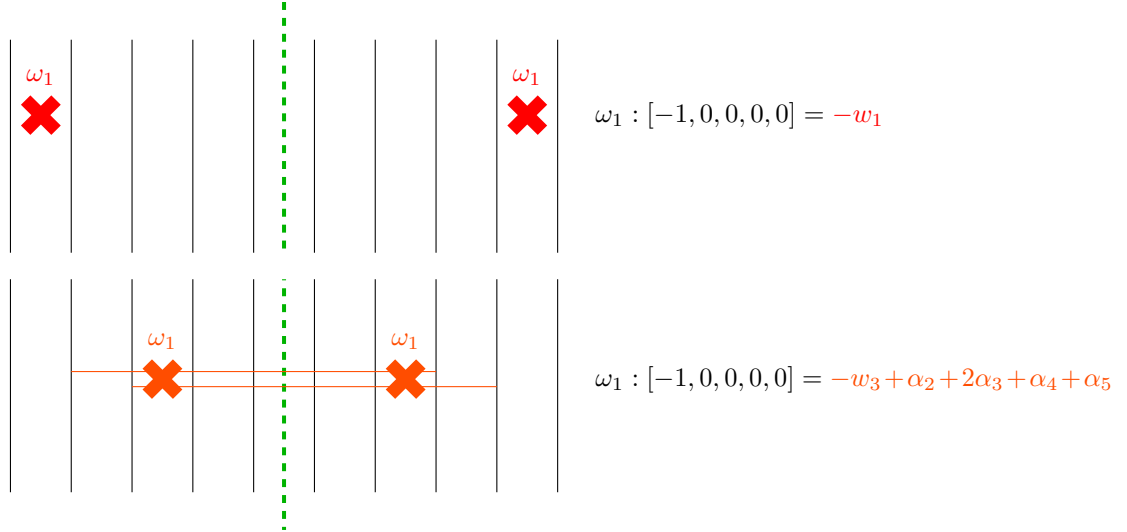

We finally come to the description of defects in the little string that happen to fall outside the parabolic subalgebra classification we have spelled out so far.\\

Suppose we pick a weight in the $i$-th fundamental representation. Unless it is the null weight, it is in the orbit of one and only one fundamental weight, say the $j$-th one. In our entire discussion so far, and in all the examples of \cite{Aganagic:2015cta}, we had $i=j$. In terms of the gauge theory, if all weights are chosen so that $i=j$, then  $T^{2d}_{m_s\rightarrow\infty}$  distinguishes a parabolic subalgebra, as explained in section \ref{ssec:3d}.
We call such a 2d theory \emph{polarized}. \footnote{The terminology here comes from the fact that the parabolic subgroup $\mathcal{P}$ in $T^*(G/\mathcal{P})$ is often called a polarization of some nilpotent orbit $\cO$, through the resolution map $T^*(G/\mathcal{P})\rightarrow \overline{\cO}$, with $\overline{\cO}$ the closure of $\cO$.}\\

However, in the $D_n$ and $E_n$ cases, it can also happen that  $i\neq j$, or that the weight we pick is the null weight. See Figures \ref{fig:cases1and2} and \ref{fig:case3}. In terms of the gauge theory, if \emph{at least one} of the weights in ${\cal W}_{\cal S}$ falls under this category, the theory $T^{2d}_{m_s\rightarrow\infty}$ does \emph{not} distinguish a parabolic subalgebra.
We call such a 2d theory \emph{unpolarized}.\\

We saw in section \ref{ssec:weightadd} that if we start with a polarized theory $T^{2d}$, then after flowing on the Higgs branch, we still end up with a polarized theory $T^{2d'}$.
What happens to unpolarized theories? If we start with a such a theory $T^{2d}$, then after moving on the Higgs branch, it is in fact always possible to end up with a theory $T^{2d'}$ that is polarized. This resulting polarized theory $T^{2d'}$ is  of course highly specialized, since some masses have to be set equal to each other as a result of the Higgs flow.

This is the viewpoint we take to analyze all unpolarized theories: we will flow on the Higgs branch until they transition to polarized theories. In practice, it means that every ``problematic'' weight in an unpolarized theory can be written as a sum of weights to give a polarized theory. Note that for $A_n$, every quiver theory $T^{2d}$ is polarized, while this is not the case for $D_n$ and $E_n$. An illustration of how one can start with an unpolarized theory and arrive at a polarized theory is shown in Figure \ref{fig:D4null} below.\\

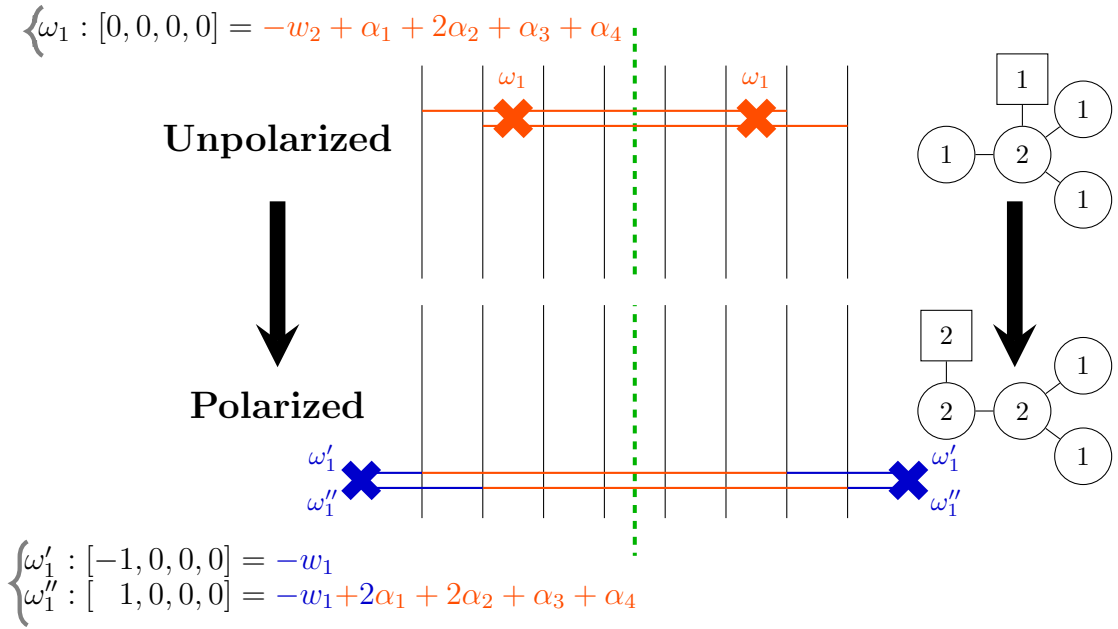
\begin{figure}[h!]
	\begin{center}
\begin{tikzpicture}[font=\footnotesize]
\draw[dashed, line width=1.5pt, green!70!black] (0,3.5) -- (0,-3.5);
\draw (-2.8,3) -- (-2.8,-3);
\draw (-2,3) -- (-2,-3);
\draw (-1.2,3) -- (-1.2,-3);
\draw (-0.4,3) -- (-0.4,-3);
\draw (0.4,3) -- (0.4,-3);
\draw (1.2,3) -- (1.2,-3);
\draw (2,3) -- (2,-3);
\draw (2.8,3) -- (2.8,-3);

\node(c1) at (3.6,-2.5) {};
\node at ($(c1)+(0.5,0.3)$) {\color{blue!80!black}$\omega_1^\prime$};
\node at ($(c1)+(0.5,-0.3)$) {\color{blue!80!black}$\omega_1^{\prime\prime}$};

\draw[-,line width=5pt, blue!80!black] ($(c1)+(-0.2,0.2)$) -- ($(c1)+(0.2,-0.2)$);
\draw[-,line width=5pt,blue!80!black] ($(c1)+(-0.2,-0.2)$) -- ($(c1)+(0.2,0.2)$);

\node(c3) at (-3.6,-2.5) {};
\node at ($(c3)+(-0.5,0.3)$) {\color{blue!80!black}$\omega_1^{\prime}$};
\node at ($(c3)+(-0.5,-0.3)$) {\color{blue!80!black}$\omega_1^{\prime\prime}$};

\draw[-,line width=5pt, blue!80!black] ($(c3)+(-0.2,0.2)$) -- ($(c3)+(0.2,-0.2)$);
\draw[-,line width=5pt, blue!80!black] ($(c3)+(-0.2,-0.2)$) -- ($(c3)+(0.2,0.2)$);

\node(c2) at (1.6,2.3) {};
\node at ($(c2)+(0,0.5)$) {\color{red!70!yellow}$\omega_1$};

\draw[-,line width=5pt, red!70!yellow] ($(c2)+(-0.2,0.2)$) -- ($(c2)+(0.2,-0.2)$);
\draw[-,line width=5pt, red!70!yellow] ($(c2)+(-0.2,-0.2)$) -- ($(c2)+(0.2,0.2)$);

\node(c4) at (-1.6,2.3) {};
\node at ($(c4)+(0,0.5)$) {\color{red!70!yellow}$\omega_1$};

\draw[-,line width=5pt, red!70!yellow] ($(c4)+(-0.2,0.2)$) -- ($(c4)+(0.2,-0.2)$);
\draw[-,line width=5pt, red!70!yellow] ($(c4)+(-0.2,-0.2)$) -- ($(c4)+(0.2,0.2)$);

\draw[red!70!yellow, thick] (-2.8,-2.4) -- (2,-2.4);
\draw[red!70!yellow, thick] (-2.8,2.4) -- (2,2.4);
\draw[red!70!yellow, thick] (-2,-2.6) -- (2.8,-2.6);
\draw[red!70!yellow, thick] (-2,2.2) -- (2.8,2.2);
\draw[blue!80!black, thick] (-3.6,-2.6) -- (-2,-2.6);
\draw[blue!80!black, thick] (-3.6,-2.4) -- (-2.8,-2.4);
\draw[blue!80!black, thick] (3.6,-2.6) -- (2.8,-2.6);
\draw[blue!80!black, thick] (3.6,-2.4) -- (2,-2.4);

\node at (5,2) {\begin{tikzpicture}[baseline]
 \begin{scope}[auto, every node/.style={minimum size=0.75cm}]
\def \spac {1cm}

\node[circle, draw](k1) at (0,0) {$1$};
\node[circle, draw](k2) at (1*\spac,0) {$2$};
\node[circle, draw](k3) at (1.8*\spac,0.6*\spac) {$1$};
\node[circle, draw](k4) at (1.8*\spac,-0.6*\spac) {$1$};

\node[draw, inner sep=0.1cm,minimum size=0.67cm](N2) at (1*\spac,\spac) {$1$};

\draw[-] (k1) to (k2);
\draw[-] (k2) to (k3);
\draw[-] (k2) to (k4);

\draw (k2) -- (N2);

\end{scope}
\end{tikzpicture}};
\node at (5,-1.4) {\begin{tikzpicture}[baseline]
 \begin{scope}[auto, every node/.style={minimum size=0.75cm}]
\def \spac {1cm}

\node[circle, draw](k1) at (0,0) {$2$};
\node[circle, draw](k2) at (1*\spac,0) {$2$};
\node[circle, draw](k3) at (1.8*\spac,0.6*\spac) {$1$};
\node[circle, draw](k4) at (1.8*\spac,-0.6*\spac) {$1$};

\node[draw, inner sep=0.1cm,minimum size=0.67cm](N1) at (0*\spac,\spac) {$2$};

\draw[-] (k1) to (k2);
\draw[-] (k2) to (k3);
\draw[-] (k2) to (k4);

\draw (k1) -- (N1);

\end{scope}
\end{tikzpicture}};
\draw[->, -stealth,  line width=0.5em](5.0,1.2) -- (5.0,-1.0);
\draw[->, -stealth,  line width=0.5em](-4.7,1.2) -- (-4.7,-1.0);
\node at(-4.7,2) {\large\textbf{Unpolarized}};
\node at(-4.7,-1.5) {\large\textbf{Polarized}};
\node[align=left] at (-4,-3.8) {\normalsize$\omega_1^{\prime\phantom{\prime}}: [-1,0,0,0]=\color{blue!80!black}-w_1$\\[0.1em]\normalsize$\omega_1^{\prime\prime}: [\phantom{-}1,0,0,0]={\color{blue!80!black}-w_1}{\color{red!70!yellow}+}{\color{blue!80!black}2}\color{red!70!yellow}\alpha_1+2\alpha_2+\alpha_3+\alpha_4$};
\node[align=left] at (-4,3.5) {\normalsize $\omega_1: [0,0,0,0]=\color{red!70!yellow}-w_2+\alpha_1+2\alpha_2+\alpha_3+\alpha_4$};

\draw[white, line width=10pt] (-2.9,0) -- (2.9,0);

\draw [decoration={brace,amplitude=0.5em},decorate,ultra thick,gray] (-7.79,3.13) -- (-7.79,3.78);
\draw [decoration={brace,amplitude=0.5em},decorate,ultra thick,gray] (-7.99,-4.42) -- (-7.99,-3.31);
\end{tikzpicture}
	\end{center}
	\caption{The brane picture for the zero weight of $D_4$ (top of the figure), which makes up an unpolarized theory at low energies. It is obtained after $\mathbb{Z}_2$ orbifolding of $A_7$. The D5 branes  sit on top of the D3 branes, and all the D3 branes are stacked together.
    After flowing on the Higgs branch, we end up with a polarized theory, with the two masses equal to each other.} 
	\label{fig:D4null}
\end{figure}

One should ask what happens to unpolarized defects in the context of Toda CFT. As we have seen in section \ref{sec:toda}, polarized defects are described by momenta that obey null state relations of the corresponding $\cal{W}(\fg)$-algebra. This is consistent with what can be found in the class $S$ literature \cite{Chacaltana:2012zy}; for instance, for the minimal puncture of $D_4$ from Figure \ref{fig:D4null}, the defect in the CFT limit is predicted to have no flavor symmetry. In particular, it is unclear what vertex operator one would write in $D_4$-Toda; indeed, in the little string formalism, the defect is the null weight, which suggests a trivial conformal block with no vertex operator insertion! To investigate this issue more carefully, it is useful to keep $m_s$ finite and work in the little string proper; there, a computation in the spirit of \cite{Aganagic:2013tta,Aganagic:2014oia,Aganagic:2015cta}, shows that the partition function of  $T^{2d}$ is in fact \emph{not} a $q$-conformal block of $D_4$ Toda, due to subtleties of certain non-cancelling fugacities. In other words, the claim that the partition function of $T^{2d}$ is a $q$-conformal block of $\fg$-type Toda fails precisely when $T^{2d}$ is an unpolarized defect, and only for those cases.

\subsection{All Codimension-Two Defects of the (2,0) Little String}

From the considerations above, we get a complete list of the D3 brane defects of the $(2,0)$ little string that are points on $\cC\times T^2$, and which preserve conformality (in a 4d sense, before $T^2$ compactification). These are the polarized and unpolarized punctures we presented. Each of them is characterized by a set of weights in $\fg$, which produce a quiver gauge theory at low energies, satisfying \ref{conformal}. Enumerating the $(2,0)$ little string defects, for a given $\fg$, is then a \emph{finite} counting problem.

\begin{figure}[hptb]
\begin{center}
\begin{tikzpicture}
\node at (-5,0) {\begin{tikzpicture}[baseline]
 \begin{scope}[auto, every node/.style={minimum size=0.75cm}]
\def \spac {1cm}

\node[circle, draw](k1) at (0,0) {$4$};
\node[circle, draw](k2) at (1*\spac,0) {$5$};
\node[circle, draw](k3) at (1.8*\spac,0.6*\spac) {$3$};
\node[circle, draw](k4) at (1.8*\spac,-0.6*\spac) {$3$};

\node[draw, inner sep=0.1cm,minimum size=0.67cm](N1) at (0*\spac,\spac) {$3$};
\node[draw, inner sep=0.1cm,minimum size=0.67cm](N3) at (2.8*\spac,0.6*\spac) {$1$};
\node[draw, inner sep=0.1cm,minimum size=0.67cm](N4) at (2.8*\spac,-0.6*\spac) {$1$};

\draw[-] (k1) to (k2);
\draw[-] (k2) to (k3);
\draw[-] (k2) to (k4);

\draw (k1) -- (N1);
\draw (k3) -- (N3);
\draw (k4) -- (N4);

\end{scope}
\end{tikzpicture}};
\node at (0,0) {\begin{tikzpicture}[baseline]
 \begin{scope}[auto, every node/.style={minimum size=0.75cm}]
\def \spac {1cm}

\node[circle, draw](k1) at (0,0) {$4$};
\node[circle, draw](k2) at (1*\spac,0) {$7$};
\node[circle, draw](k3) at (1.8*\spac,0.6*\spac) {$4$};
\node[circle, draw](k4) at (1.8*\spac,-0.6*\spac) {$4$};

\node[draw, inner sep=0.1cm,minimum size=0.67cm](N1) at (0*\spac,\spac) {$1$};
\node[draw, inner sep=0.1cm,minimum size=0.67cm](N2) at (1*\spac,\spac) {$2$};
\node[draw, inner sep=0.1cm,minimum size=0.67cm](N3) at (2.8*\spac,0.6*\spac) {$1$};
\node[draw, inner sep=0.1cm,minimum size=0.67cm](N4) at (2.8*\spac,-0.6*\spac) {$1$};

\draw[-] (k1) to (k2);
\draw[-] (k2) to (k3);
\draw[-] (k2) to (k4);

\draw (k1) -- (N1);
\draw (k2) -- (N2);
\draw (k3) -- (N3);
\draw (k4) -- (N4);

\end{scope}
\end{tikzpicture}};
\node at (5,0) {\begin{tikzpicture}[baseline]
 \begin{scope}[auto, every node/.style={minimum size=0.75cm}]
\def \spac {1cm}

\node[circle, draw](k1) at (0,0) {$3$};
\node[circle, draw](k2) at (1*\spac,0) {$6$};
\node[circle, draw](k3) at (1.8*\spac,0.6*\spac) {$4$};
\node[circle, draw](k4) at (1.8*\spac,-0.6*\spac) {$4$};

\node[draw, inner sep=0.1cm,minimum size=0.67cm](N2) at (1*\spac,\spac) {$1$};
\node[draw, inner sep=0.1cm,minimum size=0.67cm](N3) at (2.8*\spac,0.6*\spac) {$2$};
\node[draw, inner sep=0.1cm,minimum size=0.67cm](N4) at (2.8*\spac,-0.6*\spac) {$2$};

\draw[-] (k1) to (k2);
\draw[-] (k2) to (k3);
\draw[-] (k2) to (k4);

\draw (k2) -- (N2);
\draw (k3) -- (N3);
\draw (k4) -- (N4);

\end{scope}
\end{tikzpicture}};
\node at (-6,-3) {\begin{tikzpicture}[baseline]
 \begin{scope}[auto, every node/.style={minimum size=0.75cm}]
\def \spac {1cm}

\node[circle, draw](k1) at (0,0) {$3$};
\node[circle, draw](k2) at (1*\spac,0) {$5$};
\node[circle, draw](k3) at (1.8*\spac,0.6*\spac) {$3$};
\node[circle, draw](k4) at (1.8*\spac,-0.6*\spac) {$4$};

\node[draw, inner sep=0.1cm,minimum size=0.67cm](N1) at (0*\spac,\spac) {$1$};
\node[draw, inner sep=0.1cm,minimum size=0.67cm](N3) at (2.8*\spac,0.6*\spac) {$1$};
\node[draw, inner sep=0.1cm,minimum size=0.67cm](N4) at (2.8*\spac,-0.6*\spac) {$3$};

\draw[-] (k1) to (k2);
\draw[-] (k2) to (k3);
\draw[-] (k2) to (k4);

\draw (k1) -- (N1);
\draw (k3) -- (N3);
\draw (k4) -- (N4);

\end{scope}
\end{tikzpicture}};
\node at (-2,-3) {\begin{tikzpicture}[baseline]
 \begin{scope}[auto, every node/.style={minimum size=0.75cm}]
\def \spac {1cm}

\node[circle, draw](k1) at (0,0) {$3$};
\node[circle, draw](k2) at (1*\spac,0) {$5$};
\node[circle, draw](k3) at (1.8*\spac,0.6*\spac) {$4$};
\node[circle, draw](k4) at (1.8*\spac,-0.6*\spac) {$3$};

\node[draw, inner sep=0.1cm,minimum size=0.67cm](N1) at (0*\spac,\spac) {$1$};
\node[draw, inner sep=0.1cm,minimum size=0.67cm](N3) at (2.8*\spac,0.6*\spac) {$3$};
\node[draw, inner sep=0.1cm,minimum size=0.67cm](N4) at (2.8*\spac,-0.6*\spac) {$1$};

\draw[-] (k1) to (k2);
\draw[-] (k2) to (k3);
\draw[-] (k2) to (k4);

\draw (k1) -- (N1);
\draw (k3) -- (N3);
\draw (k4) -- (N4);

\end{scope}
\end{tikzpicture}};
\node at (2,-3) {\begin{tikzpicture}[baseline]
 \begin{scope}[auto, every node/.style={minimum size=0.75cm}]
\def \spac {1cm}

\node[circle, draw](k1) at (0,0) {$4$};
\node[circle, draw](k2) at (1*\spac,0) {$6$};
\node[circle, draw](k3) at (1.8*\spac,0.6*\spac) {$3$};
\node[circle, draw](k4) at (1.8*\spac,-0.6*\spac) {$4$};

\node[draw, inner sep=0.1cm,minimum size=0.67cm](N1) at (0*\spac,\spac) {$2$};
\node[draw, inner sep=0.1cm,minimum size=0.67cm](N2) at (1*\spac,\spac) {$1$};
\node[draw, inner sep=0.1cm,minimum size=0.67cm](N4) at (2.8*\spac,-0.6*\spac) {$2$};

\draw[-] (k1) to (k2);
\draw[-] (k2) to (k3);
\draw[-] (k2) to (k4);

\draw (k1) -- (N1);
\draw (k2) -- (N2);
\draw (k4) -- (N4);

\end{scope}
\end{tikzpicture}};
\node at (6,-3) {\begin{tikzpicture}[baseline]
 \begin{scope}[auto, every node/.style={minimum size=0.75cm}]
\def \spac {1cm}

\node[circle, draw](k1) at (0,0) {$4$};
\node[circle, draw](k2) at (1*\spac,0) {$6$};
\node[circle, draw](k3) at (1.8*\spac,0.6*\spac) {$4$};
\node[circle, draw](k4) at (1.8*\spac,-0.6*\spac) {$3$};

\node[draw, inner sep=0.1cm,minimum size=0.67cm](N1) at (0*\spac,\spac) {$2$};
\node[draw, inner sep=0.1cm,minimum size=0.67cm](N2) at (1*\spac,\spac) {$1$};
\node[draw, inner sep=0.1cm,minimum size=0.67cm](N3) at (2.8*\spac,0.6*\spac) {$2$};

\draw[-] (k1) to (k2);
\draw[-] (k2) to (k3);
\draw[-] (k2) to (k4);

\draw (k1) -- (N1);
\draw (k2) -- (N2);
\draw (k3) -- (N3);

\end{scope}
\end{tikzpicture}};
\end{tikzpicture}
\end{center}
\caption{All $D_4$ 2d quiver theories one obtains from a set $\cal{W}_{\cal S}$ of 5 weights, and which all denote full punctures. In the CFT limit, all these theories produce the same full puncture, denoted by the parabolic subalgebra $\fp_{\varnothing}$. In particular, the Coulomb branch of $T^{2d}_{m_s\rightarrow \infty}$ for all these theories has dimension twelve.}
\label{fig:fullpuncturesd4}
\end{figure}
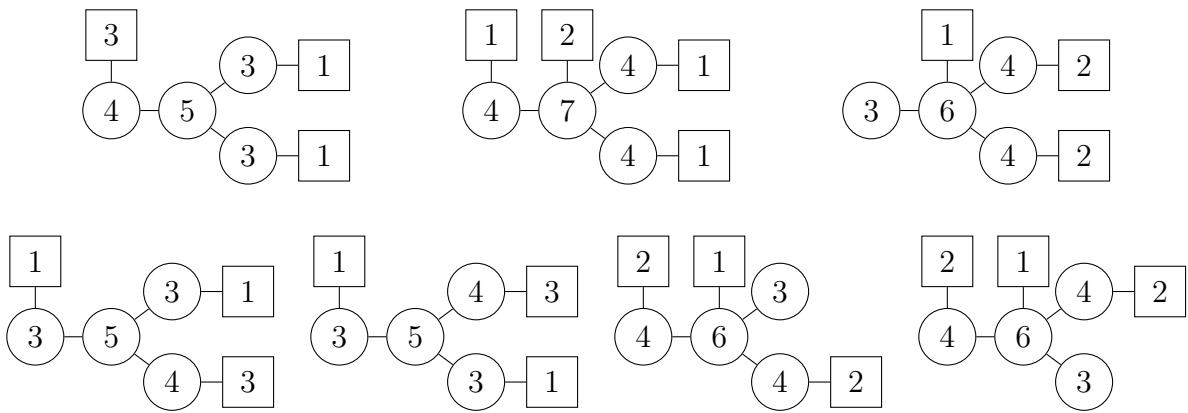

For $D_n$ and $E_n$, we find that the number of resulting theories $T^{2d}$ one obtains from specifying a set of weights, although finite, far exceeds the number of the CFT defects as enumarated in \cite{Chacaltana:2012zy}. What is happening is that in the CFT limit, many distinct defect theories $T^{2d}$ typically coalesce to one and the same defect theory $T^{2d}_{m_s\rightarrow \infty}$. The discussion in Figure \ref{fig:samequiverd4} illustrates this phenomenon. See also Figure \ref{fig:fullpuncturesd4} for the example of all theories $T^{2d}$ describing a generic full puncture of the $D_4$ little string.\\

An important point is  that even though we focused on the case of a sphere with two full punctures and an additional arbitrary puncture, the formalism we developed is automatically suited to study a sphere with an arbitrary number of defects.  

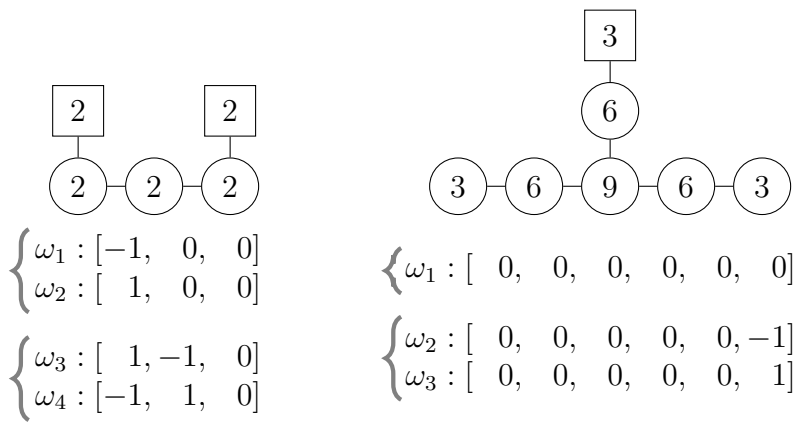
\begin{figure}[htpb]
\begin{center}
\begin{tikzpicture}
\node at (0,0) {\begin{tikzpicture}[baseline=0pt]
 \begin{scope}[auto, every node/.style={minimum size=0.75cm}]
\def \spac {1cm}

\node[circle, draw](k1) at (0,0) {$2$};
\node[circle, draw](k2) at (1*\spac,0) {$2$};
\node[circle, draw](k3) at (2*\spac,0*\spac) {$2$};

\node[draw, inner sep=0.1cm,minimum size=0.67cm](N1) at (0*\spac,\spac) {$2$};
\node[draw, inner sep=0.1cm,minimum size=0.67cm](N3) at (2*\spac,\spac) {$2$};

\draw[-] (k1) to (k2);
\draw[-] (k2) to (k3);

\draw (k1) -- (N1);
\draw (k3) -- (N3);

\end{scope}
\end{tikzpicture}};
\node at (6,0.5) {\begin{tikzpicture}[baseline=0pt]
 \begin{scope}[auto, every node/.style={minimum size=0.75cm}]
\def \spac {1cm}

\node[circle, draw](k1) at (0,0) {$3$};
\node[circle, draw](k2) at (1*\spac,0) {$6$};
\node[circle, draw](k3) at (2*\spac,0*\spac) {$9$};
\node[circle, draw](k4) at (3*\spac,0*\spac) {$6$};
\node[circle, draw](k5) at (4*\spac,0*\spac) {$3$};
\node[circle, draw](k6) at (2*\spac,1*\spac) {$6$};

\node[draw, inner sep=0.1cm,minimum size=0.67cm](N6) at (2*\spac,2*\spac) {$3$};

\draw[-] (k1) to (k2);
\draw[-] (k2) to (k3);
\draw[-] (k3) to (k4);
\draw[-] (k4) to (k5);
\draw[-] (k3) to (k6);

\draw (k6) -- (N6);

\end{scope}
\end{tikzpicture}};
\node[align=left, text width=10em] at (0.5,-2.3) {$\omega_1: [-1,\phantom{-}0,\phantom{-}0]$\\$\omega_2: [\phantom{-}1,\phantom{-}0,\phantom{-}0]$\\\vspace*{1em}$\omega_3:[\phantom{-}1,-1,\phantom{-}0]$\\$\omega_4: [-1,\phantom{-}1,\phantom{-}0]$};
\node[align=left, text width=14em] at (6.2,-2.3) {$\omega_1: [\phantom{-}0,\phantom{-}0,\phantom{-}0,\phantom{-}0,\phantom{-}0,\phantom{-}0]$\\\vspace*{1em} $\omega_2: [\phantom{-}0,\phantom{-}0,\phantom{-}0,\phantom{-}0,\phantom{-}0,-1]$\\$\omega_3: [\phantom{-}0,\phantom{-}0,\phantom{-}0,\phantom{-}0,\phantom{-}0,\phantom{-}1]$};
\draw [decoration={brace,amplitude=0.5em},decorate,ultra thick,gray] (-1.65,-2.15) -- (-1.65,-1.05);
\draw [decoration={brace,amplitude=0.5em},decorate,ultra thick,gray] (-1.65,-3.58) -- (-1.65,-2.48);
\draw [decoration={brace,amplitude=0.5em},decorate,ultra thick,gray] (3.25,-3.31) -- (3.25,-2.21);
\draw [decoration={brace,amplitude=0.5em},decorate,ultra thick,gray] (3.25,-1.9) -- (3.25,-1.30);
\end{tikzpicture}
\end{center}
\caption{Left: a four-punctured sphere of $A_3$, with two maximal (full) punctures and two minimal (simple) punctures, both denoted by the parabolic subalgebra $\fp_{\{\alpha_2, \alpha_3\}}$. The two simple punctures indicate that there are two subsets of weights in ${\cal W}_{\cal S}$ that add up to zero. In this specific example, the fact that the weights $[1,-1,0]$ and $[-1,1,0]$ denote a simple puncture can easily be seen by applying a Weyl reflection about the first simple root of $A_3$. Right: a four-punctured sphere of $E_6$, with two maximal punctures and two other punctures; the first of these is the minimal puncture, denoted by the zero weight in the 6-th fundamental representation, and is unpolarized. The second puncture is polarized, and distinguishes the parabolic subalgebra  $\fp_{\{\alpha_1, \alpha_2, \alpha_3, \alpha_4, \alpha_5\}}$.}
\label{fig:fourpoint}
\end{figure}

Simply
choose a set of weights ${\cal W}_{\cal S}$, as done before.  If there are $k$ subsets of weights which add up to zero in ${\cal W}_{\cal S}$, then the little string is in fact compactified on a sphere with $k+2$ punctures. This just follows from linearity of equation \eqref{conformal}. In particular, for the case of the sphere with 3 punctures we have been analyzing, there are then no proper subset of weights in ${\cal W}_{\cal S}$ that add up to zero. An immediate consequence is that not all quiver theories characterize a sphere with two full punctures and a third arbitrary one: some quivers represent composite arbitrary defects (and two full punctures). See Figure \ref{fig:fourpoint}.\\

As a final remark, let us mention that the techniques we used in this note to study codimension-two defects of the little string can also be applied to analyze codimension-four defects; these defects do not originate as D5 branes in the $(2,0)$ little string, but as D3 branes instead, before considering any $T^2$ compactification.

\clearpage

\section{Examples}
\label{sec:examples}

\subsection{$A_n$ Examples}
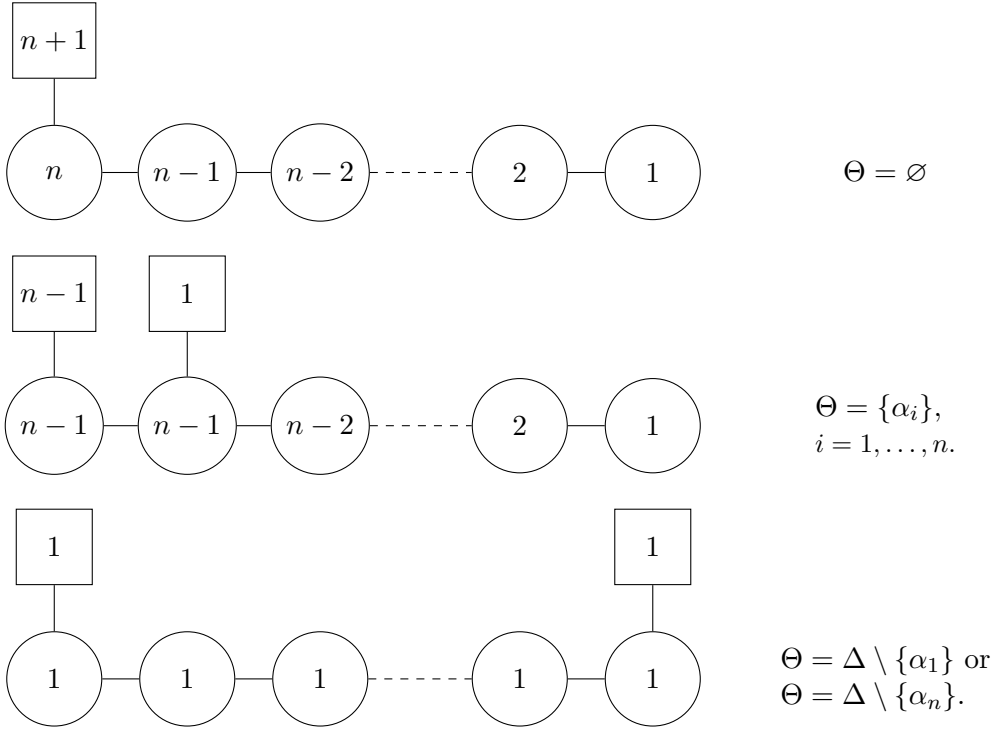
\begin{figure}[h!]
		\hspace{3em}
		\begin{tikzpicture}[align=center,font=\small, trim left]
\begin{scope}[auto, every node/.style={minimum size=1.25cm}]
\def \spac {1.75cm}
\def \spacmat {1cm}
\def \outu {0}
\def \inu {180}
\def \outl {200}
\def \inl {340}

\node[circle, draw](k1) at (0,0) {$n$};
\node[circle, draw](k2) at (1*\spac,0) {$n-1$};
\node[circle, draw](k3) at (2*\spac,0) {$n-2$};
\node[circle, draw](k4) at (3.5*\spac,0) {$2$};
\node[circle, draw](k5) at (4.5*\spac,0) {$1$};
\node[align=left] at (6.25*\spac,0) {\small $\Theta=\varnothing$};

\node[draw, inner sep=0.1cm,minimum size=1cm](N1) at (0*\spac,\spac) {$n+1$};

\draw[-] (k1) to[out=\outu,in=\inu] (k2);
\draw[-] (k2) to[out=\outu,in=\inu] (k3);
\draw[dashed] (k3) -- (k4);
\draw (k4) -- (k5);

\draw (k1) -- (N1);

\end{scope}

\end{tikzpicture}

\vspace{1em}

\hspace{3em}
\begin{tikzpicture}[align=center,font=\small, trim left]
\begin{scope}[auto, every node/.style={minimum size=1.25cm}]
\def \spac {1.75cm}
\def \spacmat {1cm}
\def \outu {0}
\def \inu {180}
\def \outl {200}
\def \inl {340}

\node[circle, draw](k1) at (0,0) {$n-1$};
\node[circle, draw](k2) at (1*\spac,0) {$n-1$};
\node[circle, draw](k3) at (2*\spac,0) {$n-2$};
\node[circle, draw](k4) at (3.5*\spac,0) {$2$};
\node[circle, draw](k5) at (4.5*\spac,0) {$1$};
\node[align=left] at (6.25*\spac,0) {$\Theta=\{\alpha_i\}$,\\ \footnotesize $i=1,\ldots,n$.};

\node[draw, inner sep=0.1cm,minimum size=1cm](N1) at (0*\spac,\spac) {$n-1$};
\node[draw, inner sep=0.1cm,minimum size=1cm](N2) at (1*\spac,\spac) {$1$};

\draw[-] (k1) to[out=\outu,in=\inu] (k2);
\draw[-] (k2) to[out=\outu,in=\inu] (k3);
\draw[dashed] (k3) -- (k4);
\draw (k4) -- (k5);

\draw (k1) -- (N1);
\draw (k2) -- (N2);

\end{scope}

\end{tikzpicture}

\vspace{1em}
\hspace{3em}
\begin{tikzpicture}[align=center,font=\small, trim left]
\begin{scope}[auto, every node/.style={minimum size=1.25cm,inner sep=1}]
\def \spac {1.75cm}
\def \spacmat {1cm}
\def \outu {0}
\def \inu {180}
\def \outl {200}
\def \inl {340}

\node[circle, draw](k1) at (0,0) {$1$};
\node[circle, draw](k2) at (1*\spac,0) {$1$};
\node[circle, draw](k3) at (2*\spac,0) {$1$};
\node[circle, draw](k4) at (3.5*\spac,0) {$1$};
\node[circle, draw](k5) at (4.5*\spac,0) {$1$};
\node[align=left] at (6.25*\spac,0) {\small $\Theta=\Delta\setminus\{\alpha_1\}$ or\\ \small $\Theta=\Delta\setminus\{\alpha_n\}$.};

\node[draw, inner sep=0.1cm,minimum size=1cm](N1) at (0*\spac,\spac) {$1$};
\node[draw, inner sep=0.1cm,minimum size=1cm](N5) at (4.5*\spac,\spac) {$1$};

\draw[-] (k1) to[out=\outu,in=\inu] (k2);
\draw[-] (k2) to[out=\outu,in=\inu] (k3);
\draw[dashed] (k3) -- (k4);
\draw (k4) -- (k5);

\draw (k1) -- (N1);
\draw (k5) -- (N5);

\end{scope}
\end{tikzpicture}
	\caption{The top quiver is the full puncture, denoted by the partition $[1^{n+1}]$. The middle quiver is the next to maximal puncture, with partition $[2,1^{n-1}]$. The bottom quiver is the simple puncture. It is denoted by the partition $[n,1]$, and has two associated parabolic subalgebras: $\fp_{\Delta\backslash \{\alpha_1\}}$ and $\fp_{\Delta\backslash \{\alpha_n\}}$.}
	\label{fig:AnExamples}
\end{figure}

We can explicitly write the parabolic subalgebras in some representation; for $A_n$,  it is customary to do so in the fundamental representation. Therefore, in what follows, the matrices are valued in $\fs\fl(n+1)$; a star $*_i$ denotes a nonzero complex number, and the label ``$i$'' stands for the positive root $e_i$. A star $*_{-i}$ denotes a nonzero complex number, and the label ``$-i$'' stands for the negative root $-e_i$. Unless specified otherwise, a partition refers to a semi-simple element denoting the Higgs field structure of the theory. These partitions are related to the nilpotent element partitions from section \ref{sec:nil} by transposition in the $A_n$ case, and more generally by the Spaltenstein map (cf.\ \cite{Collingwood:1993}). \\

\subsubsection{Maximal (``full'') Puncture}

We start with  the set  ${\cal W}_{\cal S}$ of all weights in the $n$-th fundamental representation (antifundamental). Writing $w_i$ for the highest weight of the $i$-th fundamental representation, the weights can be written as:
\begin{align*}
&\omega_1=-w_1\\
&\omega_2=-w_1+\alpha_1\\
&\omega_3=-w_1+\alpha_1+\alpha_2\\
&\vdots\;\;\;=\;\;\;\vdots\\
&\omega_{n+1}=-w_1+\alpha_1+\alpha_2+\ldots+\alpha_n,
\end{align*}
from which we read the top 2d quiver in Figure \ref{fig:AnExamples}.  This is called the full puncture. We compute the inner product of the weights with the positive roots:\\

$\omega_1\equiv[-1,0,0,\ldots,0]$ has a negative inner product with:\\
$\alpha_1\; , \;\alpha_1 + \alpha_2\; ,\; \ldots, \; \alpha_1+\alpha_2+\ldots+\alpha_n$ 
\\

$\omega_2\equiv[1,-1,0,\ldots,0]$  has a negative inner product with:\\
$\alpha_2\; , \;\alpha_2 + \alpha_3\; ,\; \ldots, \; \alpha_2+\alpha_3+\ldots+\alpha_n$ 
\\

$\omega_3\equiv[0,1,-1,0,\ldots,0]$  has a negative inner product with:\\
$\alpha_3\; , \;\alpha_3 + \alpha_4\; ,\; \ldots, \; \alpha_3+\alpha_4+\ldots+\alpha_n$ 
\\
$\vdots\qquad\qquad\qquad\qquad\qquad\qquad\vdots$\\
$\omega_{n+1}\equiv[0,\ldots,0,0,1]$ has no negative inner product with any of the positive roots.\\

Since all of the positive roots of $\fg$ have a negative inner product with some weight, they define the nilradical $\fn_{\varnothing}$.
The parabolic subalgebra is $\fp_{\varnothing}$. It is denoted by the partition $[1^{n+1}]$, which is immediately readable from the Levi subalgebra with symmetry $S(U(1)^{n+1})$.

\clearpage

The Levi decomposition gives:

\begin{align*}
\fp_\varnothing&=\begin{pmatrix}
	*&*_{1}&*_{1+2}&\cdots&*_{1+\ldots+(n-1)}&*_{1+\ldots+n}\\[5pt]
	0&* &*_{2}&\cdots&\cdots&*_{2+\ldots+n}\\
	\vdots & \ddots &\ddots&\ddots&\vdots&\vdots\\
	\vdots & \, &\ddots&\ddots&*_{(n-1)}&*_{(n-1)+n}\\
	\vdots &\,&\,&\ddots& * & *_{n}\\[5pt]
	0 &\cdots&\cdots&\cdots& 0 &*
\end{pmatrix},
\end{align*}
with $\fp_\varnothing=\fl_\varnothing\oplus\fn_\varnothing$, where
\begin{align*}
\fl_\varnothing=\begin{pmatrix}
*&0&\cdots&\cdots&\cdots&0\\
0 &*&\ddots&\,&\,&\vdots\\
\vdots&\ddots&\ddots&\ddots&\,&\vdots\\
\vdots &\,& \ddots &\ddots&\ddots&\vdots\\
\vdots & \,&\, &\ddots&*&0\\[5pt]
0 &\cdots&\cdots&\cdots& 0 &*
\end{pmatrix}\end{align*}
and 
\begin{align*}
\fn_\varnothing=
\begin{pmatrix}
0&*_{1}&*_{1+2}&\cdots&*_{1+\ldots+(n-1)}&*_{1+\ldots+n}\\[5pt]
0&0 &*_{2}&\cdots&\cdots&*_{2+\ldots+n}\\
\vdots & \ddots &\ddots&\ddots&\vdots&\vdots\\
\vdots & \, &\ddots&\ddots&*_{(n-1)}&*_{(n-1)+n}\\
\vdots &\,&\,&\ddots& 0 & *_{n}\\[5pt]
0 &\cdots&\cdots&\cdots& 0 &0
\end{pmatrix}.\\
\end{align*}

We see explicitly that the nonzero inner products $\langle e_{\gamma},\omega_i\rangle $ make up the $i$-th line of the nilradical $\fn_{\varnothing}$.\\

In this example, there is in fact one other set ${\cal W}_{\cal S}$ that singles out the nilradical $\fn_{\varnothing}$; it is the set of all weights in the first fundamental representation of $A_n$. The resulting 2d quiver is again the top one in Figure \ref{fig:AnExamples}, but with reversed orientation.\\

Now we analyze this defect from the Toda CFT perspective: starting from our set ${\cal W}_{\cal S}$ and recalling that $\beta=\sum_{i=1}^{|{\cal W}_{\cal S}|}\hat{\beta}_i w_i$,  ${\cal W}_{\cal S}$ defines the  Toda momentum vector $\beta$. We can write this momentum $\beta$ explicitly as the semi-simple element diag($\beta_1,\beta_2,\ldots,\beta_{n+1}$), where all the entries add up to 0. One checks at once that the commutant of this element is the Levi subalgebra $\fl_\varnothing$ written above.\\

The flag manifold $T^*(G/\mathcal{P})$ associated to this defect also appears as the resolution of the Higgs branch of the same quiver,
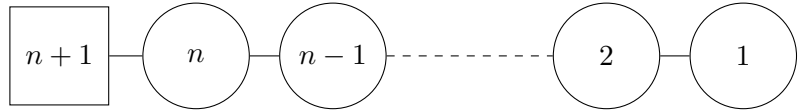
\begin{center}
\begin{tikzpicture}[baseline=0pt,font=\small]
 \begin{scope}[auto, every node/.style={minimum size=1.4cm}]
\def \spac {1.8cm}

\node[circle, draw](k1) at (0,0) {$n$};
\node[circle, draw](k2) at (1*\spac,0) {$n-1$};
\node[circle, draw](k3) at (3*\spac,0*\spac) {$2$};
\node[circle, draw](k4) at (4*\spac,0*\spac) {$1$};

\node[draw, inner sep=0.1cm,minimum size=1.3cm](N1) at (-1*\spac,0*\spac) {$n+1$};

\draw[-] (k1) to (k2);
\draw[dashed] (k2) to (k3);
\draw[-] (k3) to (k4);

\draw (k1) -- (N1);

\end{scope}
\end{tikzpicture}
\end{center}
which is an instance of mirror symmetry, since the complete flag is self-mirror.
Furthermore, it is easy to see from the method of section \ref{sec:nillevi} that the nilpotent orbit associated to this theory is the maximal nilpotent orbit of $A_n$, denoted by the partition $[n+1]$.

\subsubsection{Next to Maximal Puncture}

We start by constructing the set  ${\cal W}_{\cal S}$:
Consider all the $n+1$ weights of the $n$-th fundamental representation. For each $1\leq i\leq n$, the set contains two unique weights $\omega_i$ and $\omega_{i+1}$ such that $\alpha_i=\omega_i - \omega_{i+1}$, with  $\alpha_i$ the $i$-th simple root. Remove $\omega_i$ and $\omega_{i+1}$ from the set, and replace them  with the single weight $\omega'\equiv \omega_i+\omega_{i+1}$. $\omega'$ is always a weight in the $n-1$-th fundamental representation of $A_n$. Therefore, the set we consider is made of $n-1$ weights in the $n$-th fundamental representation, and the weight $\omega'$ in the $n-1$-th fundamental representation.  It is easy to check that the sum of these weights  is 0, so these $n$ weights define a valid set ${\cal W}_{\cal S}$. The weights once again define a 2d quiver gauge theory $T^{2d}$; it is shown in the middle of Figure \ref{fig:AnExamples}. All of the positive roots except  the $i$-th simple root $\alpha_i$ have a negative inner product with at least one weight $\omega_i\in{\cal W}_{\cal S}$, so these positive roots define the nilradical $\fn_{\{\alpha_i\}}$.\\

For a given simple root $\alpha_i$, the  parabolic subalgebra is then $\fp_{\{\alpha_i\}}$. It is denoted by the partition $[2,1^{n-1}]$, which is immediately readable from the Levi subalgebra with symmetry $S(U(2)\times U(1)^{n-1})$. 

\clearpage 

The Levi decomposition gives:

\begin{align*}
\fp_{\{\alpha_i\}}&=\begin{pmatrix}
*&*_{1}&*_{1+2}&\cdots&\cdots&\cdots&\cdots&\cdots&*_{1+\ldots+(n-1)}&*_{1+\ldots+n}\\[5pt]
0&* &*_{2}&\cdots&\cdots&\cdots&\cdots&\cdots&\cdots&*_{2+\ldots+n}\\
\vdots &\ddots &\ddots&\ddots&\,&\,& \,&\,&\vdots&\vdots\\
\vdots &\, &\ddots&\ddots&\ddots & \,&\,&\,&\vdots&\vdots\\
\vdots & \,&\, &0&*&*_{i}&\,&\,&\vdots&\vdots\\
\vdots & \,\,& &\,&*_{-i}&*&\ddots&\,&\vdots&\vdots\\
\vdots &\,&\,&\,&\,& 0 & \ddots&\ddots &\vdots&\vdots\\
\vdots &\,&\,&\,&\,&\,&\ddots&\ddots&*_{(n-1)}&*_{(n-1)+n}\\
\vdots &\,&\,&\,&\,&\,&\,&\ddots&*&*_n\\[5pt]
0 &\cdots&\cdots&\cdots&\cdots&\cdots&\cdots&\cdots& 0 &*
\end{pmatrix},
\end{align*}
with $\fp_{\{\alpha_i\}}=\fl_{\{\alpha_i\}}\oplus\fn_{\{\alpha_i\}}$, where
\begin{align*}
\fl_{\{\alpha_i\}}=\begin{pmatrix}
*&0&\cdots&\cdots&\cdots&\cdots&\cdots&0\\
0 &*&\ddots&\,&\,&\,&\,&\vdots\\
\vdots&\ddots&\ddots&0&\,&\,&\,&\vdots\\
\vdots &\,& 0 &*&*_i&\,&\,&\vdots\\
\vdots & \,&\, &*_{-i}&*&0&\,&\vdots\\
\vdots & \,&\, &\,&0&\ddots&\ddots&\vdots\\
\vdots &\,&\,&\,&\,&\ddots&*&0\\[5pt]
0 &\cdots&\cdots&\cdots&\cdots&\cdots& 0 & *
\end{pmatrix}
\end{align*}
and
\begin{align*}
\fn_{\{\alpha_i\}}=\begin{pmatrix}
0&*_{1}&*_{1+2}&\cdots&\cdots&\cdots&\cdots&\cdots&*_{1+\ldots+(n-1)}&*_{1+\ldots+n}\\[5pt]
0&0 &*_{2}&\cdots&\cdots&\cdots&\cdots&\cdots&\cdots&*_{2+\ldots+n}\\
\vdots &\ddots &\ddots&\ddots&\,&\,& \,&\,&\vdots&\vdots\\
\vdots &\, &\ddots&\ddots&*_{i-1} & \,&\,&\,&\vdots&\vdots\\
\vdots & \,&\, &\ddots&0&0&\,&\,&\vdots&\vdots\\
\vdots & \,\,& &\,&0&0&*_{i+1}&\,&\vdots&\vdots\\
\vdots &\,&\,&\,&\,& \ddots & \ddots&\ddots &\vdots&\vdots\\
\vdots &\,&\,&\,&\,&\,&\ddots&\ddots&*_{(n-1)}&*_{(n-1)+n}\\
\vdots &\,&\,&\,&\,&\,&\,&\ddots&0&*_n\\[5pt]
0 &\cdots&\cdots&\cdots&\cdots&\cdots&\cdots&\cdots& 0 &0
\end{pmatrix}.
\end{align*}

There is in fact another set ${\cal W}_{\cal S}$ that spells out the nilradical $\fn_{\{\alpha_i\}}$ for fixed $\alpha_i$; just as for the full puncture, the corresponding 2d quiver would be  the middle one in Figure \ref{fig:AnExamples}, but again with reversed orientation.\\

Now we rederive this result from the Toda CFT perspective: consider once again the set ${\cal W}_{\cal S}$. We define the momentum vector $\beta$ from $\beta=\sum_{i=1}^{|{\cal W}_{\cal S}|}\hat\beta_i \omega_i$. It is easy to check that
\[
\langle\beta,\alpha_i\rangle=0
\]
for the simple root $\alpha_i$, since $\beta$ has a unique 0 as its $i$-th Dynkin label.
This defines a null state at level 1 in the CFT. One can easily  check that there is only one other set ${\cal W}_{\cal S}$  such that $\langle\beta,\alpha_i\rangle=0$; this alternate choice gives the reflection of our 2d quiver. Also note that the commutant of the semi-simple element $\beta$ is the Levi subalgebra $\fl_{\alpha_i}$ written above in the fundamental representation.\\

We make the following important observations:

\begin{itemize}
	\item This puncture is in fact described by many sets  ${\cal W}_{\cal S}$. To obtain them, one simply considers all possible Weyl group actions that preserve the root sign: $w(\alpha_i)$ must be a positive root. Then all  possible momenta are given by $\beta'=w(\beta)$. Note that the condition $\langle\beta,\alpha_i\rangle=0$ is Weyl invariant: $\langle\beta,\alpha_i\rangle=\langle\beta',w(\alpha_i)\rangle$. Therefore, from the CFT perspective, the momentum of this different theory satisfies instead:
	\[
	\langle\beta',w(\alpha_i)\rangle=0.
	\]
	Because $w(\alpha_i)$ is a positive \textit{non}-simple root, this is strictly speaking a higher than level-1 null state condition of $A_n$-Toda. As explained in section \ref{sec:nullstates}, this higher level distinction is not relevant in the semi-classical limit $\hbar \rightarrow 0$ (or $Q \rightarrow 0$ in Toda), which is enough for our purposes. The explicit null state for all the theories obtained from the sets  ${\cal W}_{\cal S}$ can then be written at level 1, it is
\begin{equation}
\left(W^{(n+1)}_{-1}+\beta_i W^{(n)}_{-1}+\beta_i^2 W^{(n-2)}_{-1}+\cdots +\beta_i^{n-1} W^{(2)}_{-1} \right)|\vec{\beta}\rangle.
\end{equation}
Here, $W^{(j)}_{-1}$ is the mode $-1$ of the spin $j$ generator, and $\beta_i$ is the $i$-th entry of $\beta$, written in the fundamental representation, where $i$ labels the singled-out simple root $\alpha_i$. The eigenvalues of the $W^{(j)}_{0}$ modes are then functions of all the entries of $\beta$.

\item All of the many different sets  ${\cal W}_{\cal S}$  mentioned above give rise to the same 2d quiver gauge theory, in the middle of Figure \ref{fig:AnExamples}.

	\item The definition of the weight $\omega'\equiv \omega_i+\omega_{i+1}$ above is an illustration of the weight addition rule from section \ref{ssec:weightadd}. This corresponds to moving on the Higgs branch, and transitioning from the top quiver to the middle quiver in Figure \ref{fig:AnExamples}. In gauge theory terms, when  the hypermultiplet masses for $\omega_i$ and $\omega_{i+1}$ of the full puncture are set equal, one can transition from the top 2d theory to the middle 2d theory, which has a single hypermultiplet mass for $\omega'$ instead. 
\item The nilpotent orbit associated to this puncture is the unique subregular nilpotent orbit of $A_n$, with partition $[n,1]$.
\end{itemize}

The flag manifold $T^*(G/\mathcal{P})$ also appears as the resolution of the Higgs branch of the quiver

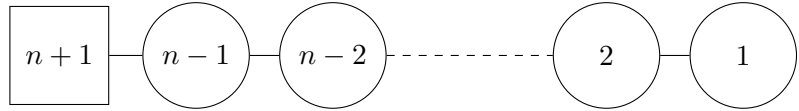
\begin{center}
\begin{tikzpicture}[baseline=0pt,font=\small]
 \begin{scope}[auto, every node/.style={minimum size=1.4cm}]
\def \spac {1.8cm}

\node[circle, draw](k1) at (0,0) {$n-1$};
\node[circle, draw](k2) at (1*\spac,0) {$n-2$};
\node[circle, draw](k3) at (3*\spac,0*\spac) {$2$};
\node[circle, draw](k4) at (4*\spac,0*\spac) {$1$};

\node[draw, inner sep=0.1cm,minimum size=1.3cm](N1) at (-1*\spac,0*\spac) {$n+1$};

\draw[-] (k1) to (k2);
\draw[dashed] (k2) to (k3);
\draw[-] (k3) to (k4);

\draw (k1) -- (N1);

\end{scope}
\end{tikzpicture}
\end{center}

which is again mirror to ours.

\subsubsection{Minimal (``simple'') Puncture}

We start by constructing the set  ${\cal W}_{\cal S}$. 
 Writing $w_i$ for the highest weight of the $i$-th fundamental representation, we define ${\cal W}_{\cal S}$ as:
\begin{align*}
&\omega_1=-w_n,\\
&\omega_2=-w_1+\alpha_1+\alpha_2+\ldots+\alpha_n.
\end{align*}
Written as above, the weights spell out the 2d quiver at the bottom of Figure \ref{fig:AnExamples}.  This is called the simple puncture. We compute the inner product of the weights with the positive roots:

$\omega_1\equiv[0,0,\ldots,0,-1]$ has a negative inner product with:\\
$\alpha_n, \;\alpha_n + \alpha_{n-1},\; \ldots, \; \alpha_n+\alpha_{n-1}+\ldots+\alpha_1$

$\omega_2\equiv[0,0,\ldots,0,\phantom{-}1]$ has no negative inner product with any of the positive roots.\\

So the only positive roots of $\fg$ that have a negative inner product with some weight $\omega_i\in{\cal W}_{\cal S}$ are $\alpha_n, \;\alpha_n + \alpha_{n-1},\; \ldots, \; \alpha_n+\alpha_{n-1}+\ldots+\alpha_1$, and they define the nilradical  $\fn_{\Delta\backslash\{\alpha_n\}}$.
The parabolic subalgebra is then $\fp_{\Delta\backslash\{\alpha_n\}}$. It is denoted by the partition $[n,1]$, which is immediately readable from the Levi subalgebra with symmetry $S(U(n)\times U(1))$. The Levi decomposition gives:

\begin{align*}
\fp_{\Delta\backslash \{\alpha_n\}}&=\begin{pmatrix}
*&*_{1}&*_{1+2}&\cdots&\cdots&*_{1+\ldots+(n-1)}&*_{1+\ldots+n}\\[5pt]
*_{-1}&* &*_{2}&\cdots&\cdots&*_{2+\ldots+(n-1)}&*_{2+\ldots+n}\\[5pt]
*_{-(1+2)} &*_{-2}&* &\cdots&\cdots&*_{3+\ldots+(n-1)}&*_{3+\ldots+n}\\
\vdots & \vdots&\vdots&\ddots&\cdots&\vdots&\vdots\\
\vdots & \vdots&\vdots&\cdots&\ddots&*_{(n-1)}&*_{(n-1)+n}\\[5pt]
*_{-(1+\ldots+(n-1))} &*_{-(2+\ldots+(n-1))}&*_{-(3+\ldots+(n-1))} & \cdots&*_{-(n-1)}& * & *_{n}\\[5pt]
0 &0 &0&\cdots&0& 0 &*
\end{pmatrix},
\end{align*}
with $\fp_{\Delta\backslash \{\alpha_n\}}=\fl_{\Delta\backslash \{\alpha_n\}}\oplus\fn_{\Delta\backslash \{\alpha_n\}}$, where
\begin{align*}
\fl_{\Delta\backslash \{\alpha_n\}}=\begin{pmatrix}
*&*_{1}&*_{1+2}&\cdots&\cdots&*_{1+\ldots+(n-1)}&0\\[5pt]
*_{-1}&* &*_{2}&\cdots&\cdots&*_{2+\ldots+(n-1)}&0\\[5pt]
*_{-(1+2)} &*_{-2}&* &\cdots&\cdots&*_{3+\ldots+(n-1)}&0\\
\vdots & \vdots&\vdots&\ddots&\cdots&\vdots&\vdots\\
\vdots & \vdots&\vdots&\cdots&\ddots&*_{(n-1)}&0\\[5pt]
*_{-(1+\ldots+(n-1))} &*_{-(2+\ldots+(n-1))}&*_{-(3+\ldots+(n-1))} & \cdots&*_{-(n-1)}& * & 0\\[5pt]
0 &0 &0&\cdots&0& 0 &*
\end{pmatrix}
\end{align*}
and
\begin{align*}
\fn_{\Delta\backslash \{\alpha_n\}}=\begin{pmatrix}
0&\cdots&\cdots&\cdots&\cdots&0&*_{1+\ldots+n}\\[5pt]
\vdots&\ddots&\,&\,&\,&\vdots&*_{2+\ldots+n}\\[5pt]
\vdots&\,&\ddots&\,&\,&\vdots&*_{3+\ldots+n}\\
\vdots & \,&\,&\ddots&\,&\vdots&\vdots\\
\vdots & \,&\,&\,&\ddots&\vdots&*_{(n-1)+n}\\[5pt]
0 &\cdots&\cdots & \cdots&\cdots& 0& *_{n}\\[5pt]
0 &\cdots &\cdots&\cdots&\cdots& 0 &0
\end{pmatrix}.
\end{align*}

We see explicitly that the non-zero inner products $\langle e_{\gamma},\omega_i\rangle $ give the last column of the nilradical $\fn_{\Delta\backslash\{\alpha_n\}}$.\\

Now we rederive this result from the CFT perspective: consider once again the set ${\cal W}_{\cal S}$. We define the momentum vector $\beta$ from $\beta=\sum_{i=1}^{|{\cal W}_{\cal S}|}\hat\beta_i \omega_i$. It is easy to check that
\[
\langle\beta,\alpha_i\rangle=0,\qquad\qquad i=1,2,\ldots,n-1
\]
since $\beta$ has a  0 as its $i$-th Dynkin label for $i=1,2,\ldots,n-1$.
This defines many  level 1 null states in the CFT.  One can easily  check that no other set ${\cal W}_{\cal S}$  satisfies the above vanishing inner product conditions. Also note that the commutant of the semi-simple element $\beta$ is the Levi subalgebra $\fl_{\Delta\backslash\{\alpha_n\}}$ written above in the fundamental representation.\\

We make the following important observations:

\begin{itemize}
	\item This puncture is in fact described by many sets  ${\cal W}_{\cal S}$. To obtain them, one simply considers all possible Weyl group actions that preserve the root sign: $w(\alpha_i)$ must be a positive root; the details are in the previous example. The upshot is once again that the explicit null states for all these 2d theories can  be written at level 1; they are:
\begin{equation}
\left(W^{(n+1)}_{-1}+\beta W^{(n)}_{-1}+\beta^2 W^{(n-2)}_{-1}+\ldots +\beta^{n-1} W^{(2)}_{-1} \right)|\vec{\beta}\rangle,
\end{equation}
and the $n-1$ derivatives of this equation with respect to $\beta$:
\begin{align*}
& \left(W^{(n)}_{-1}+2\beta W^{(n-2)}_{-1}+\ldots +(n-1)\beta^{n-2} W^{(2)}_{-1} \right)|\vec{\beta}\rangle\\
& \left(2 W^{(n-2)}_{-1}+\ldots +(n-1)(n-2)\beta^{n-3} W^{(2)}_{-1} \right)|\vec{\beta}\rangle\\
& \qquad\vdots\\
& \;\;W^{(2)}_{-1}|\vec{\beta}\rangle
\end{align*}
Here, $W^{(j)}_{-1}$ is the mode $-1$ of the spin $j$ generator, and $\vec{\beta}$=diag$(\beta,\beta,\ldots,\beta,-n\beta)$, written in the fundamental representation. The eigenvalues of the $W^{(j)}_{0}$ modes are again functions of $\beta$.

\item All the many different sets  ${\cal W}_{\cal S}$ mentioned above give rise to the same 2d quiver gauge theory, in the bottom of Figure \ref{fig:AnExamples}, and they all characterize the parabolic subalgebra $\fp_{\Delta\backslash\{\alpha_n\}}$, even if not directly readable from the positive root inner products with the Weyl reflected weights.

\item Once again, we can use the weight addition procedure to move on the Higgs branch, and transition from the top quiver to the bottom quiver in Figure \ref{fig:AnExamples}. In gauge theory terms, when  the hypermultiplet masses for $\omega_1$, $\omega_2$, $\ldots$, $\omega_{n}$ of the full puncture are set equal, one can transition from the top 2d theory to the bottom 2d theory, which has  a single hypermultiplet mass for the single weight $\omega_1+\omega_2+\ldots+\omega_{n}$ instead. Explicitly,
\[
[-1,0,0,\ldots,0]+[1,-1,0,\ldots,0]+\ldots+[0,\ldots,0,1,-1]=[0,0,\ldots,0,-1].
\]
\item The nilpotent orbit for this theory is the minimal non-trivial orbit of $A_n$, with partition $[1^{n+1}]$.
\end{itemize}

The flag manifold $T^*(G/\mathcal{P})$ associated to this defect also appears as the resolution of the Higgs branch of the quiver
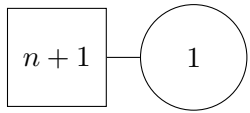
\begin{center}
\begin{tikzpicture}[baseline=0pt,font=\small]
 \begin{scope}[auto, every node/.style={minimum size=1.4cm}]
\def \spac {1.8cm}

\node[circle, draw](k1) at (0,0) {$1$};

\node[draw, inner sep=0.1cm,minimum size=1.3cm](N1) at (-1*\spac,0*\spac) {$n+1$};

\draw (k1) -- (N1);

\end{scope}
\end{tikzpicture}
\end{center}
which is  the Grassmanian $G(1,n+1)$.
Note this is again precisely mirror to our quiver theory $T^{2d}$.\\

\newpage

\subsection{$D_n$ Examples: Polarized Theories}

\subsubsection{Examples for Arbitrary $n$}

\begin{figure}[h!]
		\begin{tikzpicture}[baseline]
 \begin{scope}[auto, every node/.style={minimum size=1.5cm}]
\def \spac {1.9cm}

\node[circle, draw](k1) at (0,0) {$3$};
\node[circle, draw](k2) at (1*\spac,0) {$5$};
\node[circle, draw](k3) at (2*\spac,0) {$7$};
\node[circle, draw](k4) at (4*\spac,0) {$2n-3$};
\node[circle, draw](k5) at (4.9*\spac,0.8*\spac) {$n-1$};
\node[circle, draw](k6) at (4.9*\spac,-0.8*\spac) {$n-1$};

\node[draw, inner sep=0.1cm,minimum size=1.2cm](N1) at (0*\spac,\spac) {$1$};
\node[draw, inner sep=0.1cm,minimum size=1.2cm](N4) at (4*\spac,\spac) {$1$};
\node[draw, inner sep=0.1cm,minimum size=1.2cm](N5) at (5.9*\spac,0.8*\spac) {$1$};
\node[draw, inner sep=0.1cm,minimum size=1.2cm](N6) at (5.9*\spac,-0.8*\spac) {$1$};

\node[align=left] at (6.15*\spac,0) {\small $\Theta=\{\alpha_2,\alpha_2,\ldots,\alpha_{n-2}\}$};
\draw[-] (k1) to (k2);
\draw[-] (k2) to (k3);
\draw[dashed] (k3) to (k4);
\draw[-] (k4) to (k5);
\draw[-] (k4) to (k6);

\draw (k1) -- (N1);
\draw (k4) -- (N4);
\draw (k5) -- (N5);
\draw (k6) -- (N6);

\end{scope}
\end{tikzpicture}
\vspace{2em}
\begin{tikzpicture}[baseline]
 \begin{scope}[auto, every node/.style={minimum size=1.5cm}]
\def \spac {1.9cm}

\node[circle, draw](k1) at (0,0) {$2$};
\node[circle, draw](k2) at (1*\spac,0) {$2$};
\node[circle, draw](k3) at (2*\spac,0) {$2$};
\node[circle, draw](k4) at (4*\spac,0) {$2$};
\node[circle, draw](k5) at (4.9*\spac,0.8*\spac) {$1$};
\node[circle, draw](k6) at (4.9*\spac,-0.8*\spac) {$1$};

\node[draw, inner sep=0.1cm,minimum size=1.2cm](N1) at (0*\spac,\spac) {$2$};

\node[align=left] at (5.75*\spac,0) {\small $\Theta=\Delta\setminus\{\alpha_1\}$};
\draw[-] (k1) to (k2);
\draw[-] (k2) to (k3);
\draw[dashed] (k3) to (k4);
\draw[-] (k4) to (k5);
\draw[-] (k4) to (k6);

\draw (k1) -- (N1);

\end{scope}
\end{tikzpicture}
	\caption{The top quiver is a nontrivial puncture characterized by the parabolic subalgebra $\fp_{\{\alpha_2,\alpha_3,\ldots,\alpha_{n-2}\}}$.  It is denoted by the partition $[(n-2)^2,1^4]$ in the fundamental representation.  The bottom quiver is the simple puncture of $D_n$, characterized by the parabolic subalgebra $\fp_{\Delta\backslash\{\alpha_1\}}$. It is denoted by the partition $[2n-2,1^2]$.}
	\label{fig:DnExamples}
\end{figure}
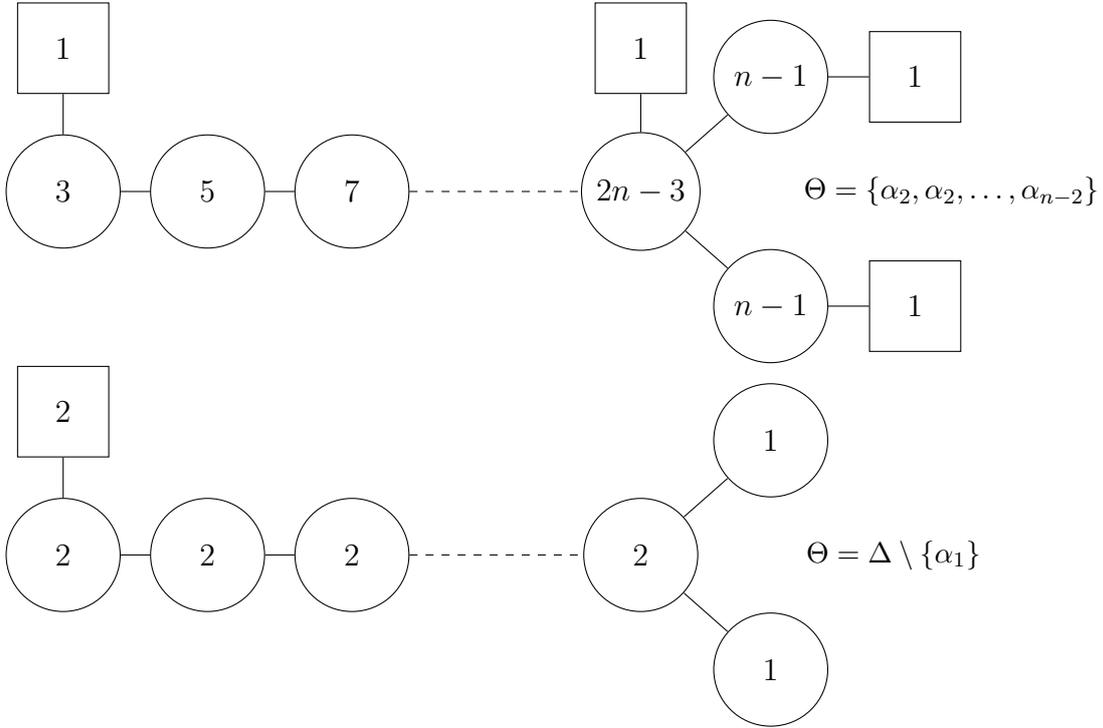

Here, we give two nontrivial $D_n$ examples.  We proceed as in the $A_n$ case and start by constructing a valid set of weights ${\cal W}_{\cal S}$:
\begin{align*}
\omega_1&\equiv[\phantom{-}1,0,\ldots,0,\phantom{-}0,\phantom{-}0],\\
\omega_2&\equiv[\phantom{-}0,0,\ldots,0,-1,\phantom{-}0],\\
\omega_3&\equiv[\phantom{-}0,0,\ldots,0,\phantom{-}0,-1],\\
\omega_4&\equiv[-1,0,\ldots,0,\phantom{-}1,\phantom{-}1].
\end{align*}
These weights obviously add up to 0, so they define a valid set ${\cal W}_{\cal S}$. Now note that:
\begin{align*}
\omega_1&=-w_1+2\alpha_1+2\alpha_2+\ldots+2\alpha_{n-2}+\alpha_{n-1}+\alpha_{n}\\
\omega_2&=-w_{n-2}+\alpha_1+3\alpha_2+5\alpha_3\ldots+(2n-5)\alpha_{n-2}+(n-2)\alpha_{n-1}+(n-2)\alpha_{n}\\
\omega_3&=-w_{n-1}\\
\omega_4&=-w_{n}
\end{align*}
This defines the 2d quiver gauge theory $T^{2d}$ shown on top of Figure \ref{fig:DnExamples}. Computing $\langle e_{\gamma},\omega_i\rangle$ for all positive roots $e_{\gamma}$, we identify the nilradical $\fn_{\{\alpha_2,\alpha_3,\ldots,\alpha_{n-2}\}}$. Therefore, we associate to ${\cal W}_{\cal S}$ the parabolic subalgebra $\fp_{\{\alpha_2,\alpha_3,\ldots,\alpha_{n-2}\}}$ from the Levi decomposition.\\

Now we rederive this result from the CFT perspective: consider once again the set ${\cal W}_{\cal S}$. We define the momentum vector $\beta$ from $\beta=\sum_{i=1}^{|{\cal W}_{\cal S}|}\hat\beta_i \omega_i$. It is easy to check that
\[
\langle\beta,\alpha_i\rangle=0,\qquad\qquad i=2,3,\ldots,n-2
\]
since $\beta$ has a  0 as its $i$-th Dynkin label for $i=2,3,\ldots,n-2$.
This defines many  level 1 null states in the CFT.  One can easily  check that no other set ${\cal W}_{\cal S}$  satisfies $\langle\beta,\alpha_i\rangle=0$. Also note that the commutant of the semi-simple element $\beta$ is the Levi subalgebra $\fl_{\{\alpha_2,\alpha_3,\ldots,\alpha_{n-2}\}}$.\\

We make the following important observations:

\begin{itemize}
\item This puncture features the instance of a new phenomenon: there are in fact many 2d quivers  associated to the parabolic subalgebra $\fp_{\{\alpha_2,\alpha_3,\ldots,\alpha_{n-2}\}}$. We just exhibited one possible 2d quiver among many valid others.
	\item Just as in the $A_n$ case, there are many different sets  ${\cal W}_{\cal S}$ for each 2d quiver, which do not directly allow us to read off the parabolic subalgebra.  The upshot is once again that the explicit null states for all these sets  ${\cal W}_{\cal S}$ can  be written at level 1; they are given by:
\begin{equation}
\label{eq:dnnull}
\left(({\tilde W}^{(n)})^2_{-1}+\beta^2 W^{(2n-2)}_{-1}+\beta^4 W^{(2n-4)}_{-1}+\cdots +\beta^{2n-2} W^{(2)}_{-1} \right)|\vec{\beta}\rangle
\end{equation}
and  derivatives of this equation with respect to $\beta$.
Here, $W^{(j)}_{-1}$ is the mode $-1$ of the spin $j$ generator. In the split representation of $\mathfrak{so}(2n)$, a generic semi-simple element is $\vec{\beta}=\text{diag}(\beta_1,\beta_2,\ldots,\beta_n,-\beta_1,-\beta_2,\ldots,-\beta_n)$. The puncture we study sets $n-2$ entries $\beta_i$ equal to each other; call them $\beta$ (and so $n-2$ entries $-\beta_i$ become $-\beta$). It is this parameter $\beta$ that appears in the null state \eqref{eq:dnnull}. 

\item We can also identify the nilpotent orbit corresponding to this theory: for even $n$, it is given by the partition $[5,3,2^{n-4}]$, and for odd $n$, the orbit has the partition $[5,3,2^{n-5},1,1]$ (this agrees with the results of \cite{Chacaltana:2011ze}.)
\end{itemize}

We now turn to the second example. We start with the set of weights:
\begin{align*}
\omega_1&\equiv[\phantom{-}1,0,0,\ldots,0]=-w_1+2\alpha_1+2\alpha_2+\ldots+2\alpha_{n-2}+\alpha_{n-1}+\alpha_{n}\\
\omega_2&\equiv[-1,0,0,\ldots,0]=-w_1
\end{align*}
These weights obviously add up to 0, so they define a valid set ${\cal W}_{\cal S}$. Written as above, they  spell out a 2d quiver theory $T^{2d}$ shown at the bottom of Figure \ref{fig:DnExamples}. Computing $\langle e_{\gamma},\omega_i\rangle$ for all positive roots $e_{\gamma}$, we identify the nilradical $\fn_{\Delta\backslash \{\alpha_1\}}$. So we associate to ${\cal W}_{\cal S}$ the parabolic subalgebra $\fp_{\Delta\backslash \{\alpha_1\}}$ from the Levi decomposition. Unlike the previous example, the 2d quiver theory associated to this puncture is unique. All other possible sets  ${\cal W}_{\cal S}$ are then obtained by Weyl reflection.

The nilpotent orbit corresponding to this theory is the minimal non-trivial orbit in $D_n$, with partition $[3,1^{2n-2}]$.\\

The corresponding space $T^*(G/\mathcal{P})$ also appears as the resolution of the Higgs branch of the quiver
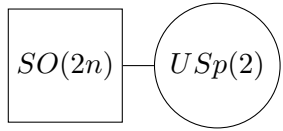
\begin{center}
\begin{tikzpicture}[baseline=0pt,font=\small]
 \begin{scope}[auto, every node/.style={minimum size=1.5cm}]
\def \spac {2.0cm}

\node[circle, draw](k1) at (0,0) {$USp(2)$};

\node[draw, inner sep=0.1cm,minimum size=1.5cm](N1) at (-1*\spac,0*\spac) {$SO(2n)$};

\draw (k1) -- (N1);

\end{scope}
\end{tikzpicture}
\end{center}
Note that this quiver theory is again mirror to ours.\\

\subsubsection{Complete $D_4$ Classification}

In Figure \ref{fig:D4allpunctures} we give the full classification of surface defects for $D_4$: the left column shows a representative  quiver $T^{2d}$ from \cite{Aganagic:2015cta} that describes each puncture. The middle column shows the subset of simple roots $\Theta$ which defines the parabolic subalgebra associated to $T^{2d}_{m_s\rightarrow\infty}$.  The right column features all the nilpotent orbits, in the notation of \cite{Chacaltana:2011ze}, as Hitchin Young diagrams. Note that lines 2 to 5 on the left denote one and the same nilpotent orbit, but different parabolic subalgebras. More subtle is the fact that lines 2 and 3 on the right feature the same Young diagram, but that is just an unfortunate misfortune in the notation: they really denote distinct nilpotent orbits and parabolic subalgebras; the Levi decompositions indeed yield two distinct nilradicals.  An asterisk is written down to differentiate those two punctures. In order to specify which of the three parabolic subalgebras the left 2d quiver of line 6 is associated to, one would need to specify explicitly the set ${\cal W}_{\cal S}$ that defines it. We omitted writing  ${\cal W}_{\cal S}$ for brevity.
\\

The nilpotent orbit classification of punctures has a disadvantage: two distinct punctures can be associated to one and the same Hitchin Young diagram (see for instance lines 2 and 3 on the right in Figure \ref{fig:D4allpunctures}), so extra data is needed to differentiate them. Classifying the CFT defects from the little string perspective, on the other hand, every polarized puncture  in the classification  is associated to a distinct parabolic subalgebra. Unpolarized punctures, however, have to be added separately. For $D_4$, there is exactly one such unpolarized puncture: the one featuring the null weight $[0,0,0,0]$; we show the explicit quiver theory $T^{2d}$ in the section \ref{sec:unpol}. It is interesting to note that special and non-special punctures in the classification of \cite{Chacaltana:2012zy} are treated on an equal footing in the little string formalism.\\

\begin{figure}
\resizebox{\textwidth}{!}{%
	\begin{tabular}[t]{lcllcl}
	2d Quiver Theory& $\Theta$ & Nilpotent orbit&2d Quiver Theory& $\Theta$ & Nilpotent orbit\\
	\cmidrule[\heavyrulewidth](r{1.5em}){1-3}\cmidrule[\heavyrulewidth](l){4-6}\addlinespace[1em]
	\begin{tikzpicture}[baseline]
 \begin{scope}[auto, every node/.style={minimum size=0.75cm}]
\def \spac {1cm}

\node[circle, draw](k1) at (0,0) {$4$};
\node[circle, draw](k2) at (1*\spac,0) {$5$};
\node[circle, draw](k3) at (1.8*\spac,0.6*\spac) {$3$};
\node[circle, draw](k4) at (1.8*\spac,-0.6*\spac) {$3$};

\node[draw, inner sep=0.1cm,minimum size=0.67cm](N1) at (0*\spac,\spac) {$3$};
\node[draw, inner sep=0.1cm,minimum size=0.67cm](N3) at (2.8*\spac,0.6*\spac) {$1$};
\node[draw, inner sep=0.1cm,minimum size=0.67cm](N4) at (2.8*\spac,-0.6*\spac) {$1$};

\draw[-] (k1) to (k2);
\draw[-] (k2) to (k3);
\draw[-] (k2) to (k4);

\draw (k1) -- (N1);
\draw (k3) -- (N3);
\draw (k4) -- (N4);

\end{scope}
\end{tikzpicture}
&$\Theta=\varnothing$&\ydiagram{7,1}\hspace*{2em}&\begin{tikzpicture}[baseline]
 \begin{scope}[auto, every node/.style={minimum size=0.75cm}]
\def \spac {1cm}

\node[circle, draw](k1) at (0,0) {$2$};
\node[circle, draw](k2) at (1*\spac,0) {$4$};
\node[circle, draw](k3) at (1.8*\spac,0.6*\spac) {$2$};
\node[circle, draw](k4) at (1.8*\spac,-0.6*\spac) {$3$};

\node[draw, inner sep=0.1cm,minimum size=0.67cm](N2) at (1*\spac,\spac) {$1$};
\node[draw, inner sep=0.1cm,minimum size=0.67cm](N4) at (2.8*\spac,-0.6*\spac) {$2$};

\draw[-] (k1) to (k2);
\draw[-] (k2) to (k3);
\draw[-] (k2) to (k4);

\draw (k2) -- (N2);
\draw (k4) -- (N4);

\end{scope}
\end{tikzpicture}
&$\Theta=\{\alpha_1,\alpha_4\}$&{\color{blue}\ydiagram{4,4}}\\[1.3cm]
		\begin{tikzpicture}[baseline]
 \begin{scope}[auto, every node/.style={minimum size=0.75cm}]
\def \spac {1cm}

\node[circle, draw](k1) at (0,0) {$2$};
\node[circle, draw](k2) at (1*\spac,0) {$4$};
\node[circle, draw](k3) at (1.8*\spac,0.6*\spac) {$3$};
\node[circle, draw](k4) at (1.8*\spac,-0.6*\spac) {$3$};

\node[draw, inner sep=0.1cm,minimum size=0.67cm](N3) at (2.8*\spac,0.6*\spac) {$2$};
\node[draw, inner sep=0.1cm,minimum size=0.67cm](N4) at (2.8*\spac,-0.6*\spac) {$2$};

\draw[-] (k1) to (k2);
\draw[-] (k2) to (k3);
\draw[-] (k2) to (k4);

\draw (k3) -- (N3);
\draw (k4) -- (N4);

\end{scope}
\end{tikzpicture}
&$\Theta=\{\alpha_1\}$&\ydiagram{5,3}&
	\begin{tikzpicture}[baseline]
 \begin{scope}[auto, every node/.style={minimum size=0.75cm}]
\def \spac {1cm}

\node[circle, draw](k1) at (0,0) {$2$};
\node[circle, draw](k2) at (1*\spac,0) {$3$};
\node[circle, draw](k3) at (1.8*\spac,0.6*\spac) {$2$};
\node[circle, draw](k4) at (1.8*\spac,-0.6*\spac) {$2$};

\node[draw, inner sep=0.1cm,minimum size=0.67cm](N1) at (0*\spac,\spac) {$1$};
\node[draw, inner sep=0.1cm,minimum size=0.67cm](N3) at (2.8*\spac,0.6*\spac) {$1$};
\node[draw, inner sep=0.1cm,minimum size=0.67cm](N4) at (2.8*\spac,-0.6*\spac) {$1$};

\draw[-] (k1) to (k2);
\draw[-] (k2) to (k3);
\draw[-] (k2) to (k4);

\draw (k1) -- (N1);
\draw (k3) -- (N3);
\draw (k4) -- (N4);

\end{scope}
\end{tikzpicture}
&$\underset{(i,j)=(1,2),(2,3),(2,4)}{\Theta=\{\alpha_i,\alpha_j\}}$&\ydiagram{3,3,1,1}$^\bigstar$\\[1.3cm]
\begin{tikzpicture}[baseline]
 \begin{scope}[auto, every node/.style={minimum size=0.75cm}]
\def \spac {1cm}

\node[circle, draw](k1) at (0,0) {$3$};
\node[circle, draw](k2) at (1*\spac,0) {$4$};
\node[circle, draw](k3) at (1.8*\spac,0.6*\spac) {$3$};
\node[circle, draw](k4) at (1.8*\spac,-0.6*\spac) {$2$};

\node[draw, inner sep=0.1cm,minimum size=0.67cm](N1) at (0*\spac,\spac) {$2$};
\node[draw, inner sep=0.1cm,minimum size=0.67cm](N3) at (2.8*\spac,0.6*\spac) {$2$};

\draw[-] (k1) to (k2);
\draw[-] (k2) to (k3);
\draw[-] (k2) to (k4);

\draw (k1) -- (N1);
\draw (k3) -- (N3);

\end{scope}
\end{tikzpicture}
&$\Theta=\{\alpha_3\}$&\ydiagram{5,3}&\begin{tikzpicture}[baseline]
 \begin{scope}[auto, every node/.style={minimum size=0.75cm}]
\def \spac {1cm}

\node[circle, draw](k1) at (0,0) {$2$};
\node[circle, draw](k2) at (1*\spac,0) {$4$};
\node[circle, draw](k3) at (1.8*\spac,0.6*\spac) {$2$};
\node[circle, draw](k4) at (1.8*\spac,-0.6*\spac) {$2$};

\node[draw, inner sep=0.1cm,minimum size=0.67cm](N2) at (1*\spac,\spac) {$2$};

\draw[-] (k1) to (k2);
\draw[-] (k2) to (k3);
\draw[-] (k2) to (k4);

\draw (k2) -- (N2);

\end{scope}
\end{tikzpicture}
&$\Theta=\{\alpha_1,\alpha_3,\alpha_4\}$&\ydiagram{3,3,1,1}\\[1.3cm]
\begin{tikzpicture}[baseline]
 \begin{scope}[auto, every node/.style={minimum size=0.75cm}]
\def \spac {1cm}

\node[circle, draw](k1) at (0,0) {$3$};
\node[circle, draw](k2) at (1*\spac,0) {$4$};
\node[circle, draw](k3) at (1.8*\spac,0.6*\spac) {$2$};
\node[circle, draw](k4) at (1.8*\spac,-0.6*\spac) {$3$};

\node[draw, inner sep=0.1cm,minimum size=0.67cm](N1) at (0*\spac,\spac) {$2$};
\node[draw, inner sep=0.1cm,minimum size=0.67cm](N4) at (2.8*\spac,-0.6*\spac) {$2$};

\draw[-] (k1) to (k2);
\draw[-] (k2) to (k3);
\draw[-] (k2) to (k4);

\draw (k1) -- (N1);
\draw (k4) -- (N4);

\end{scope}
\end{tikzpicture}
&$\Theta=\{\alpha_4\}$&\ydiagram{5,3}&\begin{tikzpicture}[baseline]
 \begin{scope}[auto, every node/.style={minimum size=0.75cm}]
\def \spac {1cm}

\node[circle, draw](k1) at (0,0) {$2$};
\node[circle, draw](k2) at (1*\spac,0) {$2$};
\node[circle, draw](k3) at (1.8*\spac,0.6*\spac) {$1$};
\node[circle, draw](k4) at (1.8*\spac,-0.6*\spac) {$1$};

\node[draw, inner sep=0.1cm,minimum size=0.67cm](N1) at (0*\spac,\spac) {$2$};

\draw[-] (k1) to (k2);
\draw[-] (k2) to (k3);
\draw[-] (k2) to (k4);

\draw (k1) -- (N1);

\end{scope}
\end{tikzpicture}
&$\Theta=\{\alpha_2,\alpha_3,\alpha_4\}$&\ydiagram{3,1,1,1,1,1}\\[1.3cm]
		\begin{tikzpicture}[baseline]
 \begin{scope}[auto, every node/.style={minimum size=0.75cm}]
\def \spac {1cm}

\node[circle, draw](k1) at (0,0) {$3$};
\node[circle, draw](k2) at (1*\spac,0) {$5$};
\node[circle, draw](k3) at (1.8*\spac,0.6*\spac) {$3$};
\node[circle, draw](k4) at (1.8*\spac,-0.6*\spac) {$3$};

\node[draw, inner sep=0.1cm,minimum size=0.67cm](N1) at (0*\spac,\spac) {$1$};
\node[draw, inner sep=0.1cm,minimum size=0.67cm](N2) at (1*\spac,\spac) {$1$};
\node[draw, inner sep=0.1cm,minimum size=0.67cm](N3) at (2.8*\spac,0.6*\spac) {$1$};
\node[draw, inner sep=0.1cm,minimum size=0.67cm](N4) at (2.8*\spac,-0.6*\spac) {$1$};

\draw[-] (k1) to (k2);
\draw[-] (k2) to (k3);
\draw[-] (k2) to (k4);

\draw (k1) -- (N1);
\draw (k2) -- (N2);
\draw (k3) -- (N3);
\draw (k4) -- (N4);

\end{scope}
\end{tikzpicture}
&$\Theta=\{\alpha_2\}$&\ydiagram{5,3}&\begin{tikzpicture}[baseline]
 \begin{scope}[auto, every node/.style={minimum size=0.75cm}]
\def \spac {1cm}

\node[circle, draw](k1) at (0,0) {$1$};
\node[circle, draw](k2) at (1*\spac,0) {$2$};
\node[circle, draw](k3) at (1.8*\spac,0.6*\spac) {$2$};
\node[circle, draw](k4) at (1.8*\spac,-0.6*\spac) {$1$};

\node[draw, inner sep=0.1cm,minimum size=0.67cm](N3) at (2.8*\spac,0.6*\spac) {$2$};

\draw[-] (k1) to (k2);
\draw[-] (k2) to (k3);
\draw[-] (k2) to (k4);

\draw (k3) -- (N3);

\end{scope}
\end{tikzpicture}
&$\Theta=\{\alpha_1,\alpha_2,\alpha_3\}$&{\color{red}\ydiagram{2,2,2,2}}\\[1.3cm]
\begin{tikzpicture}[baseline]
 \begin{scope}[auto, every node/.style={minimum size=0.75cm}]
\def \spac {1cm}

\node[circle, draw](k1) at (0,0) {$3$};
\node[circle, draw](k2) at (1*\spac,0) {$4$};
\node[circle, draw](k3) at (1.8*\spac,0.6*\spac) {$2$};
\node[circle, draw](k4) at (1.8*\spac,-0.6*\spac) {$2$};

\node[draw, inner sep=0.1cm,minimum size=0.67cm](N1) at (0*\spac,\spac) {$2$};
\node[draw, inner sep=0.1cm,minimum size=0.67cm](N2) at (1*\spac,\spac) {$1$};

\draw[-] (k1) to (k2);
\draw[-] (k2) to (k3);
\draw[-] (k2) to (k4);

\draw (k1) -- (N1);
\draw (k2) -- (N2);

\end{scope}
\end{tikzpicture}
&$\Theta=\{\alpha_3,\alpha_4\}$&\ydiagram{5,1,1,1}&\begin{tikzpicture}[baseline]
 \begin{scope}[auto, every node/.style={minimum size=0.75cm}]
\def \spac {1cm}

\node[circle, draw](k1) at (0,0) {$1$};
\node[circle, draw](k2) at (1*\spac,0) {$2$};
\node[circle, draw](k3) at (1.8*\spac,0.6*\spac) {$1$};
\node[circle, draw](k4) at (1.8*\spac,-0.6*\spac) {$2$};

\node[draw, inner sep=0.1cm,minimum size=0.67cm](N4) at (2.8*\spac,-0.6*\spac) {$2$};

\draw[-] (k1) to (k2);
\draw[-] (k2) to (k3);
\draw[-] (k2) to (k4);

\draw (k4) -- (N4);

\end{scope}
\end{tikzpicture}
&$\Theta=\{\alpha_1,\alpha_2,\alpha_4\}$&{\color{blue}\ydiagram{2,2,2,2}}\\[1.3cm]
\begin{tikzpicture}[baseline]
 \begin{scope}[auto, every node/.style={minimum size=0.75cm}]
\def \spac {1cm}

\node[circle, draw](k1) at (0,0) {$2$};
\node[circle, draw](k2) at (1*\spac,0) {$4$};
\node[circle, draw](k3) at (1.8*\spac,0.6*\spac) {$3$};
\node[circle, draw](k4) at (1.8*\spac,-0.6*\spac) {$2$};

\node[draw, inner sep=0.1cm,minimum size=0.67cm](N2) at (1*\spac,\spac) {$1$};
\node[draw, inner sep=0.1cm,minimum size=0.67cm](N3) at (2.8*\spac,0.6*\spac) {$2$};

\draw[-] (k1) to (k2);
\draw[-] (k2) to (k3);
\draw[-] (k2) to (k4);

\draw (k2) -- (N2);
\draw (k3) -- (N3);

\end{scope}
\end{tikzpicture}
&$\Theta=\{\alpha_1,\alpha_3\}$&{\color{red}\ydiagram{4,4}}\\[1.3cm]
	\end{tabular}%
	}
	\caption{Surface defects of $D_4$. 2d quiver theories from the Little String are shown in the left column. Parabolic subalgebras that arise in the CFT limit  $T^{2d}_{m_s\rightarrow\infty}$ are shown in the middle column. Nilpotent orbits from the defect classification of \cite{Chacaltana:2011ze} are shown in the right column. We omitted writing down an explicit set of weights  ${\cal W}_{\cal S}$ for each defect for brevity. The minimal nilpotent orbit is analyzed separately in section \ref{sec:unpol}.}
	\label{fig:D4allpunctures}
\end{figure}

\clearpage

\subsection{$E_n$ Examples: Polarized Theories}

Here, we give the quivers of $E_n$ with the smallest number of Coulomb moduli that describe a polarized puncture. 

\begin{figure}[h!]
	\begin{center}
\begin{tikzpicture}[baseline]
 \begin{scope}[auto, every node/.style={minimum size=1cm}]
\def \spac {1.3cm}

\node[circle, draw](k1) at (0,0) {$2$};
\node[circle, draw](k2) at (1*\spac,0) {$3$};
\node[circle, draw](k3) at (2*\spac,0) {$4$};
\node[circle, draw](k4) at (3*\spac,0) {$3$};
\node[circle, draw](k5) at (4*\spac,0) {$2$};
\node[circle, draw](k6) at (2*\spac,-1*\spac) {$2$};

\node[align=left,text width=3cm] at (8.35*\spac,0) {\small $\Theta=\Delta\setminus\{\alpha_1\}$ or $\Theta=\Delta\setminus\{\alpha_5\}$};

\node[draw, inner sep=0.1cm,minimum size=0.9cm](N1) at (0*\spac,\spac) {$1$};
\node[draw, inner sep=0.1cm,minimum size=0.9cm](N5) at (4*\spac,\spac) {$1$};

\draw[-] (k1) to (k2);
\draw[-] (k2) to (k3);
\draw[-] (k3) to (k4);
\draw[-] (k4) to (k5);
\draw[-] (k3) to (k6);

\draw (k1) -- (N1);
\draw (k5) -- (N5);

\end{scope}
\end{tikzpicture}

\begin{tikzpicture}[baseline]
 \begin{scope}[auto, every node/.style={minimum size=1cm}]
\def \spac {1.3cm}

\node[circle, draw](k1) at (0,0) {$2$};
\node[circle, draw](k2) at (1*\spac,0) {$4$};
\node[circle, draw](k3) at (2*\spac,0) {$6$};
\node[circle, draw](k4) at (3*\spac,0) {$5$};
\node[circle, draw](k5) at (4*\spac,0) {$4$};
\node[circle, draw](k6) at (5*\spac,0) {$3$};
\node[circle, draw](k7) at (2*\spac,-1*\spac) {$3$};

\node[align=left,text width=3cm] at (8.35*\spac,0) {\small $\Theta=\Delta\setminus\{\alpha_6\}$};

\node[draw, inner sep=0.1cm,minimum size=0.9cm](N6) at (5*\spac,\spac) {$2$};

\draw[-] (k1) to (k2);
\draw[-] (k2) to (k3);
\draw[-] (k3) to (k4);
\draw[-] (k4) to (k5);
\draw[-] (k5) to (k6);
\draw[-] (k3) to (k7);

\draw (k6) -- (N6);

\end{scope}
\end{tikzpicture}

\begin{tikzpicture}[baseline]
 \begin{scope}[auto, every node/.style={minimum size=1cm}]
\def \spac {1.3cm}

\node[circle, draw](k1) at (0,0) {$4$};
\node[circle, draw](k2) at (1*\spac,0) {$8$};
\node[circle, draw](k3) at (2*\spac,0) {$12$};
\node[circle, draw](k4) at (3*\spac,0) {$10$};
\node[circle, draw](k5) at (4*\spac,0) {$8$};
\node[circle, draw](k6) at (5*\spac,0) {$6$};
\node[circle, draw](k7) at (6*\spac,0) {$4$};
\node[circle, draw](k8) at (2*\spac,-1*\spac) {$6$};

\node[draw, inner sep=0.1cm,minimum size=0.9cm](N7) at (6*\spac,\spac) {$2$};

\node[align=left,text width=3cm] at (8.35*\spac,0) {\small $\Theta=\Delta\setminus\{\alpha_7\}$};
\draw[-] (k1) to (k2);
\draw[-] (k2) to (k3);
\draw[-] (k3) to (k4);
\draw[-] (k4) to (k5);
\draw[-] (k5) to (k6);
\draw[-] (k6) to (k7);
\draw[-] (k3) to (k8);

\draw (k7) -- (N7);

\end{scope}
\end{tikzpicture}
	\end{center}
	
	\caption{The top, middle, and bottom quivers are $E_6$, $E_7$, and $E_8$ 2d theories respectively. The associated parabolic subalgebras are $\fp_{\Delta\backslash \{\alpha_1\}}$, $\fp_{\Delta\backslash \{\alpha_6\}}$, and $\fp_{\Delta\backslash \{\alpha_7\}}$ respectively. These punctures all have Bala-Carter label $2 A_1$ in the classification of \cite{Chacaltana:2012zy}.}
	\label{fig:EnExamples}
\end{figure}
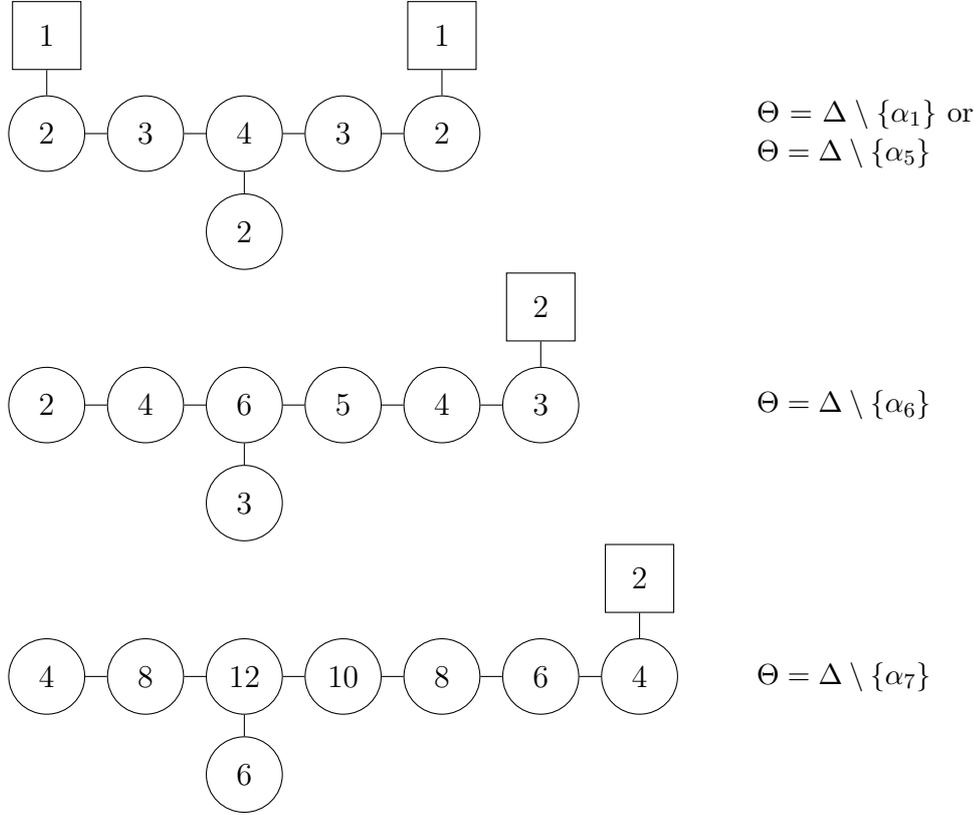

For $E_6$, we start with the set  ${\cal W}_{\cal S}$:
\begin{align*}
\omega_1&\equiv[\phantom{-}1,0,0,0,0,0]=-w_5+2\alpha_1+3\alpha_2+4\alpha_3+2\alpha_4+\alpha_5+2\alpha_6\\
\omega_2&\equiv[-1,0,0,0,0,0]=-w_1
\end{align*}
This defines a 2d theory (shown in Figure \ref{fig:EnExamples}). One checks at once from the positive roots that ${\cal W}_{\cal S}$ characterizes the nilradical  $\fn_{\Delta\backslash \{\alpha_1\}}$, so the associated parabolic subalgebra is $\fp_{\Delta\backslash \{\alpha_1\}}$.
In fact, no other set  ${\cal W}_{\cal S}$  is associated to this parabolic subalgebra. The level 1 null state condition in the $E_6$-Toda CFT is:
\[
\langle\beta,\alpha_i\rangle=0,\qquad\qquad i=2,\ldots,6
\]

The set  ${\cal W}_{\cal S}$:
\begin{align*}
\omega_1&\equiv[0,0,0,0,\phantom{-}1,0]=-w_1+2\alpha_1+3\alpha_2+4\alpha_3+2\alpha_4+\alpha_5+2\alpha_6,\\
\omega_2&\equiv[0,0,0,0,-1,0]=-w_5,
\end{align*}
produces the same 2d quiver as above, but the associated parabolic subalgebra is instead $\fp_{\Delta\backslash \{\alpha_5\}}$, and the  level 1 null state condition is:
\[
\langle\beta,\alpha_i\rangle=0,\qquad\qquad i=1,2,3,4,6.
\]

All the other possible  sets ${\cal W}_{\cal S}$ associated to $\fp_{\Delta\backslash \{\alpha_1\}}$  are obtained by Weyl reflection on the two weights (and the same is true about $\fp_{\Delta\backslash \{\alpha_5\}}$).\\

For $E_7$, we start with the set  ${\cal W}_{\cal S}$:
\begin{align*}
\omega_1&\equiv[0,0,0,0,0,\phantom{-}1,0]=-w_6+2\alpha_1+4\alpha_2+6\alpha_3+5\alpha_4+4\alpha_5+3\alpha_6+3\alpha_7\\
\omega_2&\equiv[0,0,0,0,0,-1,0]=-w_6
\end{align*}
This defines  a 2d theory (shown in the middle of Figure \ref{fig:EnExamples}). One checks at once from the positive roots that ${\cal W}_{\cal S}$ characterizes the nilradical  $\fn_{\Delta\backslash \{\alpha_6\}}$, so the associated parabolic subalgebra is $\fp_{\Delta\backslash \{\alpha_6\}}$.
In fact, no other set  ${\cal W}_{\cal S}$  is associated to this parabolic subalgebra. The level 1 null state condition in the $E_7$-Toda CFT is:
\[
\langle\beta,\alpha_i\rangle=0,\qquad\qquad i=1,2,3,4,5,7.
\]

All the other possible  sets ${\cal W}_{\cal S}$ associated to $\fp_{\Delta\backslash \{\alpha_6\}}$  are obtained by Weyl reflection on the two weights.\\

For $E_8$, we start with the set  ${\cal W}_{\cal S}$:
\begin{align*}
\omega_1&\equiv[0,0,0,0,0,0,\phantom{-}1,0]=-w_7+4\alpha_1+8\alpha_2+12\alpha_3+10\alpha_4+8\alpha_5+6\alpha_6+4\alpha_7+6\alpha_8\\
\omega_2&\equiv[0,0,0,0,0,0,-1,0]=-w_7
\end{align*}
This defines a 2d theory (shown at the bottom of Figure \ref{fig:EnExamples}). One checks at once from the positive roots that ${\cal W}_{\cal S}$ characterizes the nilradical  $\fn_{\Delta\backslash \{\alpha_7\}}$, so the associated parabolic subalgebra is $\fp_{\Delta\backslash \{\alpha_7\}}$.
In fact, no other set  ${\cal W}_{\cal S}$  is associated to this parabolic subalgebra. The level 1 null state condition in the $E_8$-Toda CFT is:
\[
\langle\beta,\alpha_i\rangle=0,\qquad\qquad i=1,2,3,4,5,6,8.
\]
All the other possible  sets ${\cal W}_{\cal S}$ that are associated to $\fp_{\Delta\backslash \{\alpha_7\}}$  are obtained by Weyl reflection on the two weights.

\subsection{Unpolarized Theories}
\label{sec:unpol}
Here we give some examples of unpolarized theories for $D_n$ and $E_n$ only, since there is no such theory for $A_n$.

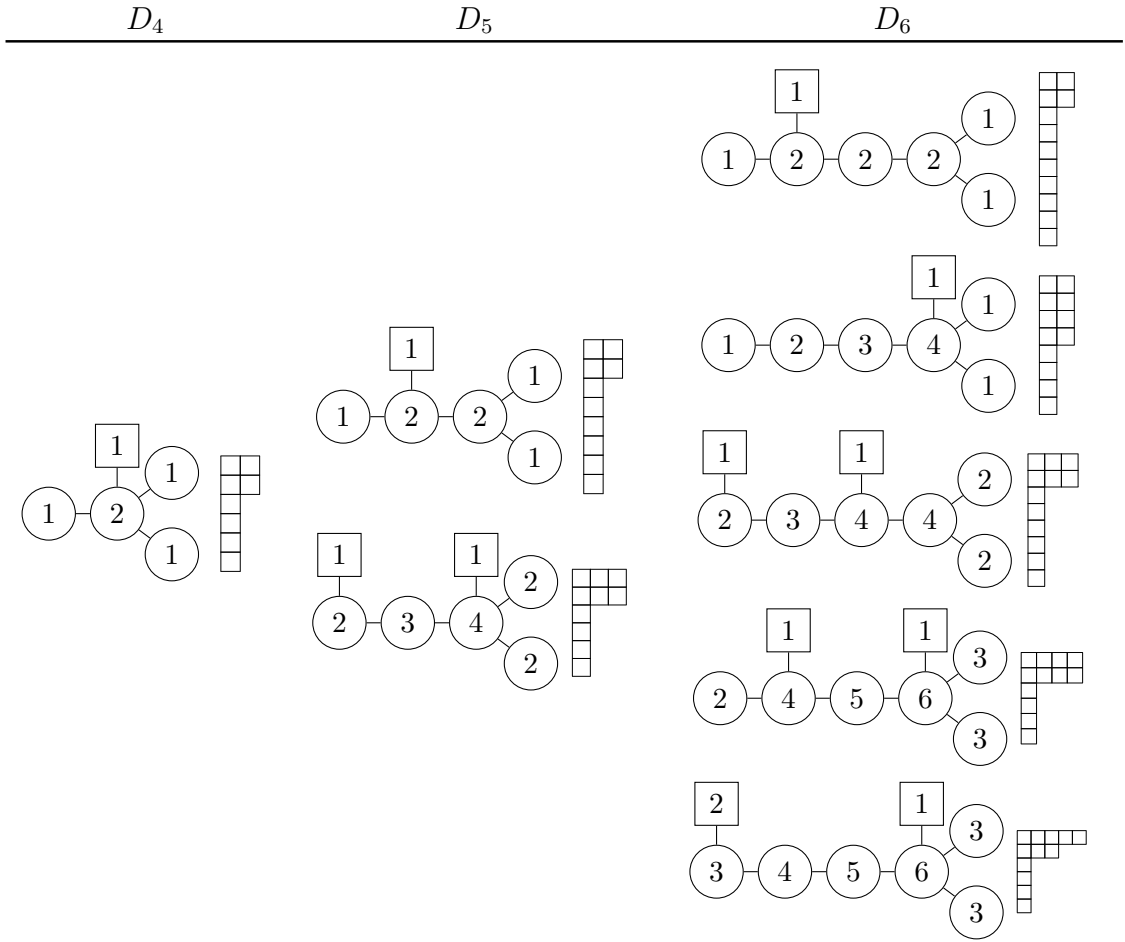
\begin{figure}[htpb]
	\begin{center}
	\ytableausetup{smalltableaux}
\begin{tabular}{ccc}
		$D_4$ & $D_5$ & $D_6$\\
		\toprule
		\begin{tikzpicture}[baseline,font=\small]
 \begin{scope}[auto, every node/.style={minimum size=0.6cm}]
\def \spac {0.9cm}

\node[circle, draw](k1) at (0,0) {$1$};
\node[circle, draw](k2) at (1*\spac,0) {$2$};
\node[circle, draw](k3) at (1.8*\spac,0.6*\spac) {$1$};
\node[circle, draw](k4) at (1.8*\spac,-0.6*\spac) {$1$};

\node[draw, inner sep=0.1cm,minimum size=0.57cm](N2) at (1*\spac,\spac) {$1$};

\draw[-] (k1) to (k2);
\draw[-] (k2) to (k3);
\draw[-] (k2) to (k4);

\draw (k2) -- (N2);

\node at (2.8*\spac,0) {\resizebox{0.5cm}{!}{\ydiagram{2,2,1,1,1,1}}};

\end{scope}
\end{tikzpicture}&
\begin{tikzpicture}[baseline]
\node at (0,1.3) {\begin{tikzpicture}[baseline,font=\small]
 \begin{scope}[auto, every node/.style={minimum size=0.6cm}]
\def \spac {0.9cm}

\node[circle, draw](k1) at (0,0) {$1$};
\node[circle, draw](k2) at (1*\spac,0) {$2$};
\node[circle, draw](k3) at (2*\spac,0) {$2$};
\node[circle, draw](k4) at (2.8*\spac,0.6*\spac) {$1$};
\node[circle, draw](k5) at (2.8*\spac,-0.6*\spac) {$1$};

\node[draw, inner sep=0.1cm,minimum size=0.57cm](N2) at (1*\spac,\spac) {$1$};

\draw[-] (k1) to (k2);
\draw[-] (k2) to (k3);
\draw[-] (k3) to (k4);
\draw[-] (k3) to (k5);

\draw (k2) -- (N2);

\node at (3.8*\spac,0) {\resizebox{0.5cm}{!}{\ydiagram{2,2,1,1,1,1,1,1}}};

\end{scope}
\end{tikzpicture}};
\node at (0,-1.3) {\begin{tikzpicture}[baseline,font=\small]
 \begin{scope}[auto, every node/.style={minimum size=0.6cm}]
\def \spac {0.9cm}

\node[circle, draw](k1) at (0,0) {$2$};
\node[circle, draw](k2) at (1*\spac,0) {$3$};
\node[circle, draw](k3) at (2*\spac,0) {$4$};
\node[circle, draw](k4) at (2.8*\spac,0.6*\spac) {$2$};
\node[circle, draw](k5) at (2.8*\spac,-0.6*\spac) {$2$};

\node[draw, inner sep=0.1cm,minimum size=0.57cm](N1) at (0*\spac,\spac) {$1$};
\node[draw, inner sep=0.1cm,minimum size=0.57cm](N3) at (2*\spac,\spac) {$1$};

\draw[-] (k1) to (k2);
\draw[-] (k2) to (k3);
\draw[-] (k3) to (k4);
\draw[-] (k3) to (k5);

\draw (k1) -- (N1);
\draw (k3) -- (N3);

\node at (3.8*\spac,0) {\resizebox{0.7cm}{!}{\ydiagram{3,3,1,1,1,1}}};

\end{scope}
\end{tikzpicture}};
\end{tikzpicture}
&\begin{tikzpicture}[baseline]
\node at (0,4.7) {\begin{tikzpicture}[baseline,font=\small]
 \begin{scope}[auto, every node/.style={minimum size=0.6cm}]
\def \spac {0.9cm}

\node[circle, draw](k1) at (0,0) {$1$};
\node[circle, draw](k2) at (1*\spac,0) {$2$};
\node[circle, draw](k3) at (2*\spac,0) {$2$};
\node[circle, draw](k4) at (3*\spac,0) {$2$};
\node[circle, draw](k5) at (3.8*\spac,0.6*\spac) {$1$};
\node[circle, draw](k6) at (3.8*\spac,-0.6*\spac) {$1$};

\node[draw, inner sep=0.1cm,minimum size=0.57cm](N2) at (1*\spac,\spac) {$1$};

\draw[-] (k1) to (k2);
\draw[-] (k2) to (k3);
\draw[-] (k3) to (k4);
\draw[-] (k4) to (k5);
\draw[-] (k4) to (k6);

\draw (k2) -- (N2);

\node at (4.8*\spac,0) {\resizebox{0.45cm}{!}{\ydiagram{2,2,1,1,1,1,1,1,1,1}}};

\end{scope}
\end{tikzpicture}};
\node at (0,2.3) {\begin{tikzpicture}[baseline,font=\small]
 \begin{scope}[auto, every node/.style={minimum size=0.6cm}]
\def \spac {0.9cm}

\node[circle, draw](k1) at (0,0) {$1$};
\node[circle, draw](k2) at (1*\spac,0) {$2$};
\node[circle, draw](k3) at (2*\spac,0) {$3$};
\node[circle, draw](k4) at (3*\spac,0) {$4$};
\node[circle, draw](k5) at (3.8*\spac,0.6*\spac) {$1$};
\node[circle, draw](k6) at (3.8*\spac,-0.6*\spac) {$1$};

\node[draw, inner sep=0.1cm,minimum size=0.57cm](N4) at (3*\spac,\spac) {$1$};

\draw[-] (k1) to (k2);
\draw[-] (k2) to (k3);
\draw[-] (k3) to (k4);
\draw[-] (k4) to (k5);
\draw[-] (k4) to (k6);

\draw (k4) -- (N4);

\node at (4.8*\spac,0) {\resizebox{0.45cm}{!}{\ydiagram{2,2,2,2,1,1,1,1}}};

\end{scope}
\end{tikzpicture}};
\node at (0,0) {\begin{tikzpicture}[baseline,font=\small]
 \begin{scope}[auto, every node/.style={minimum size=0.6cm}]
\def \spac {0.9cm}

\node[circle, draw](k1) at (0,0) {$2$};
\node[circle, draw](k2) at (1*\spac,0) {$3$};
\node[circle, draw](k3) at (2*\spac,0) {$4$};
\node[circle, draw](k4) at (3*\spac,0) {$4$};
\node[circle, draw](k5) at (3.8*\spac,0.6*\spac) {$2$};
\node[circle, draw](k6) at (3.8*\spac,-0.6*\spac) {$2$};

\node[draw, inner sep=0.1cm,minimum size=0.57cm](N1) at (0*\spac,\spac) {$1$};
\node[draw, inner sep=0.1cm,minimum size=0.57cm](N3) at (2*\spac,\spac) {$1$};

\draw[-] (k1) to (k2);
\draw[-] (k2) to (k3);
\draw[-] (k3) to (k4);
\draw[-] (k4) to (k5);
\draw[-] (k4) to (k6);

\draw (k1) -- (N1);
\draw (k3) -- (N3);

\node at (4.8*\spac,0) {\resizebox{0.65cm}{!}{\ydiagram{3,3,1,1,1,1,1,1}}};

\end{scope}
\end{tikzpicture}};
\node at (0,-2.3) {\begin{tikzpicture}[baseline,font=\small]
 \begin{scope}[auto, every node/.style={minimum size=0.6cm}]
\def \spac {0.9cm}

\node[circle, draw](k1) at (0,0) {$2$};
\node[circle, draw](k2) at (1*\spac,0) {$4$};
\node[circle, draw](k3) at (2*\spac,0) {$5$};
\node[circle, draw](k4) at (3*\spac,0) {$6$};
\node[circle, draw](k5) at (3.8*\spac,0.6*\spac) {$3$};
\node[circle, draw](k6) at (3.8*\spac,-0.6*\spac) {$3$};

\node[draw, inner sep=0.1cm,minimum size=0.57cm](N2) at (1*\spac,\spac) {$1$};
\node[draw, inner sep=0.1cm,minimum size=0.57cm](N4) at (3*\spac,\spac) {$1$};

\draw[-] (k1) to (k2);
\draw[-] (k2) to (k3);
\draw[-] (k3) to (k4);
\draw[-] (k4) to (k5);
\draw[-] (k4) to (k6);

\draw (k2) -- (N2);
\draw (k4) -- (N4);

\node at (4.85*\spac,0) {\resizebox{0.8cm}{!}{\ydiagram{4,4,1,1,1,1}}};

\end{scope}
\end{tikzpicture}};
\node at (0,-4.6) {\begin{tikzpicture}[baseline,font=\small]
 \begin{scope}[auto, every node/.style={minimum size=0.6cm}]
\def \spac {0.9cm}

\node[circle, draw](k1) at (0,0) {$3$};
\node[circle, draw](k2) at (1*\spac,0) {$4$};
\node[circle, draw](k3) at (2*\spac,0) {$5$};
\node[circle, draw](k4) at (3*\spac,0) {$6$};
\node[circle, draw](k5) at (3.8*\spac,0.6*\spac) {$3$};
\node[circle, draw](k6) at (3.8*\spac,-0.6*\spac) {$3$};

\node[draw, inner sep=0.1cm,minimum size=0.57cm](N1) at (0*\spac,\spac) {$2$};
\node[draw, inner sep=0.1cm,minimum size=0.57cm](N4) at (3*\spac,\spac) {$1$};

\draw[-] (k1) to (k2);
\draw[-] (k2) to (k3);
\draw[-] (k3) to (k4);
\draw[-] (k4) to (k5);
\draw[-] (k4) to (k6);

\draw (k1) -- (N1);
\draw (k4) -- (N4);

\node at (4.9*\spac,0) {\resizebox{0.9cm}{!}{\ydiagram{5,3,1,1,1,1}}};

\end{scope}
\end{tikzpicture}};
\end{tikzpicture}
		\end{tabular}
	\end{center}
	\caption{Exhaustive list of unpolarized quiver gauge theories $T^{2d}$ for $D_4$, $D_5$, and $D_6$. The nilpotent orbit in the classification of \cite{Chacaltana:2011ze} is also written for reference.}
	\label{fig:Dntype2}
\end{figure}

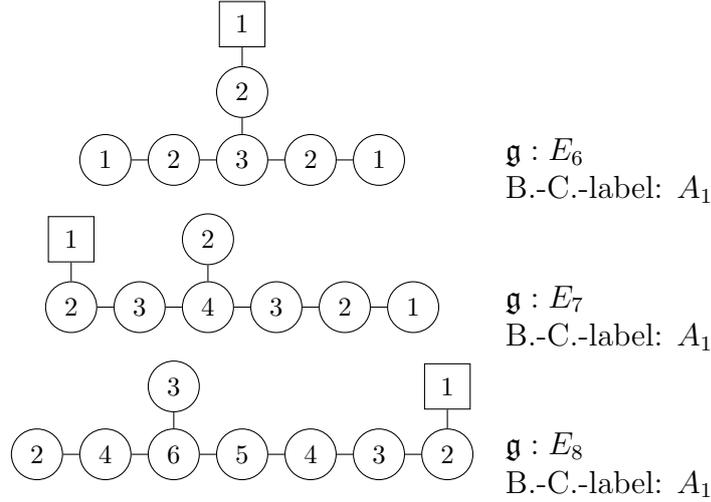
\begin{figure}[htpb]
	\begin{center}
	\renewcommand{\arraystretch}{1.3}
\begin{tabular}{cp{5cm}}
\begin{tikzpicture}[baseline=0pt,font=\footnotesize]
 \begin{scope}[auto, every node/.style={minimum size=0.5cm}]
\def \spac {0.9cm}

\node[circle, draw](k1) at (0,0) {$1$};
\node[circle, draw](k2) at (1*\spac,0) {$2$};
\node[circle, draw](k3) at (2*\spac,0*\spac) {$3$};
\node[circle, draw](k4) at (3*\spac,0*\spac) {$2$};
\node[circle, draw](k5) at (4*\spac,0*\spac) {$1$};
\node[circle, draw](k6) at (2*\spac,1*\spac) {$2$};

\node[draw, inner sep=0.1cm,minimum size=0.6cm](N6) at (2*\spac,2*\spac) {$1$};

\draw[-] (k1) to (k2);
\draw[-] (k2) to (k3);
\draw[-] (k3) to (k4);
\draw[-] (k4) to (k5);
\draw[-] (k3) to (k6);

\draw (k6) -- (N6);

\end{scope}
\end{tikzpicture}&$\fg: E_6$\newline B.-C.-label: $A_1$\\
\begin{tikzpicture}[baseline=0pt,font=\footnotesize]
 \begin{scope}[auto, every node/.style={minimum size=0.5cm}]
\def \spac {0.9cm}

\node[circle, draw](k1) at (0,0) {$2$};
\node[circle, draw](k2) at (1*\spac,0) {$3$};
\node[circle, draw](k3) at (2*\spac,0*\spac) {$4$};
\node[circle, draw](k4) at (3*\spac,0*\spac) {$3$};
\node[circle, draw](k5) at (4*\spac,0*\spac) {$2$};
\node[circle, draw](k6) at (5*\spac,0*\spac) {$1$};
\node[circle, draw](k7) at (2*\spac,1*\spac) {$2$};

\node[draw, inner sep=0.1cm,minimum size=0.6cm](N1) at (0*\spac,\spac) {$1$};

\draw[-] (k1) to (k2);
\draw[-] (k2) to (k3);
\draw[-] (k3) to (k4);
\draw[-] (k4) to (k5);
\draw[-] (k5) to (k6);
\draw[-] (k3) to (k7);

\draw (k1) -- (N1);

\end{scope}
\end{tikzpicture}&$\fg: E_7$\newline B.-C.-label: $A_1$\\
\begin{tikzpicture}[baseline=0pt,font=\footnotesize]
 \begin{scope}[auto, every node/.style={minimum size=0.5cm}]
\def \spac {0.9cm}

\node[circle, draw](k1) at (0,0) {$2$};
\node[circle, draw](k2) at (1*\spac,0) {$4$};
\node[circle, draw](k3) at (2*\spac,0*\spac) {$6$};
\node[circle, draw](k4) at (3*\spac,0*\spac) {$5$};
\node[circle, draw](k5) at (4*\spac,0*\spac) {$4$};
\node[circle, draw](k6) at (5*\spac,0*\spac) {$3$};
\node[circle, draw](k7) at (6*\spac,0*\spac) {$2$};
\node[circle, draw](k8) at (2*\spac,1*\spac) {$3$};

\node[draw, inner sep=0.1cm,minimum size=0.6cm](N7) at (6*\spac,\spac) {$1$};

\draw[-] (k1) to (k2);
\draw[-] (k2) to (k3);
\draw[-] (k3) to (k4);
\draw[-] (k4) to (k5);
\draw[-] (k5) to (k6);
\draw[-] (k6) to (k7);
\draw[-] (k3) to (k8);

\draw (k7) -- (N7);

\end{scope}
\end{tikzpicture}&$\fg: E_8$\newline B.-C.-label: $A_1$
\end{tabular}
\end{center}
	\caption{Examples of unpolarized quiver gauge theories for $E_n$. The ones shown here have the smallest Coulomb branch dimension. The Bala Carter label $A_1$ in the defect classification of  \cite{Chacaltana:2012zy} is also written for reference.}
	\label{fig:Entype2}
\end{figure}

The simplest case of an unpolarized quiver gauge theory arises when only a single fundamental hypermultiplet is present, so there is only one mass. The corresponding weight is then the null weight, which is obviously not in the orbit of any fundamental weight. For instance, such a scenario occurs for the unique unpolarized theory of $D_4$, where the weight $[0,0,0,0]$ is indeed in the second fundamental representation; see Figure \ref{fig:Dntype2} for examples in the $D_n$ case, and Figure \ref{fig:Entype2} for examples in the $E_n$ case.\\

As explained in section \ref{sec:types}, unpolarized theories can also have more than one weight: for example, looking at $D_5$, it is possible to choose weights in the third fundamental representation that actually belong to the orbit of the first fundamental weight instead. One can then construct the bottom $D_5$ quiver of Figure \ref{fig:Dntype2}. An example of two weights  that make up such a quiver is $[1,0,0,0,0]$, chosen in the first fundamental representation, and $[-1,0,0,0,0]$, chosen in the third fundamental representation. If one wishes, it is always possible to flow on the Higgs branch and make these defects polarized, \ref{sec:types}.

\section*{Acknowledgements}
We first want to thank Mina Aganagic for her guidance and insights throughout this project. We also thank Aswin Balasubramanian, Oscar Chacaltana, Sergey Cherkis, Jacques Distler, Amihay Hanany, Peter Koroteev,  Noppadol Mekareeya, Hiraku Nakajima, Shamil Shakirov and Alex Takeda for their time to discuss various points and their willingness to answer our questions. The research of N. H. and C. S. is supported in part by the Berkeley Center for Theoretical
Physics,  by  the  National  Science  Foundation  (award PHY-1521446) and by the
US Department of Energy under Contract DE-AC02-05CH11231.
\newpage
\mciteSetMidEndSepPunct{}{}{} 
\bibliography{summary}
\bibliographystyle{utphysmcite}

\end{document}